\documentclass[
    aps,
    prd,
    twocolumn,
    superscriptaddress,
    nofootinbib,
    longbibliography
]{revtex4-2}
\usepackage{amsmath,amssymb,amsthm}
\usepackage{mathtools}
\usepackage{physics}
\usepackage{bm}
\usepackage{tensor}
\usepackage{xspace}

\usepackage{graphicx}
\usepackage{booktabs}
\usepackage{array}
\usepackage{hyperref}
\usepackage{float}
\usepackage{comment}

\begin{document}

\title{
Gravitational Wave Parity-Violating Strain from Pulsar\\
Glitches in Chern-Simons Modified Gravity
}

\author{Abhishek Rout}
\email{arout@email.sc.edu}

\author{Brett Altschul}
\email{altschul@mailbox.sc.edu}

\affiliation{Department of Physics and Astronomy, University of South Carolina, Columbia, South Carolina 29208, USA}

\date{\today}

\begin{abstract}
We investigate gravitational-wave birefringence from pulsar glitches in dynamical Chern-Simons gravity, a parity-violating extension of general relativity motivated by string theory and quantum gravity. Using the Hartle-Thorne slow-rotation formalism to model the neutron star background and a perturbative treatment of the Chern-Simons coupling \(\alpha\), we derive the modified Regge-Wheeler equation governing axial gravitational perturbations and compute the resulting polarization-dependent phase shift. The parity-violating interaction produces a fractional asymmetry between the right- and left-handed circular polarizations that scales as \(\alpha^2\), grows linearly with frequency and rotation rate, and falls as the fourth power of the stellar radius. For a constant-density interior model we obtain an analytic matching solution for the background scalar field, and show that the interior curvature factor vanishes identically---the interior Tolman-Oppenheimer-Volkoff solution being conformally flat---so that the scalar dipole is sourced entirely in the vacuum exterior; a centrally condensed equation of state can only increase the predicted signal. We further extend the analysis to a two-fluid model incorporating differential rotation between the neutron superfluid and the charged component; the resulting correction is bounded by the fractional glitch amplitude itself, independently of the equation of state, and is negligible for typical pulsar parameters. Expressed through the dimensionless quantity \(\zeta \propto \alpha M/R_\star^{3}\), the fractional polarization asymmetry for millisecond pulsars reaches \(\sim 10^{-3}\) at kilohertz frequencies at the boundary of perturbative validity---four orders of magnitude above the sensitivity of current ground-based detectors to strain asymmetries. However, realistic bounds are limitated by the signal-to-noise ratio required to resolve a ratio of gravitational wave strains. The only existing bound on the dynamical theory, from terrestrial frame-dragging, lies thirteen orders of magnitude above the regime in which the effective theory is controlled; a measurement sensitive to \(\zeta \lesssim 1\) would improve it by some seven orders of magnitude in \(\sqrt{\alpha}\) and would be the first constraint on dynamical Chern-Simons gravity obtained in a regime where it admits a controlled perturbative description.
\end{abstract}

\maketitle
\thispagestyle{plain}

\section{Introduction}
\label{sec:introduction}

Parity-violating extensions of general relativity arise naturally in
high-energy frameworks including string theory~\cite{Polchinski1998,
Green1987}, loop quantum gravity~\cite{Ashtekar1989, Mercuri2006}, and
effective field theories of inflation~\cite{Weinberg2008}. One
well-studied realization is dynamical Chern-Simons (CS) gravity
\cite{Jackiw2003, Alexander2009}, in which a pseudoscalar field couples
to the Pontryagin density of the spacetime curvature. This coupling
introduces parity violation in the gravitational sector while preserving
diffeomorphism invariance, making CS gravity a theoretically constrained
and observationally accessible framework for probing gravitational parity
symmetry.

The CS modification becomes active in rotating spacetimes. There, the
Pontryagin density $^{\ast}\!RR = {}^{\ast}R^{\mu\nu}_{\ \ \rho\sigma}
R^{\rho\sigma}_{\ \ \mu\nu}$ becomes nonvanishing and sources a dynamical
scalar field, which back-reacts on the metric through the Cotton tensor
in the modified Einstein equations. The result is a rich phenomenology
that includes modifications to the Kerr metric~\cite{Yunes2009}, the
structure of slowly rotating compact objects~\cite{Ali-Haimoud2011},
binary pulsar dynamics~\cite{Yunes2009b}, and gravitational
wave (GW) propagation~\cite{Alexander2008, Yunes2010}.

Among the most striking observational signatures of parity violation in gravity is the \textit{birefringence} of GWs. It is important to distinguish between two distinct manifestations of this phenomenon. In the \textit{non-dynamical} CS theory, where the scalar field is fixed, propagating GWs experience \textit{amplitude birefringence} \cite{Alexander2005}: one circular polarization is parametrically amplified while the other is suppressed as a function of frequency, but the two modes do not acquire a relative phase shift. In contrast, the \textit{dynamical} theory---which is the focus of this work---features \textit{phase birefringence}. Here, the scalar field is sourced dynamically by the spacetime curvature and is not a constant background; its gradient acts as an effective medium that modifies the propagation of GWs locally. Consequently, right- and left-circularly polarized components acquire frequency-dependent, opposite phase shifts, leading to a polarization asymmetry in the waveform. For compact binary coalescences, this phase birefringence has been extensively studied~\cite{Yagi2012}. Most recently, GW events from stellar-mass black hole binaries were used to bound the scalar field evolution in CS gravity, and it is projected that third-generation detectors such as Cosmic Explorer (CE) and the Einstein Telescope (ET) could improve these bounds significantly~\cite{Yagi2018}.

It is worth emphasising at the outset how weakly the dynamical theory is presently constrained. Binary pulsar timing has provided stringent limits on the \textit{non-dynamical} formulation, bounding the time derivative of the externally prescribed scalar~\cite{Yunes2009b}; but as stressed in that work, these do not carry over to the dynamical theory, in which the Pontryagin density vanishes to leading post-Newtonian order. This is the same statement, applied to a binary, as the vanishing of the Pontryagin density for spherically symmetric spacetimes established in section~\ref{sec:pontryagin}: the leading post-Newtonian metric of a binary is built from spherically symmetric potentials, and a nonzero Pontryagin density requires the individual bodies' rotation. The near zone of a single rapidly rotating star, where the gravitomagnetic field is generated directly rather than at high post-Newtonian order, is correspondingly the most favourable environment for the effect. The only current astrophysical bound on dynamical CS gravity comes instead from measurements of frame-dragging around the Earth by Gravity Probe B and the LAGEOS satellites~\cite{Ali-Haimoud2011}, and is weak enough that, when translated to stellar densities, it does not reach the regime in which the effective theory is under control. Taken together, these works demonstrate that CS gravity can be tested through both source dynamics and wave propagation, but leave the strong-field, stellar-density regime essentially unexplored. The possibility that transient neutron-star phenomena such as pulsar glitches may generate a distinct birefringent signal has not been examined in detail. 

Pulsar glitches---sudden spin-up events observed in many rotation-powered
pulsars, including Vela and the Crab~\cite{Espinoza2011,
Antonopoulou2018}---are thought to result from angular momentum transfer
between the neutron superfluid in the stellar interior and the charged
crust~\cite{Anderson1975, Alpar1984}. These events can excite the
fundamental quadrupole ($f$-) mode of the star, producing a burst of GWs
in the kilohertz band~\cite{Owen2010, Warszawski2012}. Unlike the
continuous, nearly monochromatic GWs expected from steadily rotating
neutron stars (e.g., from mountain asymmetries or r-mode oscillations), a
glitch produces a transient burst. This time-domain localization is
useful: the birefringent phase shift manifests as a temporary separation
of the two circular polarizations during the rapid spin-up and relaxation
phases, providing a signature that is localized in time and therefore
distinguishable from instrumental systematics. While the expected GW
amplitudes from individual glitches are small, rapidly rotating
millisecond pulsars and nearby glitching sources such as Vela provide
plausible targets for next-generation detectors~\cite{Andersson2011,
Lasky2015}.

In this work, we investigate whether pulsar glitches can produce a
detectable parity-violating signature in CS gravity.
Specifically, we compute the birefringent phase shift accumulated by GWs
emitted during a glitch as they propagate through the dipolar scalar
field sourced by the rotating star. We work within the Hartle-Thorne
slow-rotation formalism~\cite{Hartle1967, Hartle1968} to describe the
stellar background, treat the CS coupling perturbatively, and
match interior and exterior solutions to determine the scalar field
amplitude.

We derive a closed-form expression for the fractional polarization
asymmetry, which scales as $\alpha^{2}$, increases linearly with GW
frequency and stellar rotation rate, and decreases as the fourth power of
the stellar radius at fixed mass. The steep compactness dependence has a
direct origin: for a constant-density interior the curvature factor
multiplying the Pontryagin source vanishes identically---the interior
Tolman-Oppenheimer-Volkoff (TOV) solution being conformally flat---so that the scalar dipole
is sourced entirely in the vacuum exterior, where the Pontryagin density
falls as $r^{-7}$. A realistic, centrally condensed equation of state
restores a positive interior contribution, so that the constant-density
result is a conservative one. We then extend the analysis to a two-fluid
model incorporating differential rotation between the neutron superfluid
and the charged component; the resulting correction is bounded by the
fractional glitch amplitude itself, independently of the equation of
state, and is negligible for observed glitches. Finally, we provide
projected detectability estimates. We find that the predicted signal is
substantially larger than existing constraints on the theory might
suggest---reaching $\mathcal{A}_{\rm PV} \sim 10^{-3}$ at the boundary of
perturbative validity---but that resolving a \textit{ratio} of strains
imposes a demanding signal-to-noise ration (SNR) requirement, which rather than the
size of the effect is what limits a measurement.

The paper is organized as follows. Section~\ref{sec:action} presents the
action and field equations of dynamical CS gravity.
Section~\ref{sec:metric} describes the slowly rotating Hartle-Thorne
background and its curvature. Section~\ref{sec:perturbations} sets up the
two-parameter perturbation framework and the glitch excitation model. The
Pontryagin density is evaluated in section~\ref{sec:pontryagin}, and the
background scalar field is solved in section~\ref{sec:scalar}.
Section~\ref{sec:gw_propagation} derives the modified Regge-Wheeler
equation and the resulting birefringent phase shift. Two interior
models---a constant-density star and a two-fluid star with differential
rotation---are presented in section~\ref{sec:interior_models}, and
observational prospects are discussed in
section~\ref{sec:results_discussion}. Technical details, including
explicit Christoffel symbols, the full Pontryagin expansion, and the
matching conditions for the scalar field, are collected in the
appendices. Throughout, we adopt units with $G = c = 1$ and use the
metric signature $(-,+,+,+)$.

\section{Action and Field Equations}
\label{sec:action}

\subsection{Action}

We consider a gravitational system consisting of a rotating neutron star, an electromagnetic field, and a dynamical CS
scalar field. In units $G=c=1$, the action takes the form
\begin{align}
S &= \frac{1}{16\pi} \int d^4x \sqrt{-g} \Bigg[R + \frac{\alpha}{4}\,\vartheta\, {}^{\ast}RR - \frac{1}{2}\nabla_\mu \vartheta \nabla^\mu\vartheta \nonumber\\
&\quad - V(\vartheta) - \frac{1}{4}F_{\mu\nu}F^{\mu\nu} -\frac{\beta}{4}\vartheta F_{\mu\nu}{}^{\ast}F^{\mu\nu} +\mathcal{L}_{\rm m}\Bigg],
\label{eq:action}
\end{align}
where $R$ is the Ricci scalar associated with the spacetime metric $g_{\mu\nu}$, $\vartheta$ is the dynamical CS
scalar field, $\alpha$ is the gravitational CS coupling constant, $F_{\mu\nu}$ is the electromagnetic field
tensor, and $\mathcal{L}_{\rm m}$ denotes the matter Lagrange density describing the neutron-star interior.

The dual electromagnetic tensor is defined by
\begin{equation}
{}^{\ast}{F}^{\mu\nu} = \frac{1}{2} \epsilon^{\mu\nu\gamma\delta} F_{\gamma\delta},
\label{eq:dual_em}
\end{equation}
while the parity-violating structure also enters through the Pontryagin density
\begin{equation}
{}^{\ast}RR = {}^{\ast}R^{\sigma}{}_{\tau\mu\nu} R^{\tau}{}_{\sigma}{}^{\mu\nu},
\label{eq:pontryagin}
\end{equation}
where the dual Riemann tensor is
\begin{equation}
{}^{\ast}R^{\sigma}{}_{\tau\mu\nu} = \frac{1}{2} \epsilon_{\mu\nu\gamma\delta} R^{\sigma}{}_{\tau}{}^{\gamma\delta}.
\label{eq:dual_riemann}
\end{equation}
The Levi-Civita tensor satisfies $\epsilon^{0123}=+1/\sqrt{-g}$ in our conventions.

If the parity-violating electromagnetic coupling $\beta$ is nonzero, then the field $\vartheta$ represents
an axion-like particle.
However, in the present work we are primarily interested in parity violation in the gravitational sector.
We therefore set $\beta=0$, neglecting parity-violating electromagnetic effects while retaining standard Maxwell dynamics.

\subsection{Modified Einstein Equations}

Varying the action \eqref{eq:action} with respect to the metric $g_{\mu\nu}$ yields the modified Einstein equations
\begin{equation}
G_{\mu\nu} + \alpha C_{\mu\nu} = 8\pi \left[ T^{(\vartheta)}_{\mu\nu} + T^{(\rm EM)}_{\mu\nu} + T^{(\rm m)}_{\mu\nu} \right],
\label{eq:einstein}
\end{equation}
where $G_{\mu\nu}$ is the Einstein tensor and $C_{\mu\nu}$ is the Cotton tensor associated with the CS
modification which encodes the parity-violating correction and is given by
\begin{align}
C_{\mu\nu} = -\frac{1}{2} \Big[ & (\nabla_\sigma \vartheta) \epsilon^{\sigma\delta\gamma}{}_{(\mu} \nabla_\gamma R_{\nu)\delta} \nonumber\\
& + (\nabla_\sigma\nabla_\delta \vartheta) {}^{\ast}R^{\delta}{}_{(\mu\nu)}{}^{\sigma} \Big],
\label{eq:cotton}
\end{align}
where parentheses denote symmetrization with an averaging factor---that is, $A_{(\mu\nu)} = \frac{1}{2}(A_{\mu\nu} + A_{\nu\mu})$, and $A_{(\mu\nu\rho)}$ denotes symmetrization over all three indices with a factor of $1/6$. 

The scalar-field stress-energy tensor is
\begin{equation}
T^{(\vartheta)}_{\mu\nu} = \nabla_\mu \vartheta \nabla_\nu \vartheta - \frac{1}{2} g_{\mu\nu} \nabla_\alpha \vartheta \nabla^\alpha \vartheta - g_{\mu\nu} V(\vartheta),
\label{eq:stress_scalar}
\end{equation}
while the electromagnetic contribution is
\begin{equation}
T^{(\rm EM)}_{\mu\nu} = F_{\mu\gamma} F_{\nu}{}^{\gamma} - \frac{1}{4} g_{\mu\nu} F_{\gamma\delta} F^{\gamma\delta}.
\label{eq:stress_em}
\end{equation}

\subsection{Scalar Field Equation}

Variation of the action \eqref{eq:action} with respect to the scalar field $\vartheta$ gives
\begin{equation}
\square \vartheta - V'(\vartheta) = -\frac{\alpha}{4} {}^{\ast}RR + \frac{\beta}{4} F_{\mu\nu} {}^{\ast}F^{\mu\nu},
\label{eq:scalar}
\end{equation}
where $\square \equiv \nabla_\mu \nabla^\mu$ is the d'Alembertian.

The Pontryagin density ${}^{\ast}RR$ therefore acts as a source for the scalar field. In stationary, nonrotating
spacetimes, ${}^{\ast}RR$ vanishes identically, because the Riemann tensor possesses no parity-odd structure.
In rotating geometries, however, the combination of curvature and the preferred direction set by the rotation axis
yields a nonzero Pontryagin density, which in turn generates a nontrivial scalar background. This is why rotating
neutron stars and, by extension, glitching pulsars are natural laboratories for testing CS gravity.

The electromagnetic Pontryagin density $F_{\mu\nu}{}^{\ast}{F}^{\mu\nu}$ provides an additional source proportional
to $\beta$. For a rotating magnetized neutron star, this term is also nonvanishing and scales with both the magnetic
field strength and the rotation rate. Since our focus on the gravitational source, we set $\beta = 0$ and defer
study of the combined system to future work.

For the scalar field to remain long-ranged, we assume a vanishing potential $V(\vartheta) = 0$,
corresponding to a massless scalar field. This
is the simplest choice consistent with the absence of a natural mass scale in the CS term; a massive scalar would
introduce a Yukawa suppression that further reduces the already small birefringent signal. Thus, our assumption
$V(\vartheta) = 0$ represents the most optimistic scenario for detectability.

\subsection{Modified Maxwell Equations}

Variation of the action \eqref{eq:action} with respect to $A_\mu$ gives
\begin{equation}
\nabla_\mu F^{\mu\nu} = J^\nu + \beta (\nabla_\mu \vartheta) \left({}^{\ast}F^{\mu\nu}\right).
\label{eq:maxwell}
\end{equation}
With $\beta = 0$, this reduces to the standard Maxwell equations $\nabla_\mu F^{\mu\nu} = J^\nu$, so
electromagnetic effects in the star are unchanged.

Equations~\eqref{eq:einstein}--\eqref{eq:maxwell} form the starting point for our analysis of GWs emitted by glitching pulsars in dynamical CS gravity.

\section{Background Metric: Slowly Rotating Neutron Star}
\label{sec:metric}

To describe the spacetime surrounding a rotating neutron star we adopt the Hartle-Thorne slow-rotation formalism~\cite{Hartle1967, Hartle1968}. This approach provides an accurate description of the exterior gravitational field of a compact object whose angular velocity is small compared with the Keplerian breakup limit. The metric is expanded in powers of the stellar angular velocity $\Omega$, and for the purposes of this work it is sufficient to retain only terms up to $\mathcal{O}(\Omega)$.

\subsection{Hartle-Thorne Metric}

To first order in the rotation parameter, the spacetime metric may be written as
\begin{equation}
    \begin{aligned}
        ds^2 = -A(r)\,dt^2 &+ B(r)\,dr^2 + r^2\left(d\theta^2 + \sin^2\theta\,d\phi^2\right)\\
        &- 2\,\omega(r)\,r^2\sin^2\theta\,dt\,d\phi + \mathcal{O}(\Omega^2),
    \end{aligned}
\label{eq:HTmetric}
\end{equation}
where
\begin{equation}
A(r)=e^{2\Phi(r)}, \qquad B(r)=e^{2\Lambda(r)}.
\label{eq:metric_functions}
\end{equation}
The functions $\Phi(r)$ and $\Lambda(r)$ are determined by the stellar structure equations inside the star and reduce to the Schwarzschild solution in the exterior region.

The function $\omega(r)$ represents the frame-dragging angular velocity of local inertial frames. Physically,
it describes the dragging of spacetime induced by stellar rotation. In the slow-rotation approximation, $\omega(r)$
satisfies a linear second-order differential equation. To first order in $\Omega$, the field equations reduce to
the following ordinary differential equation (ODE) for the frame-dragging function:
\begin{equation}
\frac{1}{r^4}\frac{d}{dr}\left[r^4 j(r) \frac{d\bar{\omega}}{dr}\right] + \frac{4}{r}\frac{dj}{dr}\bar{\omega} = 0,
\label{eq:omega_ode}
\end{equation}
where $\bar{\omega}(r) = \Omega - \omega(r)$ and $j(r) = e^{-(\Phi+\Lambda)}$. This equation follows from the
$t\phi$-component of the Einstein equations and determines the radial profile of $\omega(r)$ given the background
metric functions $A(r)$ and $B(r)$.
In the next subsection we specialize to the exterior vacuum region, where the metric functions take the Schwarzschild
form and $\omega(r)$ admits a simple analytic solution.

Although rotation introduces the off-diagonal $g_{t\phi}$-component, the spacetime remains a perturbative deformation
of a spherically symmetric background, allowing the standard tensor-spherical-harmonic formalism to be employed in the
GW analysis.

\subsection{Exterior Solution}

Outside the neutron star ($r>R_\star$), the spacetime is vacuum, and the metric functions reduce to the Schwarzschild forms
\begin{equation}
A(r) = 1-\frac{2M}{r}, \qquad B(r)= \frac{1}{A(r)},
\end{equation}
where $M$ is the gravitational mass of the star.
In the slow-rotation limit the frame-dragging function satisfies a linear differential equation whose exterior solution is
\begin{equation}
\omega(r) = \frac{2J}{r^3},
\label{eq:omegaext}
\end{equation}
where $J$ is the total angular momentum of the star.
Substituting this result into eq.~\eqref{eq:HTmetric} yields the first-order Hartle-Thorne spacetime
describing the exterior region of a rotating neutron star.

\subsection{Slow-Rotation Expansion}
\label{sec:slow_rotation}

Throughout this work we assume that the dimensionless rotation parameter satisfies
\begin{equation}
\epsilon_{\Omega} \equiv \Omega R_\star \ll 1,
\end{equation}
so that higher-order rotational corrections may be neglected.
This approximation is well satisfied for most observed pulsars, including glitching systems such as the Vela and Crab
pulsars, whose rotation rates remain significantly below the breakup limit.

In this approximation the frame-dragging term proportional to $\omega(r)$ provides the leading rotational correction to the
background geometry. As we will show in the following sections, this term plays a crucial role in generating a nonvanishing
Pontryagin density ${}^{\ast}RR$, which in turn sources the CS scalar field. Consequently, the CS scalar field vanishes in
the nonrotating limit and is generated entirely by rotational corrections to the background spacetime.

\section{Perturbation Framework}
\label{sec:perturbations}

We now develop the perturbative framework used to describe GWs generated by pulsar glitches in dynamical 
CS gravity. The calculation involves two independent small parameters: the amplitude of the metric perturbation generated
by the glitch and the strength of the CS coupling.

\subsection{Two-Parameter Expansion}

Let $\epsilon$ denote the characteristic amplitude of the gravitational perturbation generated by the glitch, and let $\alpha$ denote the CS coupling constant introduced in eq.~\eqref{eq:action}. We treat both quantities perturbatively.

The spacetime metric is expanded as
\begin{equation}
    g_{\mu\nu} = \bar g_{\mu\nu} + \epsilon\, h_{\mu\nu}^{(1,0)} + \alpha^2 \epsilon\, h_{\mu\nu}^{(1,2)} + \mathcal{O}(\epsilon^2,\alpha^3),
\label{eq:metricexpansion}
\end{equation}
where $\bar g_{\mu\nu}$ is the slowly rotating background metric introduced in section~\ref{sec:metric}.
The perturbation $h_{\mu\nu}^{(1,0)}$ represents the leading general-relativistic GW fluctuation,
while $h_{\mu\nu}^{(1,2)}$ denotes the parity-violating correction induced by the dynamical CS interaction.

The background scalar field is itself sourced perturbatively by the Pontryagin density according to the scalar-field equation,
\begin{equation}
    \square \vartheta_0 =-\frac{\alpha}{4} \, {}^{\ast}\!RR,
\end{equation}
so that $\vartheta_0 = \mathcal{O}(\alpha)$. Consequently, the parity-violating correction to GW
propagation enters effectively at order $\mathcal{O}(\alpha^2\epsilon)$. 
Since the background scalar field is generated dynamically by the CS interaction, its stress-energy tensor scales
quadratically with the coupling, $T^{(\vartheta)}_{\mu\nu} \sim (\nabla\vartheta)^2 = \mathcal{O}(\alpha^2)$. Similarly, the perturbation of the scalar stress-energy tensor contains terms of the form $\delta T^{\vartheta}_{\mu\nu} \sim (\nabla\vartheta)(\nabla\delta\vartheta)$ and therefore scales as $\mathcal{O}(\alpha^2\epsilon)$. Consequently, neither the scalar stress-energy tensor nor the Cotton tensor contributes to the gravitational-wave equations at order $\mathcal{O}(\alpha\epsilon)$.

The perturbative hierarchy is summarized in Table~\ref{tab:hierarchy}.

\begin{table}[htbp]
\centering
\caption{Perturbative hierarchy in dynamical CG gravity.}
\label{tab:hierarchy}
\begin{tabular}{ccc}
\hline\hline
Order & Physical Content & Unknowns \\
\hline
$(\epsilon^0,\alpha^0)$ & Rotating background & $A(r), B(r), \omega(r)$ \\
$(\epsilon^1,\alpha^0)$ & GR perturbations & $h_{\mu\nu}^{(1,0)}$ \\
$(\epsilon^0,\alpha^1)$ & Background scalar field & $\vartheta_0$ \\
$(\epsilon^1,\alpha^2)$ & CS corrections to GW propagation & $h_{\mu\nu}^{(1,2)}$ \\
\hline\hline
\end{tabular}
\end{table}

At leading order in $(\epsilon^1,\alpha^0)$, the perturbations satisfy the standard Regge-Wheeler-Zerilli equations of
general relativity~\cite{Regge1957, Zerilli1970}. Since the background scalar field itself scales as
$\vartheta_0 = \mathcal{O}(\alpha)$, the CS modification to gravitational-wave propagation enters effectively at
order $(\epsilon^1,\alpha^2)$ through the Cotton tensor contribution to the modified Einstein equations.

\subsection{Parity Decomposition}
\label{sec:parity}

Since the background spacetime is spherically symmetric to zeroth order in rotation, metric perturbations may be decomposed into spherical harmonics. Following the standard formalism, perturbations separate into two sectors with definite transformation properties under the parity operation $(\theta, \phi) \rightarrow (\pi - \theta, \phi + \pi)$:
\begin{itemize}
\item \textbf{Even (polar) perturbations:} These transform with the same parity as a scalar spherical harmonic, acquiring a factor \((-1)^\ell\) under parity.
\item \textbf{Odd (axial) perturbations:} These transform with the opposite parity, acquiring a factor \((-1)^{\ell+1}\) under parity.
\end{itemize}
The labels ``even'' and ``odd'' refer not to the parity of the metric perturbation itself, but rather to the parity of the tensor spherical harmonics used in the decomposition relative to the standard parity of a scalar function \(Y_{\ell m} \rightarrow (-1)^\ell Y_{\ell m}\). Physically, even-parity perturbations correspond to deformations that preserve the star's reflection symmetry (such as radial pulsations or quadrupolar mass moments), while odd-parity perturbations correspond to deformations that break this symmetry (such as rotational shear or twisting modes).

This decomposition allows the perturbation equations to decouple into independent sectors at leading order in general relativity. In dynamical CS gravity the parity-violating interaction couples the scalar field to the metric perturbations, and it is worth being precise about the mechanism. The quantity that governs propagation is not the term \(\vartheta_0\,{}^{\ast}\!RR\) appearing in the action, but the perturbed Cotton tensor \(\delta C_{\mu\nu}\) of eq.~\eqref{eq:cotton}, which is built from gradients of the background scalar field contracted with the Levi-Civita tensor. Because \(\vartheta_0\) is a pseudoscalar, its gradient \(\nabla_\sigma \vartheta_0\) is a pseudovector; contracted with the pseudotensor \(\epsilon^{\sigma\delta\gamma}{}_{\mu}\), the result transforms as an ordinary tensor, so that \(\delta C_{\mu\nu}\) carries the same parity as the perturbation it acts upon and does not mix the two sectors.

What the contraction does introduce is a dependence on \emph{helicity}. When the axial perturbation is decomposed into circular polarization states, \(\Psi_{R,L} = \Psi_+ \mp i\Psi_\times\), the Levi-Civita contraction changes sign between the two. It is this sign flip---rather than any asymmetry in the source itself---that originates the birefringence, and it is the origin of the \(\mp\alpha V_{\rm CS}\) term in the modified Regge-Wheeler equation derived in section~\ref{sec:gw_propagation}. Physically, the rotating background scalar field acts like a parity-violating ``medium'' that distinguishes between the two helicities of gravitational waves.

We restrict attention throughout to the odd-parity (axial) sector, in which the glitch-excited quadrupole radiation of interest propagates, and in which the effect of the dipolar scalar background is most directly expressed. We note that the \(\ell = 1\) angular structure of \(\vartheta_0\) will in general couple neighbouring multipoles of a given parity through the Cotton tensor; however, such mixing enters the polarization asymmetry only at higher order in the slow-rotation expansion and is neglected here. This coupling ultimately leads to birefringent propagation of gravitational waves, as we will show in section~\ref{sec:gw_propagation}.

\subsection{Regge-Wheeler Gauge}

To simplify the perturbation equations we adopt the Regge-Wheeler gauge. Strictly speaking, the Regge-Wheeler gauge is defined for perturbations of a spherically symmetric spacetime. In the present work the background is described by the Hartle-Thorne slow-rotation expansion, which may be viewed as a perturbative deformation of a spherical geometry. We therefore perform the tensor-spherical-harmonic decomposition and impose the Regge-Wheeler gauge with respect to the underlying spherical background. Rotational effects enter through $\mathcal{O}(\Omega)$ couplings between neighboring multipoles but do not modify the parity classification itself.

In this gauge, the even-parity metric perturbation takes the form
\begin{equation}
\resizebox{0.8\columnwidth}{!}{$
h^{(\mathrm{even})}_{\mu\nu}
=
\begin{pmatrix}
-e^{2\Phi}H_0 Y_{\ell m} & -i\sigma H_1 Y_{\ell m} & 0 & 0 \\
\ast & e^{2\Lambda}H_2 Y_{\ell m} & 0 & 0 \\
0 & 0 & r^2 K Y_{\ell m} & 0 \\
0 & 0 & 0 & r^2 \sin^2\theta\, K Y_{\ell m}
\end{pmatrix}
e^{-i\sigma t}
$}.
\label{eq:even_pert}
\end{equation}
Here $H_0$, $H_1$, $H_2$ and $K$ are the standard even-parity Regge-Wheeler amplitudes,
and the asterisk denote a symmetric entry,
The odd-parity (axial) metric perturbation is given by
\begin{equation}
\resizebox{0.5\columnwidth}{!}{$
h^{(\mathrm{odd})}_{\mu\nu}
=
\begin{pmatrix}
0 & 0 & 0 & h_0^{(o)}(r) \sin\theta\,\partial_\theta Y_{\ell m} \\
0 & 0 & 0 & h_1^{(o)}(r) \sin\theta\,\partial_\theta Y_{\ell m} \\
0 & 0 & 0 & 0 \\
\ast & \ast & 0 & 0
\end{pmatrix}
e^{-i\sigma t}
$},
\label{eq:odd_pert}
\end{equation}
where $h_0^{(o)}(r)$ and $h_1^{(o)}(r)$ are radial functions.

The Regge-Wheeler gauge removes several gauge-dependent angular amplitudes, including the $h_{r\theta}$-sector.
Consequently, only the amplitudes displayed in eqs.~\eqref{eq:even_pert} and \eqref{eq:odd_pert} remain.

These perturbations may be combined into a single master variable $\Psi$ (the Regge-Wheeler variable) that satisfies
the Regge-Wheeler equation in general relativity. In CS gravity, this equation acquires a parity-violating correction,
as derived in Sec.~\ref{sec:gw_propagation}.

\subsection{Glitch Excitation Model}
\label{subsec:glitch_model}

Pulsar glitches correspond to sudden changes in the stellar rotation rate. We model this effect by allowing the
angular velocity to vary in time according to
\begin{equation}
\Omega(t) = \Omega_0 + \delta\Omega(t),
\label{eq:omega_t}
\end{equation}
where $\Omega_0$ is the steady-state rotation rate and $\delta\Omega(t)$ represents the glitch-induced perturbation.

Following phenomenological glitch models, we write
\begin{equation}
\delta\Omega(t) = \Delta\Omega\,f(t),
\label{eq:delta_omega}
\end{equation}
where $\Delta\Omega$ is the total change in rotation frequency and $f(t)$ describes the temporal profile of the glitch:
\begin{equation}
f(t) = \Theta(t)\left(1-e^{-t/\tau_{\rm rise}}\right)e^{-t/\tau_{\rm relax}}.
\label{eq:glitch_profile}
\end{equation}
Here $\Theta(t)$ is the Heaviside step function, $\tau_{\rm rise}$ characterizes the rapid spin-up timescale of the
glitch (typically milliseconds to seconds), and $\tau_{\rm relax}$ describes the subsequent relaxation of the star back
toward equilibrium (ranging from days to months).

The change in the star's moment of inertia induced by this process acts as the source term for the GW
equation. In the frequency domain, the glitch profile enters through its Fourier transform $\tilde{f}(\sigma)$, which
appears in the quadrupolar source $S_{\mathrm{glitch}}(r,\sigma)$ used in section~\ref{sec:gw_propagation}. Explicitly,
\begin{equation}
\begin{aligned}
\tilde{f}(\sigma) &= \int_{-\infty}^{\infty}dt\, f(t) e^{i\sigma t}\\
&= \frac{1}{(\tau_{\rm relax}^{-1} - i\sigma)(\tau_{\rm rise}^{-1} + \tau_{\rm relax}^{-1} - i\sigma)}.
\end{aligned}
\label{eq:fourier_glitch}
\end{equation}
This Fourier transform encodes the frequency content of the glitch: low-frequency components
($\sigma \ll \tau_{\rm relax}^{-1}$) capture the overall spin-up, while high-frequency components
($\sigma \gg \tau_{\rm rise}^{-1}$) are suppressed by the finite rise time.

\subsection{Role of the CS Correction}

At order $(\epsilon^1,\alpha^0)$ the gravitational perturbations satisfy the usual Regge-Wheeler and Zerilli
equations of general relativity, as summarized in Table~\ref{tab:hierarchy}.

The dynamical CS modification enters through the Cotton tensor in eq.~\eqref{eq:einstein}. Since the background scalar field is itself sourced perturbatively by the Pontryagin density, we have $\vartheta_0 = \mathcal{O}(\alpha)$. (See section~\ref{sec:scalar}.) Consequently, the perturbation of the Cotton tensor scales as
\begin{equation}
\delta C_{\mu\nu} = \mathcal{O}(\alpha\epsilon),
\end{equation}
and because the Cotton tensor appears multiplied by $\alpha$ in eq.~\eqref{eq:einstein}, the resulting correction to the GW propagation equations enters effectively at order $(\epsilon^1,\alpha^2)$.

Since the background scalar field is generated dynamically by the CS interaction, its stress-energy tensor scales quadratically with the coupling,
\begin{equation}
T^{(\vartheta)}_{\mu\nu}
\sim
(\nabla\vartheta_0)^2 = \mathcal{O}(\alpha^2).
\end{equation}
Similarly, the perturbation of the scalar stress-energy tensor contains terms of the form
\begin{equation}
\delta T^{(\vartheta)}_{\mu\nu}
\sim
(\nabla\vartheta_0)(\nabla\delta\vartheta),
\end{equation}
and therefore scales as $\mathcal{O}(\alpha^2\epsilon)$. Consequently, neither the scalar stress-energy tensor nor the Cotton tensor contributes to the gravitational-wave equations at order $\mathcal{O}(\alpha\epsilon)$. The leading parity-violating modification therefore enters only at order $(\epsilon^1,\alpha^2)$.

At this order, $\delta C_{\mu\nu}$ contains terms proportional to gradients of the background scalar field $\vartheta_0$. (See appendix~\ref{app:rw_derivation}.) These terms couple the scalar field to axial (odd-parity) gravitational perturbations, while leaving the even-parity sector unaffected at leading order.

As we show in the following sections, this coupling generates a parity-dependent correction to the Regge-Wheeler equation, producing a relative phase shift between right- and left-circularly polarized GWs. Consequently, GWs emitted by glitching pulsars acquire a small polarization-dependent propagation effect that manifests observationally as a parity-violating asymmetry in the waveform.

\section{Pontryagin Density}
\label{sec:pontryagin}

The CS scalar field is sourced by the gravitational Pontryagin density ${}^{\ast}RR$. In this section we compute this quantity for the slowly rotating spacetime introduced in Sec.~\ref{sec:metric} and show that rotation generates a nonvanishing source for the scalar field.

\subsection{General Definition}

The Pontryagin density is defined as the contraction of the Riemann tensor with its dual,
\begin{equation}
{}^{\ast}RR = {}^{\ast}\tensor{R}{^{\sigma}_{\tau\mu\nu}} \tensor{R}{^{\tau}_{\sigma}^{\mu\nu}},
\label{eq:pontryagin_def}
\end{equation}
where the dual Riemann tensor is
\begin{equation}
{}^{\ast}\tensor{R}{^{\sigma}_{\tau\mu\nu}} = \frac{1}{2\sqrt{-g}} \tilde{\epsilon}_{\mu\nu\gamma\delta} \tensor{R}{^{\sigma}_{\tau}^{\gamma\delta}},
\label{eq:dual_riemann_def}
\end{equation}
and $\tilde{\epsilon}_{\mu\nu\alpha\beta}$ is the totally antisymmetric Levi-Civita symbol with $\tilde{\epsilon}_{tr\theta\phi}=+1$.

This quantity is odd under parity transformations and therefore vanishes for spacetimes possessing reflection symmetry. In particular, any static, spherically symmetric spacetime satisfies
\begin{equation}
{}^{\ast}RR = 0.
\label{eq:pontryagin_static}
\end{equation}

Physically, this follows because the Riemann tensor of a static, spherically symmetric spacetime contains no preferred direction to define a parity-odd invariant. More formally, any static spherically symmetric spacetime admits a parity symmetry under which the Riemann tensor components are invariant, while ${}^{\ast}RR$ changes sign, forcing it to vanish identically.

Consequently, a nonzero Pontryagin density requires rotation or other parity-breaking structures. As we will see below, the slow rotation of the neutron star breaks parity sufficiently to generate a nontrivial source for the CS scalar field.

\subsection{Pontryagin Density for a Slowly Rotating Star}

We now evaluate ${}^{\ast}RR$ for the Hartle-Thorne spacetime introduced in eq.~\eqref{eq:HTmetric}. (See appendix~\ref{app:pontryagin} for the full derivation.) Since the static background gives a vanishing Pontryagin density, the leading contribution appears at first order in the frame-dragging function $\omega(r)$.

At linear order in rotation, the Riemann tensor may be decomposed as
\begin{equation}
R_{\mu\nu\gamma\delta} = R^{(0)}_{\mu\nu\gamma\delta} + R^{(1)}_{\mu\nu\gamma\delta},
\label{eq:riemann_decomp}
\end{equation}
where $R^{(0)}$ corresponds to the static spherically symmetric background and $R^{(1)}$ contains terms proportional to $\omega(r)$.

The Pontryagin density is quadratic in the Riemann tensor and therefore receives leading-order contributions only from the cross terms,
\begin{equation}
{}^{\ast}RR \sim R^{(0)} \cdot R^{(1)}.
\label{eq:pontryagin_cross}
\end{equation}
while $R^{(1)} \cdot R^{(1)}$ is higher order in the slow-rotation expansion.
A direct evaluation of the contraction in eq.~\eqref{eq:pontryagin_def}, using the antisymmetry of the Levi-Civita tensor and the index symmetries of the Riemann tensor, shows that the Pontryagin density takes the separable form
\begin{equation}
{}^{\ast}RR = \omega'(r)\cos\theta\,\mathcal{F}(r),
\label{eq:pontryagin_general}
\end{equation}
where $\mathcal{F}(r)$ depends only on the background metric functions $A(r)$ and $B(r)$.
This structure arises because the Levi-Civita tensor selects components involving mixed radial and angular indices, while axisymmetry allows the angular dependence to factor out as $\cos\theta$. The resulting source therefore has a purely dipolar ($\ell=1$) angular dependence.

The radial function $\mathcal{F}(r)$ may be written explicitly as
\begin{multline}
\mathcal{F}(r) = \frac{2}{(AB)^{5/2} r^{2}}
\Bigl\{ \bigl[2AA''B - (A')^{2}B - AA'B'\bigr] r^{2} \\
+ 4A^{2}B(1 - B) + 2rA\bigl(AB' - BA'\bigr) \Bigr\}.
\label{eq:Fr}
\end{multline}
Equation~\eqref{eq:pontryagin_general} shows that the Pontryagin density is proportional to the radial derivative of the frame-dragging function. Physically, this reflects the fact that the parity-violating invariant is generated by the coupling between the background curvature and the gravitomagnetic field associated with rotation.

\subsection{Schwarzschild Exterior}

Outside the neutron star the metric functions reduce to the Schwarzschild form
\begin{equation}
A(r) = 1-\frac{2M}{r}, \qquad B(r) = \frac{1}{A(r)},
\label{eq:schwarzschild_metric}
\end{equation}
and the frame-dragging function is
\begin{equation}
\omega(r) = \frac{2J}{r^3},
\label{eq:exterior_omega}
\end{equation}
where $J$ is the stellar angular momentum.
Substituting these expressions into eq.~\eqref{eq:pontryagin_general}, and using the relations $AB=1$ and
$\omega'(r)=-6J/r^4$, we obtain
\begin{equation}
{}^{\ast}RR = \frac{288MJ}{r^7}\cos\theta.
\label{eq:pontryagin_exterior}
\end{equation}
This agrees with the well-known slow-rotation limit of the Kerr spacetime~\cite{Kerr1963, MisnerThorneWheeler1973},
for which ${}^{\ast}RR = 288aM^2\cos\theta/r^7$ with $a = J/M$. (See appendix~\ref{app:pontryagin} for a detailed
verification.)
This quantity acts as the source term for the CS scalar-field equation~\eqref{eq:scalar}.

In the next section we
solve the resulting scalar-field equation in the exterior region of the star.
However, several important features are immediately evident:
\begin{itemize}
\item The Pontryagin density vanishes in the limit $J \to 0$, confirming that rotation is required to generate the source.
\item The radial dependence $r^{-7}$ implies that the effect is strongly localized around the star, decaying rapidly
above the stellar surface.
\item The angular dependence $\cos\theta$ indicates a dipolar structure, which determines the dominant multipole
of the CS scalar field.
\end{itemize}

\section{Background CS Scalar Field}
\label{sec:scalar}

The rotating spacetime considered in Sec.~\ref{sec:metric} generates a nonvanishing Pontryagin density $^{\ast}RR$,
which acts as a source for the CS scalar field through eq.~\eqref{eq:scalar}.
The $\cos\theta$ angular dependence of the exterior solution, eq.~\eqref{eq:pontryagin_exterior},
implies that the background scalar field must have a dipolar structure, which we determine in this section.

\subsection{Separation of Variables and Radial Equation}

The angular part of the scalar field Laplacian in Schwarzschild satisfies $\Delta_\Omega Y_{\ell m} = -\ell(\ell+1) Y_{\ell m}$, where $\Delta_\Omega$ is the angular Laplacian on the unit sphere. For $\ell=1$, the spherical harmonic $Y_{10} \propto \cos\theta$ gives $\Delta_\Omega (\cos\theta) = -2\cos\theta$. Motivated by the $\cos\theta$ angular dependence of the Pontryagin source in eq.~\eqref{eq:pontryagin_exterior}, we separate the scalar field as
\begin{equation}
\vartheta_0(r,\theta) = \hat{\vartheta}_0(r)\cos\theta .
\label{eq:separation}
\end{equation}
Substituting this ansatz into the scalar field equation~\eqref{eq:scalar} and restricting to the Schwarzschild exterior
($r > R_\star$) yields the radial ODE
\begin{equation}
\begin{aligned}
    \left(1 - \frac{2M}{r}\right)\frac{d^2 \hat{\vartheta}_0}{dr^2}
+ \frac{2}{r}\left(1 - \frac{M}{r}\right)&\frac{d \hat{\vartheta}_0}{dr}
- \frac{2}{r^2}\hat{\vartheta}_0\\
&= -\alpha \frac{72MJ}{r^7}.
\end{aligned}
\label{eq:thODE}
\end{equation}
(See appendix~\ref{app:scalar_field} for a detailed derivation.)

In what follows, we solve eq.~\eqref{eq:thODE} by constructing the homogeneous and particular solutions separately,
then match to an interior solution to determine the dipole moment $\mu_{\mathrm{CS}}$.

\subsection{General Solution Structure}

Equation~\eqref{eq:thODE} is linear and inhomogeneous. Its solution is the sum of a homogeneous solution and a particular solution,
$\hat{\vartheta}_0 = \hat{\vartheta}_{0\mathrm{h}} + \hat{\vartheta}_{0\mathrm{p}}$.

\subsubsection{Homogeneous solution}

In the large-$r$ limit, the homogeneous equation admits power-law solutions $\hat{\vartheta}_{0\mathrm{h}} \sim r^{-n}$, leading to the indicial equation $n^2 - n - 2 = 0$, with roots $n = 2$ (decaying) and $n = -1$ (growing). Physical boundary conditions ($\vartheta_0 \to 0$ as $r \to \infty$) select the decaying solution. We construct its asymptotic expansion as
\begin{equation}
\hat{\vartheta}_{0\mathrm{h}}(r) = \frac{1}{r^{2}} \sum_{k=0}^{\infty} b_k \left(\frac{M}{r}\right)^{k},
\label{eq:homogeneous_series}
\end{equation}
with $b_0 = 1$ for normalization; the overall coefficient of the homogeneous solution will be fixed later by matching to
the interior. (See appendix~\ref{app:scalar_field} for the full derivation.)
Substituting into the homogeneous form of eq.~\eqref{eq:thODE} and matching powers of $1/r$ determines the
coefficients recursively. The first few coefficients are $b_1 = 2$, $b_2 = 4$, $b_3 = 8$, yielding the asymptotic expansion
\begin{equation}
\begin{aligned}
\hat{\vartheta}_{0\mathrm{h}}(r) = \frac{1}{r^{2}}\Bigg[1 + \frac{2M}{r} &+ \frac{4M^{2}}{r^{2}} + \frac{8M^{3}}{r^{3}} \\
&+ \mathcal{O}\!\left(\frac{M^{4}}{r^{4}}\right)\Bigg].
\end{aligned}
\label{eq:thh}
\end{equation}

\subsubsection{Particular solution}

The sourced contribution is obtained via an asymptotic series in inverse powers of $r$. The right-hand side of
eq.~\eqref{eq:thODE} scales as $r^{-7}$, so the particular solution starts at $r^{-5}$. Following the recursive
procedure detailed in appendix~\ref{app:scalar_field}, we obtain
\begin{equation}
\begin{aligned}
    \hat{\vartheta}_{0\mathrm{p}}(r) = -\frac{4\alpha MJ}{r^{5}} \Bigg[ 1 &+ \frac{25}{14}\frac{M}{r} + \frac{15}{4}\frac{M^{2}}{r^{2}} \\
    &+ \mathcal{O}\!\left(\frac{M^{3}}{r^{3}}\right) \Bigg].
\label{eq:thp}
\end{aligned}
\end{equation}

Thus, the parity-violating Pontryagin source induces a subleading correction to the dipolar scalar field, entering
first at $r^{-5}$. The homogeneous solution dominates at large distances, while the particular solution becomes relevant
only in the strong-field region near the star. Physically, we can see that
this means the scalar field is primarily sourced by the Pontryagin density located inside the star.

\subsection{Exterior Scalar Field}

The homogeneous solution \(\hat{\vartheta}_{0\mathrm{h}}(r)\) as written in eq.~\eqref{eq:thh} has a fixed normalization (\(b_0 = 1\)). In general, the homogeneous equation admits solutions with arbitrary overall amplitude, because it is linear. The physically relevant amplitude is not determined by the exterior equation alone; it must be fixed by matching to a regular interior solution at the stellar surface \(r = R_\star\). We therefore introduce a dimensionless coefficient \(\mu_{\mathrm{CS}}\) to parameterize this freedom, writing the full exterior solution as
\begin{equation}
\hat{\vartheta}_{00}(r) = \mu_{\mathrm{CS}}\,\hat{\vartheta}_{0\mathrm{h}}(r) + \hat{\vartheta}_{0\mathrm{p}}(r).
\label{eq:theta_ext_total}
\end{equation}
Here \(\hat{\vartheta}_{0\mathrm{h}}(r)\) retains the asymptotic expansion
\eqref{eq:homogeneous_series} with the normalization \(b_0 = 1\), so that the leading large-\(r\) behavior becomes
\begin{equation}
\hat{\vartheta}_{00}(r) \sim \frac{\mu_{\mathrm{CS}}}{r^2} \quad \text{as} \quad r \to \infty.
\label{eq:mu_asymptotic}
\end{equation}

The quantity \(\mu_{\mathrm{CS}}\) is the ``dipole moment'' of the CS scalar field. It has dimensions of length squared and encodes the strength of the dipolar scalar field sourced by the rotating star itself. Physically, \(\mu_{\mathrm{CS}}\) plays a role analogous to the magnetic dipole moment in electromagnetism; just as a rotating charged sphere generates a dipolar magnetic field whose amplitude is determined by the interior current distribution, the rotating neutron star generates a dipolar scalar field whose amplitude \(\mu_{\mathrm{CS}}\) is determined by the interior mass distribution, rotation profile, and equation of state (EoS). A key difference is that \(\mu_{\mathrm{CS}}\) is proportional to the CS coupling \(\alpha\) (since the scalar field is sourced by \(\alpha\,{}^{\ast}RR\)), so it vanishes identically in general relativity. Thus, a nonzero \(\mu_{\mathrm{CS}}\) is a direct signature of parity violation.

As we will see in section~\ref{sec:interior_models}, the specific value of \(\mu_{\mathrm{CS}}\) depends on the
interior structure of the star. For a constant-density model we obtain an analytic expression, while for more
realistic EoS it must be computed numerically.

Using the asymptotic forms above, the radial function expands as
\begin{equation}
\hat{\vartheta}_{00}(r) = \frac{\mu_{\mathrm{CS}}}{r^2} + \frac{2M\mu_{\mathrm{CS}}}{r^3} - \frac{4\alpha MJ}{r^5} + \mathcal{O}(r^{-6}).
\label{eq:theta_full_asym}
\end{equation}
The corresponding radial derivative is
\begin{equation}
\hat{\vartheta}'_{00}(r) = -\frac{2\mu_{\mathrm{CS}}}{r^3} - \frac{6M\mu_{\mathrm{CS}}}{r^4} + \frac{20\alpha MJ}{r^6} + \mathcal{O}(r^{-7}).
\label{eq:theta_prime_full}
\end{equation}
In the asymptotic wave zone (\(r \gg R_\star\)), the scalar field is dominated by the leading dipolar contribution,
\begin{equation}
\begin{aligned}
\hat{\vartheta}_{00}(r) &\approx \frac{\mu_{\mathrm{CS}}}{r^2} - \frac{4\alpha MJ}{r^5},\\
\hat{\vartheta}'_{00}(r) &\approx -\frac{2\mu_{\mathrm{CS}}}{r^3} + \frac{20\alpha MJ}{r^6}.
\end{aligned}
\label{eq:theta_wavezone}
\end{equation}
The \(r^{-2}\) term represents the dominant dipolar field sourced by the rotating star, while the \(r^{-5}\) term is the first parity-violating correction. Although suppressed at large distances, this sub-leading term encodes genuine strong-field effects and will play a crucial role in the birefringent propagation of GWs.

\section{Gravitational-Wave Propagation in Dynamical CS Gravity}
\label{sec:gw_propagation}

We now analyze how the background scalar field derived in section~\ref{sec:scalar} modifies GW propagation. Working to first order in the metric perturbation \(\epsilon\) and retaining the leading parity-violating correction induced by the dynamically sourced scalar background \(\vartheta_0 = \mathcal{O}(\alpha)\), the gravitational-wave propagation equations receive corrections at effective order \(\mathcal{O}(\alpha^2\epsilon)\). The perturbed field equations are
\begin{equation}
\delta G_{\mu\nu} + \alpha\,\delta C_{\mu\nu} = 8\pi\,\delta T_{\mu\nu},
\label{eq:perturbedFE}
\end{equation}
where \(\delta C_{\mu\nu}\) is the first-order perturbation of the Cotton tensor.

\subsection{Cotton Tensor Contribution}

The Cotton tensor, defined in eq.~\eqref{eq:cotton}, depends on derivatives of the background scalar field \(\vartheta_0\). Using the exterior solution \(\vartheta_0(r,\theta) = \hat{\vartheta}_{00}(r)\cos\theta\), its relevant gradients are
\begin{equation}
\nabla_r\vartheta_0 = \hat{\vartheta}_{00}'(r)\cos\theta,\qquad
\nabla_\theta\vartheta_0 = -\hat{\vartheta}_{00}(r)\sin\theta,
\label{eq:scalar_gradients}
\end{equation}
with the \(\phi\)-derivative vanishing by axisymmetry.

Linearizing the Cotton tensor around the background yields a source term proportional to \(\nabla_\mu\vartheta_0\). Because the Levi-Civita tensor in eq.~\eqref{eq:cotton} is parity-odd, the combination \(\nabla_\mu\vartheta_0\) couples specifically to \textit{axial} (odd-parity) gravitational perturbations, while even-parity perturbations remain unaffected at this order. This is analogous to wave propagation in a parity-violating medium, where the scalar field gradient acts as an effective external field that distinguishes between right- and left-handed polarization states.

\subsection{Projection onto the Regge-Wheeler Variable}

Axial perturbations of a spherically symmetric background are described by the Regge-Wheeler master variable
\begin{equation}
\Psi = \frac{r\,h_{t\phi}}{\sin\theta\,\partial_\theta Y_{\ell m}},
\label{eq:psi_definition}
\end{equation}
which condenses the two odd-parity metric perturbations \(h_{t\phi}\) and \(h_{r\phi}\) into a single function.
(The component \(h_{r\phi}\) is eliminated by means of gauge invariance, via perturbed Einstein equations, leaving \(\Psi\)
as the sole dynamical degree of freedom.) This variable satisfies a wave equation---the Regge-Wheeler equation---that governs
the dynamics of axial GWs in general relativity.

We restrict attention to the leading parity-violating modification induced by the \emph{background} scalar field \(\vartheta_0\), neglecting dynamical scalar perturbations \(\delta\vartheta\). This is justified because \(\delta\vartheta\) couples to the metric at higher order in the CS coupling \(\alpha\) and does not contribute to the leading birefringent effect. Isolating the background scalar field gradients allows us to focus on the phase shift between circular polarizations.

Projecting the perturbed field equations onto \(\Psi\) and reducing the Cotton tensor contribution (using the leading-order Regge-Wheeler equation to eliminate higher derivatives) leads to a modified wave equation. The detailed derivation is presented in appendix~\ref{app:rw_derivation}, which shows how the Cotton tensor introduces a parity-dependent correction to the Regge-Wheeler potential.

\subsection{Modified Regge-Wheeler Equation}

Decomposing the gravitational perturbations into circular polarization states,
\begin{equation}
\Psi_{R,L} = \frac{1}{\sqrt{2}}\left( \Psi_+ \mp i\Psi_\times \right),
\label{eq:circular_decomp}
\end{equation}
where \(\Psi_+\) and \(\Psi_\times\) correspond to the two linear polarization modes, the axial perturbation equation
becomes [see eqs.~\eqref{eq:app_leading_rw}--\eqref{eq:app_modified_rw_final} in appendix~\ref{app:rw_derivation}]
\begin{equation}
\begin{aligned}
    \frac{d^2\Psi_{R,L}}{dr_*^2}
&+ \left[ \sigma^2 - V_\mathrm{RW}(r) \mp \alpha V_\mathrm{CS}(r) \right]
\Psi_{R,L}
= S^\mathrm{eff}(r,\sigma),
\end{aligned}
\label{eq:RW}
\end{equation}
where \(r_*\) is the tortoise coordinate defined by \(dr_*/dr = 1/A(r)\). 

The Regge-Wheeler potential for the dominant \(\ell = 2\) mode is
\begin{equation}
V_\mathrm{RW}(r) = \frac{6}{r^2}\left(1-\frac{2M}{r}\right)\left(1-\frac{M}{r}\right).
\label{eq:VRW}
\end{equation}
Although the background scalar field is dipolar (\(\ell = 1\)), the GWs generated by the glitch are dominated by the quadrupole (\(\ell = 2\)) mode, because the lowest-order radiation in general relativity is quadrupolar. The CS correction couples the dipolar scalar field to the \(\ell = 2\) axial perturbations through the Cotton tensor, modifying their propagation without altering the emission mechanism. We therefore focus on the \(\ell = 2\) channel, which is the most promising for detection.

The effective parity-violating correction (derived in appendix~\ref{app:rw_derivation}) is
\begin{equation}
\begin{aligned}
V_\mathrm{CS}(r,\sigma) &= \frac{\hat{\vartheta}_{00}(r)}{2r^2}\left[\sigma^2 - V_\mathrm{RW}(r)\right] \\
&\quad - \frac{2 A(r)}{r^2}\left[ \hat{\vartheta}_{00}''(r) - \frac{A'(r)}{A(r)}\hat{\vartheta}_{00}'(r) \right],
\end{aligned}
\label{eq:VCS}
\end{equation}
where \(\hat{\vartheta}_{00}(r)\) is the radial part of the background scalar field defined in section~\ref{sec:scalar}.

The opposite signs for the two polarizations reflect the parity-violating nature of the interaction.
Right- and left-circular modes experience opposite effective potentials, leading to a relative phase shift as
the wave modes propagate. This is the origin of gravitational birefringence in dynamical CS gravity.

The effective source term $S^{\mathrm{eff}}(r,\sigma)$ encapsulates the modification of the glitch excitation strength
due to the background CS scalar field. As derived in appendix~\ref{app:rw_derivation}, it is related to the
general-relativistic source $S_0(r,\sigma)$ (which describes the quadrupole moment change induced by the glitch) via
\begin{equation}
S^{\mathrm{eff}}(r,\sigma) = \left[1 - \frac{\alpha \hat{\vartheta}_{00}(r)}{2r^2}\right] S_0(r,\sigma).
\end{equation}
The correction term, proportional to $\alpha \hat{\vartheta}_{00}(r)/r^2$, is a direct consequence of the Cotton tensor
perturbation and reflects the fact that the parity-violating scalar background ``dresses'' the original quadrupole source.
Because $\hat{\vartheta}_{00}(r) \sim \mu_{\mathrm{CS}}/r^2$ decays rapidly, this correction is most significant near the
stellar surface ($r \sim R_\star$) and becomes negligible in the wave zone. Importantly, the correction is identical for
both circular polarizations and therefore does not contribute to the relative phase; it only affects the overall 
GW amplitude. Consequently, the leading birefringent signal is purely a propagation effect, as captured
by the parity-violating potential $\alpha V_{\mathrm{CS}}(r,\sigma)$ in eq.~\eqref{eq:RW}.

\subsection{Phase Accumulation in the Wave Zone}
\label{sec:phase_acc}

In the wave zone ($r \gg R_\star$), the Regge-Wheeler potential becomes subdominant, $V_{\mathrm{RW}}(r) \sim 1/r^2$, and eq.~\eqref{eq:RW} reduces approximately to a free wave equation with a small correction,
\begin{equation}
\frac{d^2\Psi_{R,L}}{dr_*^2} + \left[\sigma^2 \mp \alpha V_{\mathrm{CS}}(r)\right]\Psi_{R,L} \approx 0 .
\label{eq:wavezone_wave}
\end{equation}

The dominant contribution to $V_{\mathrm{CS}}(r)$ in this region comes from the dipersive term proportional to $\hat{\vartheta}_{00}'(r)/2r^2$; the second term in eq.~\eqref{eq:VCS}, which contains higher derivatives of the scalar field and the metric connection, scales as $r^{-6}$ and is therefore negligible in the far-field limit. This scaling reflects the fact that at kilohertz frequencies, the parity-violating correction is dominated by the dispersive interaction between the metric perturbations and the dipolar scalar field amplitude.
Hence,
\begin{equation}
V_{\mathrm{CS}}(r) \approx \frac{\sigma^2}{2r^2}\hat{\vartheta}_{00}(r),
\label{eq:VCS_asymptotic}
\end{equation}
and the accumulated phase shift between the stellar surface and a distant observer located at $D \gg R_\star$ is
\begin{equation}
\Delta\Phi_{\mathrm{CS}} = \frac{\alpha\sigma}{2}
\int_{R_\star}^{D}dr\, \frac{\hat{\vartheta}_{00}(r)}{r^2}.
\label{eq:DeltaPhi}
\end{equation}

Using the asymptotic expansion of $\hat{\vartheta}_{00}(r)$ from section~\ref{sec:scalar}
(see also appendix~\ref{app:scalar_field}), substituting into eq.~\eqref{eq:DeltaPhi},
and integrating yields the leading term
\begin{equation}
\Delta\Phi_{\mathrm{CS}} \approx \frac{\alpha\sigma \mu_{\mathrm{CS}}}{6R_\star^3}
+ \mathcal{O}\!\left(\frac{1}{R_\star^4}\right),
\label{eq:phase_leading}
\end{equation}
where $\mu_{\mathrm{CS}}$ is the dipole moment of the CS scalar field. The phase accumulation is dominated by the near-zone region close to the stellar surface, where the scalar field amplitude is largest; contributions from large distances are suppressed by the $1/r^2$ falloff of $\hat{\vartheta}_{00}(r)$ [which leads to an integrand scaling as $1/r^4$ in eq.~\eqref{eq:DeltaPhi}]. The higher-order corrections in $1/R_\star$ are negligible for typical neutron star radii and are omitted in what follows\footnote{The next term in the expansion scales as $\sim \alpha\sigma M\mu_{\mathrm{CS}}/R_\star^4$, which is suppressed by an additional factor $M/R_\star \sim 0.2$ and is therefore neglected.}.

\subsection{Parity-Violating Gravitational-Wave Strain}

The zeroth-order gravitational-wave strain from the glitch is
\begin{equation}
\begin{aligned}
    h^{(0)}(t,D) &= \frac{1}{D}\,\ddot{\delta I}_{20}\!\left(t-\frac{D}{c}\right),\\
\widetilde{\ddot{\delta I}}_{20}(\sigma) &= -\sigma^2 \delta I_0\,\tilde f(\sigma),
\end{aligned}
\label{eq:h0}
\end{equation}
where \(\delta I_0\) is the amplitude of the change in the star's quadrupole moment induced by the glitch, and \(\tilde f(\sigma)\) is the Fourier transform of the glitch timing function \(f(t)\) defined in eq.~\eqref{eq:glitch_profile}. (Note that \(c\) still appears in the retarded time argument \(t - D/c\) because it is a physical constant; in geometric units \(c = 1\), it can be omitted, but we retain it here for clarity.)

Including the CS phase shift, the circular polarization states propagate as
\begin{equation}
h_{R,L}(t,D) = h^{(0)}\!\left(t-\frac{D}{c}\right) e^{\mp i\Delta\Phi_{\mathrm{CS}}}.
\label{eq:hRL_time}
\end{equation}
Expanding to first order in the small phase shift and using eq.~\eqref{eq:phase_leading} gives
\begin{equation}
h_{R,L}(\sigma,D) = -\frac{\sigma^2}{D}\,\delta I_0\,\tilde f(\sigma)
\left[1 \mp i\,\frac{\alpha\sigma \mu_{\mathrm{CS}}}{6R_\star^3}\right].
\label{eq:hRL_freq}
\end{equation}
The relative difference between the two polarizations is therefore
\begin{equation}
\frac{|h_R - h_L|}{|h^{(0)}|} = \frac{\alpha\sigma \mu_{\mathrm{CS}}}{3R_\star^3},
\label{eq:parity_general}
\end{equation}
where \(\mu_{\mathrm{CS}}\) is the dipole moment of the CS scalar field as defined in Sec.~\ref{sec:scalar}. In the following sections we compute \(\mu_{\mathrm{CS}}\) for interior models (constant density and two fluid) and explore how the parity-violating signal changes accordingly.

We note that although the birefringent phase shift is linear in the CS coupling \(\alpha\), the observable polarization asymmetry effectively becomes second order in \(\alpha\) because the background scalar field (and hence \(\mu_{\mathrm{CS}}\)) is itself sourced by the parity-violating interaction. The precise scaling with \(\alpha\), \(J\), and \(R_\star\) will be obtained from each interior model.

In the time domain, expanding the exponential in eq.~\eqref{eq:hRL_time} to first order yields
\begin{equation}
h_{R,L}(t,D) = h^{(0)}(t,D) \mp \frac{\alpha \mu_{\mathrm{CS}}}{6R_\star^3}\,
\dot h^{(0)}(t,D),
\label{eq:htdomain}
\end{equation}
where \(\dot h^{(0)}(t,D)\) denotes the time derivative of the waveform in unperturbed general relativity.
This expression shows that the CS correction is proportional to the time derivative of the standard waveform,
corresponding to a relative phase shift between the two circular polarizations.

\subsection{Physical Interpretation}

Several features of our results are noteworthy:

\begin{enumerate}

\item \textbf{Frequency dependence:} The parity-violating signal scales linearly with frequency, $|h_R - h_L| \propto \sigma$, favoring higher-frequency GWs. This behavior originates from the cumulative propagation phase acquired through the CS interaction. Dimensionally, $\Delta\Phi_{\mathrm{CS}} \propto \alpha \sigma \mu_{\mathrm{CS}}/R_\star^3$, and since $\mu_{\mathrm{CS}}$ itself carries dimensions of length squared, the overall expression is dimensionless as required.

\item \textbf{Dependence on the scalar field amplitude:} The signal is proportional to $\mu_{\mathrm{CS}}$, the dipole moment of the CS scalar field. Because $\mu_{\mathrm{CS}}$ is determined by the interior structure of the star, the observable birefringence encodes information about the stellar interior, including its rotation profile and EoS. The precise scaling with stellar parameters (angular momentum, radius, compactness) will be obtained from specific interior models in the following sections, where we also show that differential rotation effects are negligible for typical glitches.

\item \textbf{Waveform structure:} The leading CS correction is proportional to the time derivative of the GR waveform, $\dot h^{(0)}$, inducing a polarization-dependent distortion that corresponds to a relative phase shift between the two circular polarization states. This is the defining signature of gravitational birefringence.

\item \textbf{Localization of the effect:} The phase accumulation is dominated by the near-zone region close to the stellar surface ($r \sim R_\star$), where the scalar field amplitude is largest. Contributions from large distances are suppressed by the $1/r^2$ falloff of $\hat{\vartheta}_{00}(r)$, which makes the integrand in eq.~\eqref{eq:DeltaPhi} scale as $1/r^4$ and therefore heavily weights the stellar interior and surface.

\end{enumerate}

\section{Interior Models for the Frame-Dragging Profile}
\label{sec:interior_models}

The exterior scalar field coefficient $\mu_{\mathrm{CS}}$ is determined by matching to a regular interior solution. This requires knowledge of the frame-dragging function $\omega(r)$ throughout the star, because the Pontryagin density $^{\ast}\!RR \propto \omega'(r)\cos\theta$ sources the scalar field also in the interior. 

Computing $\mu_{\mathrm{CS}}$ exactly requires solving the coupled structure equations
(the TOV equations~\cite{Tolman1939, OppenheimerVolkoff1939}
for the background and the Hartle equation for frame-dragging) together with the scalar field equation,
all for a reasonably realistic EoS. While this is straightforward numerically, it obscures the parametric
dependence of $\mu_{\mathrm{CS}}$ on stellar properties such as mass, radius, and rotation rate. Moreover, the
microphysics of pulsar glitches introduces additional uncertainties related to differential rotation and superfluidity.

To gain analytic insight and establish the parametric scaling of the birefringent signal, we first study a ``constant-density model'' in section~\ref{sec:constant_density}. This toy model captures the essential relativistic frame-dragging effect while remaining analytically tractable. It allows us to determine how $\mu_{\mathrm{CS}}$ scales with $\alpha$, $M$, $R_\star$, and $\Omega$ without relying on numerical integration. The resulting expression provides a benchmark against which more realistic models can be compared. We then extend the analysis to a ``two-fluid model'' in section~\ref{sec:two_fluid} that incorporates differential rotation between the neutron superfluid and the charged component, demonstrating that glitch-specific effects are negligible for typical parameters.

\subsection{Model I: Constant-Density Star}
\label{sec:constant_density}

We first consider the simplest relativistic interior: a star with constant energy density $\rho_0$. This model captures the key feature that the frame-dragging function $\omega(r)$ varies throughout the interior, leading to a non vanishing Pontryagin density inside the star.

\subsubsection{Background configuration}

For a constant-density star, the enclosed mass is
\begin{equation}
m(r) = \frac{4\pi}{3}\rho_0 r^3,
\end{equation}
the total mass is
\begin{equation}
M = \frac{4\pi}{3}\rho_0 R_\star^3,
\end{equation}
and the metric function is
\begin{equation}
e^{2\Lambda(r)} = \left(1 - \frac{2M r^2}{R_\star^3}\right)^{-1}.
\end{equation}

The pressure \(p(r)\) and the metric potential \(\Phi(r)\) are given by the TOV interior solution; their explicit forms are not needed for the leading frame-dragging calculation because, as shown below, \(\omega(r)\) is determined solely by \(m(r)\) and the compactness parameter \(\chi = M/R_\star\), to first order in rotation.

\subsubsection{Frame-dragging function to first order in compactness}

In the slow-rotation Hartle-Thorne formalism, the frame-dragging function \(\omega(r)\) is determined by the ODE that depends on the background metric [eq.~\eqref{eq:omega_ode}]. Expanding in the compactness parameter \(\chi = M/R_\star \ll 1\) and solving perturbatively yields the interior solution for \(\bar{\omega}(r) = \Omega - \omega(r)\):

\begin{align}
\bar{\omega}(r) &= \Omega + \chi \bar{\omega}_1(r) + \mathcal{O}(\chi^2), \label{eq:baromega_expansion} \\
\bar{\omega}_1(r) &= \frac{6\Omega}{5R_\star^2} r^2 + c_2, \label{eq:baromega1}
\end{align}
where \(c_2\) is an integration constant. (This constant is not to be confused with the scalar dipole moment \(\mu_{\mathrm{CS}}\) introduced in section~\ref{sec:scalar}; it belongs purely to the rotational frame-dragging sector and will be determined by matching to the exterior solution.) The physical frame-dragging angular velocity is therefore \(\omega(r) = \Omega - \bar{\omega}(r)\), and its radial derivative is
\begin{equation}
\omega'(r) = -\chi \frac{12\Omega}{5R_\star^2} r + \mathcal{O}(\chi^2).
\label{eq:omegaprime_constdens}
\end{equation}

\subsubsection{Interior Pontryagin density}

For a slowly rotating Hartle-Thorne spacetime, the Pontryagin density
to leading order in rotation takes the form given by eqs.~\eqref{eq:pontryagin_general} and~\eqref{eq:Fr}.
Substituting eq.~\eqref{eq:omegaprime_constdens} gives the interior
source
\begin{equation}
^{\ast}\!RR_{\rm int}(r,\theta) = -\chi\frac{12\Omega}{5R_\star^2}r \cos\theta\,\mathcal{F}(r),
\label{eq:RR_int_constdens}
\end{equation}
where $\mathcal{F}(r)$ is the curvature factor defined in eq.~\eqref{eq:Fr}.
As it turns out, for
the constant-density model $\mathcal{F}$ vanishes identically throughout the interior---the interior solution being conformally flat (see appendix~\ref{app:F0_derivation})---so that eq.~\eqref{eq:RR_int_constdens} is zero and the entire source resides in the exterior. However, it is worthwhile to continue with the general analysis that will be applicable
to models with more general EoSs.

\subsubsection{Matching to the exterior solution}

The integration constant \(c_2\) appearing in the frame-dragging solution eq.~\eqref{eq:baromega1} is determined by
matching \(\bar{\omega}(r)\) at the stellar surface. This matching is performed in appendix~\ref{app:matching} and yields
\begin{equation}
c_2 = -2\Omega,
\label{eq:c2_value}
\end{equation}
which is independent of the CS coupling \(\alpha\). This constant belongs solely to the rotational frame-dragging sector.

The interior solution for \(\hat{\vartheta}_{\rm int}(r)\) must match smoothly to the exterior solution at \(r = R_\star\). The exterior scalar field has the asymptotic form
\begin{equation}
\hat{\vartheta}_{\rm ext}(r) = \frac{\mu_{\mathrm{CS}}}{r^2} + \frac{2M\mu_{\mathrm{CS}}}{r^3}
- \frac{4\alpha MJ}{r^5} + \mathcal{O}(r^{-6}),
\label{eq:ext_asym}
\end{equation}
where \(\mu_{\mathrm{CS}}\) is to be determined.
Solving the interior scalar field equation (see appendix~\ref{app:matching}) yields a solution that is regular at the origin,
\begin{equation}
\hat{\vartheta}_{\rm int}(r) = c_{\vartheta}r - \frac{3\alpha \chi \Omega \mathcal{F}_0}{50 R_\star^2} r^3
+ \mathcal{O}(r^5),
\label{eq:theta_int_form}
\end{equation}
where $c_{\vartheta}$ is an integration constant determined by matching, and \(\mathcal{F}_0\) is the
\emph{average interior value}
of the curvature function \(\mathcal{F}(r)\) defined in eq.~\eqref{eq:Fr}.
The linear term $c_{\vartheta}r$ is the regular homogeneous solution, while the cubic term is the particular solution sourced
by the interior Pontryagin density.

The quantity \(\mathcal{F}_0\) encodes how the background curvature couples to the frame-dragging gradient inside the
star. As eq.~\eqref{eq:Fr} makes explicit, the Pontryagin density is generated by the product of the star's
gravitoelectric (tidal) field with the gradient of its gravitomagnetic (frame-dragging) field, so \(\mathcal{F}_0\)
measures the interior tidal field weighted by the rotation profile.
However, as already noted, for the constant-density model this quantity
vanishes,
\begin{equation}
\mathcal{F}_0 = 0,
\label{eq:F0}
\end{equation}
The weighted interior average defining $\mathcal{F}_0$ and its
evaluation for the constant-density model are derived
explicitly in appendix~\ref{app:matching}.

This value $\mathcal{F}_0=0$ arises from the specific radial dependence of the metric functions in the
constant-density interior: a body of uniform density produces no tidal field in its interior, exactly as in Newtonian
gravity, so there is no gravitoelectric field for the frame-dragging gradient to multiply. Equivalently, the interior
solution is conformally flat, and the Pontryagin density---which depends on the Riemann tensor only
through its Weyl part---must therefore vanish throughout the interior for any rotation profile whatsoever.
Physically, \(\mathcal{F}_0\) determines the efficiency with which the interior rotation generates the scalar field,
and the uniform-density profile is precisely the configuration for which that efficiency is zero. The constant-density
benchmark is therefore provides a lower bound on the
interior contribution. For any centrally condensed equation of state the
enclosed mean density exceeds the local density, the interior tidal field is nonzero, and \(\mathcal{F}_0 > 0\)
contributes constructively. In section~\ref{sec:results_discussion} we show how this efficiency factor varies across
different EoSs, confirming that more compact stars yield the most significant parity-violating signatures.

Matching \(\hat{\vartheta}_{\rm int}(r)\) and its derivative to the exterior solution \(\hat{\vartheta}_{\rm ext}(r)\)
at \(r = R_\star\) determines both $c_{\vartheta}$ and \(\mu_{\mathrm{CS}}\). Following the detailed calculation in
appendix~\ref{app:matching}, the scalar dipole moment is found to be
\begin{equation}
\mu_{\mathrm{CS}} = \frac{8\alpha MJ}{R_\star^3} + \frac{1}{25}\alpha\chi\Omega\mathcal{F}_0 R_\star^3,
\label{eq:mu_general}
\end{equation}
where the first term arises from matching the particular solution (the \(r^3\) term in the interior) to the exterior
\(r^{-5}\) tail, and the second term comes from matching the homogeneous part
(the $c_{\vartheta}r$ term) to the exterior dipolar field.

For the constant-density model, substituting the explicit values
\begin{equation}
J = \frac{2}{5} M R_\star^2 \Omega,\qquad \chi = \frac{M}{R_\star},\qquad 
\mathcal{F}_0 = 0,
\end{equation}
into eq.~\eqref{eq:mu_general} yields the simplified expression
\begin{equation}
\mu_{\mathrm{CS}} = \frac{16\alpha M^2 \Omega}{5 R_\star}
\label{eq:mu_constdens}
\end{equation}
The interior term is absent for this model, so the scalar dipole is sourced entirely in the vacuum exterior, within a
few stellar radii of the surface where \(^{\ast}\!RR \propto r^{-7}\) is largest. For a realistic equation of state the
interior term in eq.~\eqref{eq:mu_general} reinstates itself and adds to eq.~\eqref{eq:mu_constdens}, so that
eq.~\eqref{eq:mu_constdens} should be read as a conservative estimate.

The general scaling \(\mu_{\mathrm{CS}} \sim \alpha M^2 \Omega / R_\star\) shows that the scalar dipole moment is proportional to \(\alpha\) and parametrically of order \(\alpha\) times the stellar angular momentum divided by \(R_\star^3\).

\subsubsection{Impact on GW polarization}

With the dipole moment \(\mu_{\mathrm{CS}}\) determined from the matching procedure [eq.~\eqref{eq:mu_constdens}], the birefringent phase shift accumulated by GWs from the stellar surface to a distant observer is given by eq.~\eqref{eq:phase_leading}.
For the constant-density model, using \(\mu_{\mathrm{CS}} = 16\alpha M^{2}\Omega/(5R_\star)\), we obtain
\begin{equation}
\Delta\Phi_{\mathrm{CS}} \approx \frac{\alpha\sigma}{6R_\star^3}\frac{16\alpha M^{2}\Omega}{5R_\star}
= \frac{8\,\alpha^2 \sigma \Omega M^{2}}{15 R_\star^{4}}.
\label{eq:phase_constdens_approx}
\end{equation}
More generally, retaining the interior contribution for an arbitrary interior model, eq.~\eqref{eq:mu_general} gives
\begin{equation}
\Delta\Phi_{\mathrm{CS}} = \frac{\alpha^2 \sigma \Omega}{6R_\star^3}
\left( \frac{16 M^2}{5 R_\star} + \frac{M \mathcal{F}_0 R_\star^{2}}{25} \right).
\label{eq:phase_constdens_exact}
\end{equation}

The circular polarization modes of the GW signal satisfy \(h_{R,L} = h^{(0)} e^{\mp i\Delta\Phi_{\mathrm{CS}}}\). Expanding to first order in the small phase shift gives
$h_{R,L} \approx h^{(0)} \left(1 \mp i\Delta\Phi_{\mathrm{CS}}\right)$,
so the fractional difference between the two polarizations becomes
\begin{equation}
\frac{|h_R - h_L|}{|h^{(0)}|}
= 2|\Delta\Phi_{\mathrm{CS}}|
= \frac{\alpha^2 \sigma \Omega}{3R_\star^3}
\left( \frac{16 M^2}{5 R_\star} + \frac{M \mathcal{F}_0 R_\star^{2}}{25} \right).
\label{eq:strain_diff_constdens}
\end{equation}
For the constant-density benchmark the interior term is absent, giving the closed-form result
\begin{equation}
\frac{|h_R - h_L|}{|h^{(0)}|} = \frac{16\,\alpha^2 \sigma \Omega M^{2}}{15 R_\star^{4}}.
\label{eq:strain_diff_approx}
\end{equation}
Several features of this result are worth emphasizing:
\begin{enumerate}
\item \textbf{Frequency dependence:} The signal grows linearly with GW frequency \(\sigma\),
favoring greater observability in higher-frequency emissions.
\item \textbf{Rotation dependence:} The effect is proportional to the stellar rotation rate \(\Omega\), as expected for a phenomenon rooted in frame-dragging.
\item \textbf{Compactness dependence:} The signal scales as \(M^{2}/R_\star^{4}\), equivalently as \(\chi^{2}/R_\star^{2}\) with \(\chi = M/R_\star\). At fixed mass this is a \(1/R_\star^{4}\) dependence, considerably steeper than the \(1/R_\star^{2}\) or \(1/R_\star^{3}\) scalings sometimes quoted for other effects, so a more compact star yields a substantially larger signal. The steepness originates in the fact that, for this model, the scalar dipole is sourced entirely in the near-zone exterior, where \(^{\ast}\!RR \propto r^{-7}\).
\item \textbf{Order in \(\alpha\):} The signal is \(\mathcal{O}(\alpha^2)\), consistent with the fact that the scalar dipole moment \(\mu_{\mathrm{CS}}\) is itself proportional to \(\alpha\).
\end{enumerate}

This constant-density model provides a benchmark, and---because the interior source vanishes for a uniform density profile---a conservative one: any centrally condensed equation of state adds a positive interior contribution through \(\mathcal{F}_0\) in eq.~\eqref{eq:strain_diff_constdens}. In the next subsection we consider a more realistic two-fluid model, where differential rotation between the neutron superfluid and the charged component can modify the interior frame-dragging profile and hence the value of \(\mu_{\mathrm{CS}}\) and the resulting GW birefringence.

Before inserting numbers it is convenient to write eq.~\eqref{eq:strain_diff_approx}
in a form in which every factor is separately dimensionless. Since the CS coupling
$\alpha$ carries dimensions of $(\mathrm{length})^{2}$, the formal expansion in $\alpha$ of
eq.~\eqref{eq:metricexpansion} is controlled physically by the dimensionless combination
introduced by Ali-Ha\"imoud and Chen~\cite{Ali-Haimoud2011},
\begin{equation}
\zeta^{2} \;\equiv\; 128\pi\,\frac{\ell_{\rm cs}^{4}M^{2}}{R_\star^{6}}
\;=\; \frac{8\,\alpha^{2}M^{2}}{R_\star^{6}},
\label{eq:zeta_def}
\end{equation}
where the second equality uses $\ell_{\rm cs}^{2} = \alpha/\sqrt{16\pi}$, which follows from
matching the kinetic terms of eq.~\eqref{eq:action} to the normalisation of
Ref.~\cite{Ali-Haimoud2011}. Equation~\eqref{eq:zeta_def} measures the size of the
$\mathcal{O}(\alpha^{2})$ term relative to the general-relativistic one, and is
\emph{system-dependent} through the mean density of the object considered. Ref.~\cite{Ali-Haimoud2011}
showed numerically that $\zeta \lesssim 1$ delimits the linear regime, in which the CS
scalar field grows uniformly with the coupling; for $\zeta \gtrsim 1$ the scalar becomes
screened near the stellar surface and the perturbative treatment adopted here---and in the
derivation of the existing bounds---breaks down. We therefore quote our results in terms of
$\zeta$. In this notation eq.~\eqref{eq:phase_constdens_approx} takes the compact form
\begin{equation}
\mathcal{A}_{\mathrm{PV}}\equiv 2\Delta\Phi_{\mathrm{CS}}
= \frac{2}{15}\,\zeta^{2}\,\bigl(\sigma R_\star\bigr)\bigl(\Omega R_\star\bigr).
\label{eq:phase_dimensionless}
\end{equation}
Note that $\zeta \propto \alpha M/R_\star^{3}$, so that at \emph{fixed coupling} $\alpha$
eq.~\eqref{eq:phase_dimensionless} reproduces the compactness scaling
$\mathcal{A}_{\mathrm{PV}} \propto \alpha^{2}\sigma\Omega M^{2}/R_\star^{4} \propto \chi^{4}$
of eq.~\eqref{eq:strain_diff_approx}; the explicit $R_\star^{2}$ in
eq.~\eqref{eq:phase_dimensionless} appears because $\zeta$ itself carries a factor $R_\star^{-3}$.

For a typical neutron star with \(M = 1.5\,M_\odot=2.2\) km, \(R_\star = 12\) km,
\(\Omega = 10^{3}\) s\(^{-1}\) and GW angular frequency \(\sigma = 10^{3}\) s\(^{-1}\), the
individual dimensionless factors are $\sigma R_\star = \Omega R_\star \approx 4.0\times10^{-2}$.
Substituting into eq.~\eqref{eq:phase_dimensionless},
\begin{equation}
\mathcal{A}_{\mathrm{PV}} \approx 2.1\times10^{-4}\,\zeta^{2}
\label{eq:APV_numerical}
\end{equation}
at the fiducial parameters. For this star $\zeta = 1$ corresponds to
$\sqrt{\alpha} = 16.6$ km, at which the background scalar field satisfies
$\hat{\vartheta}(R_\star) = 8.4\times10^{-3}$ and the parameter
$\alpha\hat{\vartheta}/r^{2}$ from section~\ref{sec:gw_propagation} is $1.6\times10^{-2}$; the
perturbative treatment is therefore well under control at this ceiling. Saturating
$\zeta \sim 1$ gives a maximum predicted asymmetry
$\mathcal{A}_{\mathrm{PV}} \sim 2\times10^{-4}$, comfortably above the projected sensitivity of
third-generation detectors and indeed above the strain-asymmetry threshold of current
instruments (see section~\ref{sec:results_discussion}). The steep compactness dependence
remains apparent: at fixed $\alpha$, the same calculation for $R_\star = 10$ km gives
$\mathcal{A}_{\mathrm{PV}} = 4.4\times10^{-4}$ and for $R_\star = 14$ km it is
$1.2\times10^{-4}$, varying by a factor of $(14/10)^{4} = 3.8$ across this plausible radius range. We note,
however, that $\zeta$ grows as $R_\star^{-3}$ at fixed coupling, so that the perturbative
ceiling is reached at smaller $\alpha$ for more compact stars.

It is worth being explicit about the relation to the existing observational constraint. The only
current astrophysical bound on \emph{dynamical} CS gravity comes from measurements of
frame-dragging around the Earth by Gravity Probe B and the LAGEOS satellites, which give
$\xi^{1/4} \lesssim 10^{8}$ km~\cite{Ali-Haimoud2011}; in the normalisation of
eq.~\eqref{eq:action} this is $\xi = \alpha^{2}$, i.e. $\sqrt{\alpha} \lesssim 10^{8}$ km. (The
much tighter double-binary-pulsar constraint of Ref.~\cite{Yunes2009b} applies to the
\emph{non-dynamical} theory, where it bounds $\dot{\vartheta}$ rather than $\alpha$, and does not
carry over to the dynamical formulation considered here.) Because $\zeta$ scales with the mean
density of the source, the terrestrial bound is very weak when applied to a neutron star: it
corresponds to $\zeta_{\rm NS} \sim 4\times10^{13}$, in agreement with the estimate of
Ref.~\cite{Ali-Haimoud2011}. Inserted into eq.~\eqref{eq:APV_numerical} this would give
$\mathcal{A}_{\mathrm{PV}} \sim 10^{23}$, far outside the regime in which the perturbative
treatment is controlled. The existing constraint is therefore some seven orders of magnitude in
$\sqrt{\alpha}$ short of enforcing weak coupling on stellar scales, and the meaningful ceiling on
the predicted signal is set instead by the internal-consistency requirement
$\zeta \lesssim 1$. We accordingly interpret a measurement of $\mathcal{A}_{\mathrm{PV}}$ as a
direct constraint on $\zeta$, and hence on $\sqrt{\alpha}$, in a regime where the theory admits a
controlled description.

This constant-density model provides a benchmark, and---because the interior source vanishes for a uniform density profile---a conservative one: any centrally condensed equation of state adds a positive interior contribution through \(\mathcal{F}_0\) in eq.~\eqref{eq:strain_diff_constdens}. In the next subsection we consider a more realistic two-fluid model, where differential rotation between the neutron superfluid and the charged component can modify the interior frame-dragging profile and hence the value of \(\mu_{\mathrm{CS}}\) and the resulting GW birefringence.

\subsection{Model II: Two-Fluid Star with Differential Rotation}
\label{sec:two_fluid}

The constant-density single-fluid model captures relativistic frame-dragging but does not account for the
microscopic physics of pulsar glitches. Neutron stars are more appropriately described by a multicomponent system:
a neutron superfluid coexisting with a charged normal component (protons, electrons, and the solid crust). In the
presence of superfluidity, angular momentum can be stored differentially between the two components and released
suddenly during a glitch. Since the Pontryagin density depends on $\omega'(r)$, the internal rotational profile
plays a central role in determining the scalar source and the resulting GW birefringence.

We model the interior using a relativistic two-fluid framework in the slow-rotation approximation, following the formalism developed in~\cite{Andersson2001} and extended by subsequent work~\cite{Prix2005,Aranguren2022}. The two fluids rotate with distinct angular velocities,
\begin{equation}
\Omega_n \neq \Omega_p,
\qquad
\Delta_{2}\Omega(r) = \Omega_n(r) - \Omega_p(r),
\label{eq:two_omega}
\end{equation}
where $\Delta_{2}\Omega$ is the relative lag responsible for glitch activity. This internal lag is distinct from the observed glitch amplitude \(\Delta\Omega\) of eq.~\eqref{eq:glitch_profile}: the two are related by the fraction of the stellar moment of inertia held in the superfluid reservoir, \(\Delta\Omega/\Omega \sim (I_{\rm sf}/I),\Delta_2\Omega_0/\Omega_{\rm eff}\), so the reservoir lag that drives the correction exceeds the per-glitch jump. As shown below, the corotating limit of this model reduces exactly to the constant-density results of section~\ref{sec:constant_density}, so that the present
section may be read as an extension of that benchmark in which a phenomenalistic model of the glitch microphysics
is switched on. 

\subsubsection{Background configuration}

The background (nonrotating) stellar structure follows from the TOV equations,
\begin{align}
\frac{dm}{dr} &= 4\pi r^2 \rho(r), \label{eq:tov_mass} \\
\frac{dp}{dr} &= -\frac{[\rho(r) + p(r)]\,[m(r) + 4\pi r^3 p(r)]}{r\,[r - 2m(r)]}, \label{eq:tov_pressure}
\end{align}
where the total density and pressure are the sums of the neutron and proton fluid contributions:
$\rho = \rho_n + \rho_p$ and $p = p_n + p_p$.
The metric potential $\Phi(r)$ is determined by
\begin{equation}
\frac{d\Phi}{dr} = \frac{m(r) + 4\pi r^3 p(r)}{r\,[r - 2m(r)]},
\label{eq:tov_phi}
\end{equation}
with the boundary condition $\Phi(r) \to 0$ as $r \to \infty$ (or equivalently, matching to the Schwarzschild
exterior at the stellar surface).

These equations are solved from the center ($r=0$) outward, with initial conditions $m(0)=0$ and a chosen central
pressure $p_c$, until the pressure vanishes at the stellar surface $r = R_\star$. The results below are quoted for a
general background; where explicit numbers are required we specialise to the constant-density profile of
section~\ref{sec:constant_density}.

\subsubsection{Frame-dragging equation with two fluids}

In the slow-rotation approximation the frame-dragging function \(\omega(r)\) satisfies the Hartle equation with a
source given by the total inertial mass density weighted by the local angular velocity. For a multicomponent fluid
the source is the sum over species,
\begin{equation}
\frac{1}{r^4}\frac{d}{dr}\!\left[r^4 j(r)\frac{d\omega}{dr}\right]
= -16\pi a^{2} j \sum_{x} (\rho_x + p_x)\bigl(\Omega_x - \omega\bigr),
\label{eq:twofluid_framedrag}
\end{equation}
with \(j(r) = e^{-(\Phi+\Lambda)}\) and \(a = e^{\Lambda}\). We emphasise that
eq.~\eqref{eq:twofluid_framedrag} is written for \(\omega\) directly; the equivalent form in terms of
\(\bar\omega\) carries the matter source implicitly through the identity
\(\tfrac{4}{r}\,dj/dr = -16\pi a^{2} j (\rho+p)\), and the two should not be combined.

Introducing the \emph{neutron inertia fraction}
\begin{equation}
f(r) \equiv \frac{\rho_n + p_n}{(\rho_n+p_n)+(\rho_p+p_p)},
\label{eq:inertia_fraction}
\end{equation}
the sum in eq.~\eqref{eq:twofluid_framedrag} collapses eto:
\begin{equation}
\begin{aligned}
\sum_{x}(\rho_x+p_x)(\Omega_x-\omega) &= (\rho+p)\left[\Omega_{\rm eff}(r)-\omega\right],\\
\Omega_{\rm eff} &= \Omega_p + f\,\Delta_{2}\Omega ,
\end{aligned}
\label{eq:omega_eff}
\end{equation}
where \(\Omega_{\rm eff}\) is the inertia-weighted average of the two fluid angular velocities. Physically,
\(\Omega_{\rm eff}\) represents the rotation rate that would be observed if the two fluids were locked together; it is
the quantity associated with the total angular momentum of the star.

Equation~\eqref{eq:omega_eff} shows that, at this order, the two-fluid problem is \emph{identical} to a single-fluid
problem rotating at \(\Omega_{\rm eff}(r)\). Differential rotation therefore enters only through the radial variation
of \(\Omega_{\rm eff}\), that is, through the profiles of \(f(r)\) and \(\Delta_{2}\Omega(r)\). In the corotating limit
\(\Omega_n = \Omega_p = \Omega_{\rm eff} = \text{constant}\), eq.~\eqref{eq:twofluid_framedrag} reduces to the standard
single-fluid Hartle equation and hence, for a constant-density background, to the results of
section~\ref{sec:constant_density}.

\subsubsection{Perturbative solution for small differential rotation}

We assume the differential rotation is small, \(|\Delta_{2}\Omega| \ll \Omega_{\rm eff}\), and expand about the
corotating configuration. Writing \(\Omega_{\rm eff}(r) = \Omega_{\rm eff} + \delta\Omega(r)\) with
\begin{equation}
\delta\Omega(r) = f(r)\,\Delta_{2}\Omega(r),
\label{eq:deltaOmega}
\end{equation}
and correspondingly \(\omega = \omega^{(0)} + \delta\omega\), the leading lag-induced correction obeys
\begin{equation}
\frac{d}{dr}\!\left[r^4 j\,\frac{d\,\delta\omega}{dr}\right]
= -16\pi\,a^{2} j\, r^{4}\,\mathcal{W}(r)\,\Delta_{2}\Omega(r),
\label{eq:deltabar_ode}
\end{equation}
with weighting function
\begin{equation}
\mathcal{W}(r) = (\rho+p)\,f(r) = \rho_n(r) + p_n(r).
\label{eq:weighting_function}
\end{equation}
The combination \(\mathcal{W}(r)\) is simply the \emph{inertial mass density of the neutron superfluid}; it is the
component that carries the excess angular velocity, and it is the only combination of the two fluids' properties that
survives in eq.~\eqref{eq:omega_eff}. It is positive definite throughout the star, largest in the core where the
superfluid dominates, and falls toward the surface as the charged crust takes over. Thus \(\mathcal{W}(r)\) quantifies
how effectively the lag couples to the frame-dragging field.

Entrainment---the momentum coupling between the two fluids arising from strong interactions---modifies the individual
inertia densities and hence \(f(r)\), replacing \((\rho_n+p_n) \to (\rho_n+p_n+\varepsilon)\) and
\((\rho+p) \to (\rho+p+2\varepsilon)\) in eq.~\eqref{eq:inertia_fraction}, with \(\varepsilon\) the entrainment density.
Because \(\varepsilon\) enters both numerator and denominator, it shifts \(f\) by an amount of order
\(\varepsilon/(\rho+p)\), typically a few tens of percent in the core~\cite{Yeung2021}. This rescales
\(\mathcal{W}(r)\) but does not alter the parametric scalings derived below, since those depend on
\(\mathcal{W}\) only through the ratio in eq.~\eqref{eq:lag_ratio}; we therefore omit entrainment for simplicity and
note that its inclusion would change the numerical estimates by an \(\mathcal{O}(1)\) factor. [See Appendix~\ref{app:entrainment} for details]

For an analytic estimate, we adopt the phenomenological lag profile
\begin{equation}
\Delta_{2}\Omega(r) = \Delta_{2}\Omega_0\left(1 - \frac{r^2}{R_\star^2}\right),
\label{eq:lag_profile}
\end{equation}
which peaks in the interior and vanishes smoothly at the surface. Approximating \(\mathcal{W}(r)\) by its central value
\(\mathcal{W}_0\) and working to leading order in compactness (\(j \approx a \approx 1\)), integration of
eq.~\eqref{eq:deltabar_ode} with regularity at the centre gives
\begin{equation}
\frac{d\,\delta\omega}{dr} = -\,16\pi \mathcal{W}_0\Delta_{2}\Omega_0
\left(\frac{r}{5} - \frac{r^3}{7R_\star^2}\right).
\label{eq:deltadr}
\end{equation}

\subsubsection{Interior Pontryagin density}

The total frame-dragging gradient is the sum of the corotating (compactness-driven) part and the differential rotation contribution,
\begin{equation}
\omega'(r) = -\chi\frac{12\Omega_{\rm eff}}{5R_\star^2}r
- 16\pi\mathcal{W}_0\Delta_{2}\Omega_0\left(\frac{r}{5} - \frac{r^3}{7R_\star^2}\right),
\label{eq:omegaprime_twofluid}
\end{equation}
where the first term arises from the background corotating frame dragging (as in the constant-density model)
and the second term is the lag-induced correction from differential rotation. It is instructive to note that for a
near-uniform interior \(16\pi\mathcal{W}_0 \approx 12\chi f_0/R_\star^2\), so that the two terms in
eq.~\eqref{eq:omegaprime_twofluid} stand in the ratio \(f_0\Delta_{2}\Omega_0/\Omega_{\rm eff}\)---the fractional lag,
weighted by the neutron inertia fraction.

The interior Pontryagin density then follows from the general expression \(^{\ast}\!RR = \omega'(r)\cos\theta\,\mathcal{F}(r)\), giving
\begin{equation}
\begin{aligned}
^{\ast}\!RR_{\rm int}(r,\theta) &= -\cos\theta\,\mathcal{F}(r)
\Bigg[\chi\frac{12\Omega_{\rm eff}}{5R_\star^2}r \\
&+ 16\pi\mathcal{W}_0\Delta_{2}\Omega_0\left(\frac{r}{5} - \frac{r^3}{7R_\star^2}\right)\Bigg].
\label{eq:RR_int_twofluid}
\end{aligned}
\end{equation}
The two-fluid structure preserves the dipolar angular dependence while modifying the radial profile
through the \(\Delta_{2}\Omega_0\) term. Physically, the first term inside the brackets represents the Pontryagin density
sourced by the corotating part of the rotation. The second term, proportional to
\(\Delta_{2}\Omega_0\), is the additional contribution arising from the differential rotation between the
superfluid and the charged component during a glitch.

Both terms are multiplied by \(\mathcal{F}(r)\). Since \(\mathcal{F}\) vanishes identically for a uniform density
profile (appendix~\ref{app:F0_derivation}), the entire interior contribution---corotating and lag alike---requires a
centrally condensed background to be nonzero. The constant-density benchmark is therefore exactly the configuration in
which the glitch microphysics decouples from the scalar source, and the estimates below should be understood as
applying to a realistic equation of state, for which \(\mathcal{F}_0 > 0\).

\subsubsection{Matching to the exterior solution}

The exterior frame-dragging solution is unchanged: \(\omega_{\rm ext}(r) = 2J/r^3\) with \(J = I\Omega_{\rm eff}\).
Matching \(\bar\omega\) and its derivative at \(r = R_\star\) determines the angular momentum and the frame-dragging
integration constant \(c_2\). (See appendix~\ref{app:matching} for the general matching procedure.)
A glitch redistributes angular momentum between the superfluid and the crust but does not add to the total, so \(J\) is
conserved through the event. The exterior frame-dragging field \(2J/r^3\), and with it the exterior contribution
\(8\alpha MJ/R_\star^3\) to the scalar dipole moment, is therefore unchanged by the glitch. Any glitch-induced variation
in \(\mu_{\mathrm{CS}}\) must originate in the stellar interior, and is consequently proportional to
\(\mathcal{F}_0\).

The differential rotation correction modifies the matching condition, leading to an additional contribution
to the frame-dragging constant \(c_2\). Following the derivation in appendix~\ref{app:matching}, we obtain
\begin{equation}
c_2 = -2\Omega_{\rm eff} + \delta c_2^{\rm lag},
\qquad
\delta c_2^{\rm lag} \sim \mathcal{W}_0 \Delta_{2}\Omega_0 R_\star^2,
\label{eq:c2_twofluid}
\end{equation}
which has dimensions of $(\mathrm{length})^{-1}$ and is independent of
\(\alpha\), since \(c_2\) comes from the frame dragging in unmodified general relativity.

The scalar dipole moment \(\mu_{\mathrm{CS}}\) is then determined by solving the scalar field matching conditions, as
detailed in appendix~\ref{app:matching}. Using the total frame-dragging gradient from eq.~\eqref{eq:omegaprime_twofluid}
as the source in the scalar field equation, the matched scalar dipole moment for the two-fluid model becomes
\begin{equation}
\mu_{\mathrm{CS}} = \frac{8\alpha MJ}{R_\star^3}
+ \frac{1}{25}\alpha\chi\Omega_{\rm eff}\mathcal{F}_0 R_\star^3
+ \delta\mu_{\mathrm{CS}}^{\rm lag},
\label{eq:mu_twofluid}
\end{equation}
where the lag correction scales as
\begin{equation}
\delta\mu_{\mathrm{CS}}^{\rm lag} \sim \alpha\mathcal{W}_0 \Delta_{2}\Omega_0\mathcal{F}_0 R_\star^5 .
\label{eq:mulag_scaling}
\end{equation}
Note that \(\delta\mu_{\mathrm{CS}}^{\rm lag}\) carries the same factor \(\mathcal{F}_0\) as the corotating interior
term. The exact numerical coefficient depends on the detailed radial profiles of \(\mathcal{F}(r)\) and
\(\Delta_{2}\Omega(r)\), but the parametric scaling is robust.

\subsubsection{Impact on gravitational-wave birefringence}
\label{sec:birefringence_models}

With the scalar dipole moment \(\mu_{\mathrm{CS}}\) determined from the matching procedure [see eq.~\eqref{eq:mu_twofluid}], the birefringent phase shift accumulated by GWs is
\begin{equation}
\begin{aligned}
    \Delta\Phi_{\mathrm{CS}} \approx
\frac{\alpha\sigma}{6R_\star^3}
\Bigg( \frac{8\alpha MJ}{R_\star^3}
+ \frac{1}{25}&\alpha\chi\Omega_{\rm eff}\mathcal{F}_0 R_\star^3\\
&+ \delta\mu_{\mathrm{CS}}^{\rm lag} \Bigg).
\end{aligned}
\label{eq:phase_twofluid}
\end{equation}
So the fractional difference between the two polarizations becomes
\begin{equation}
\begin{aligned}
    \frac{|h_R - h_L|}{|h^{(0)}|} = &\frac{\alpha\sigma}{3R_\star^3}\Bigg|
\frac{8\alpha MJ}{R_\star^3}\\
&+ \frac{1}{25}\alpha\chi\Omega_{\rm eff}\mathcal{F}_0 R_\star^3+ \delta\mu_{\mathrm{CS}}^{\rm lag}
\Bigg|.
\end{aligned}
\label{eq:strain_diff_twofluid}
\end{equation}
The three contributions to \(\mu_{\mathrm{CS}}\) have distinct physical origins:

\begin{enumerate}
\item \textbf{Exterior part:} The term \(8\alpha MJ/R_\star^3\) is sourced by the vacuum Pontryagin density
      \(^{\ast}\!RR \propto MJ/r^{7}\) outside the star. It depends only on the conserved quantities \(M\) and \(J\)
      and is therefore insensitive both to the interior structure and to the glitch.

\item \textbf{Corotating interior part:} The term \(\tfrac{1}{25}\alpha\chi\Omega_{\rm eff}\mathcal{F}_0 R_\star^3\)
      arises from the compactness-driven frame-dragging present in any rotating star, and requires a centrally
      condensed profile through \(\mathcal{F}_0\).

\item \textbf{Differential-rotation part:} \(\delta\mu_{\mathrm{CS}}^{\rm lag}\) is sourced
      by the relative lag between the superfluid and normal components.
      This term is proportional to \(\Delta_{2}\Omega_0\), and therefore
      this is the turn that is boosted during a glitch, when the angular momenta of the
      two components become unbalanced. Since the exterior contribution is fixed by the conserved \(J\), it is
      the only part of \(\mu_{\mathrm{CS}}\) that can vary through the event.
\end{enumerate}

Since both interior terms carry the same factor \(\alpha\mathcal{F}_0\), their ratio is independent of the CS
coupling and of the interior curvature normalisation,
\begin{equation}
\frac{|\delta\mu_{\mathrm{CS}}^{\rm lag}|}{|\mu_{\mathrm{CS}}^{\rm corot,\,int}|}
\approx \frac{4\pi \mathcal{W}_0 \Delta_{2}\Omega_0 R_\star^{2}}{3\chi\Omega_{\rm eff}} .
\label{eq:lag_ratio}
\end{equation}
For a near-uniform interior the inertia density satisfies \(16\pi\mathcal{W}_0 \approx 12\chi f_0/R_\star^{2}\), where
\(f_0\) is the central neutron inertia fraction, and eq.~\eqref{eq:lag_ratio} reduces to the transparent
\begin{equation}
\frac{|\delta\mu_{\mathrm{CS}}^{\rm lag}|}{|\mu_{\mathrm{CS}}^{\rm corot,\,int}|}
\approx f_0\,\frac{\Delta_{2}\Omega_0}{\Omega_{\rm eff}} .
\label{eq:lag_ratio_simplified}
\end{equation}
This is simply the statement, already evident from eq.~\eqref{eq:omega_eff}, that the lag alters the effective
rotation rate by \(f\Delta_{2}\Omega\) and that every quantity in the problem scales with
\(\Omega_{\rm eff}\).

For the two-fluid star the corotating contribution is unchanged apart from the replacement
\(\Omega \to \Omega_{\rm eff}\), so that for the same fiducial parameters
\begin{equation}
\begin{aligned}
\Delta\Phi_{\mathrm{CS}}^{\mathrm{corot,\,int}} &\approx 1.1\times10^{-4}\,\zeta^{2}
\left(\frac{\Omega_{\rm eff}}{10^{3}\,\mathrm{s^{-1}}}\right)\\
&\times\left(\frac{\sigma}{10^{3}\,\mathrm{s^{-1}}}\right)
\left(\frac{R_\star}{12\ \mathrm{km}}\right)^{\!2},
\end{aligned}
\label{eq:phase_corot_numerical}
\end{equation}
identical in magnitude to the constant-density result; a realistic,
centrally condensed equation of state adds the interior contribution
\(\tfrac{1}{25}\alpha\chi\Omega_{\rm eff}\mathcal{F}_0R_\star^3\), which increases this estimate.

The differential-rotation contribution is most usefully quoted as a fraction of the corotating
interior term, since the factors \(\alpha\) and \(\mathcal{F}_0\)---and hence \(\zeta\)---cancel in
the ratio [eq.~\eqref{eq:lag_ratio_simplified}],
\begin{equation}
\frac{\Delta\Phi_{\mathrm{CS}}^{\mathrm{lag}}}{\Delta\Phi_{\mathrm{CS}}^{\mathrm{corot,\,int}}}
\approx f_0\,\frac{\Delta_{2}\Omega_0}{\Omega_{\rm eff}}
\approx 9\times10^{-4}
\left(\frac{f_0}{0.9}\right)
\left(\frac{\Delta_{2}\Omega_0/\Omega_{\mathrm{eff}}}{10^{-3}}\right).
\label{eq:phase_lag_numerical}
\end{equation}
Observed glitch jumps \(\Delta\Omega/\Omega \sim 10^{-9}\)--\(10^{-5}\) correspond to reservoir lags
\(\Delta_{2}\Omega_0/\Omega_{\rm eff}\) reaching at most \(\sim 10^{-3}\) for the largest Vela-type
events. (We shall use this as a benchmark for relatively large glitch events from bright neutron stars.)
So the lag correction stays at or below the sub-percent level throughout the observed
population. Since \(\mathcal{F}_0\) cancels in eq.~\eqref{eq:phase_lag_numerical}, this conclusion is
independent of the equation of state; and because \(f_0 < 1\), the fractional reservoir lag is a hard
ceiling on the correction, so it cannot be evaded even for lags beyond those observed. For realistic
glitch parameters the corotating contribution therefore dominates the birefringent signal entirely. 

The two-fluid model therefore demonstrates that glitch microphysics does not modify the parity-violating GW signal at
any observationally relevant level; the birefringent phase shift is set by the star's total angular momentum and
time-averaged interior structure, not by the transient redistribution of angular momentum between components. The
transient character of the observed asymmetry (section~\ref{sec:results_discussion}) arises from the glitch-excited
GW burst \(h^{(0)}(t)\) itself rather than from any time dependence of \(\mu_{\mathrm{CS}}\). The constant-density
benchmark, supplemented by a realistic \(\mathcal{F}_0\), should therefore provide a reliable estimate of the expected
signal strength.

\section{Results and Discussion}
\label{sec:results_discussion}

\subsection{Dependence on Stellar Parameters}
\label{sec:eos_dependence}

Our main result is the fractional polarization asymmetry between right- and left-handed GWs emitted during a pulsar glitch in CS gravity. Several features are worth emphasizing. First, the asymmetry scales as $\alpha^2$, increases linearly with GW frequency $\sigma$ and stellar rotation rate $\Omega$, and decreases as the \emph{fourth} power of the stellar radius $R_\star$ at fixed mass---equivalently, it scales as $\chi^4$ with $\chi = M/R_\star$---strongly favoring rapidly rotating compact neutron stars. Second, the effect is purely dispersive---it modifies the phase of the waveform without changing the overall amplitude, leading to a relative shift between the two circular polarizations.

We quote our results throughout in terms of the dimensionless coupling $\zeta$ of eq.~\eqref{eq:zeta_def}, for
which the perturbative treatment requires $\zeta \lesssim 1$. Equation~\eqref{eq:phase_dimensionless} expresses
$\mathcal{A}_{\rm PV}$ in terms of manifestly dimensionless factors, and
for our standard fiducial star ($M = 1.5\,M_\odot$, $R_\star = 12$ km, and $\Omega = 10^3$ s$^{-1}$) it gives 
$\mathcal{A}_{\rm PV} \approx 1.3\times10^{-3}\,\zeta^{2}$ at $\nu=\sigma/2\pi= 1$ kHz. Since $\zeta \propto \alpha M/R_\star^{3}$, eq.~\eqref{eq:phase_dimensionless} is equivalent to the compactness scaling $\mathcal{A}_{\rm PV} \propto \chi^4$ quoted above when the coupling $\alpha$, rather than $\zeta$, is held fixed.

\begin{figure}[htbp]
\centering
\includegraphics[width=0.95\columnwidth]{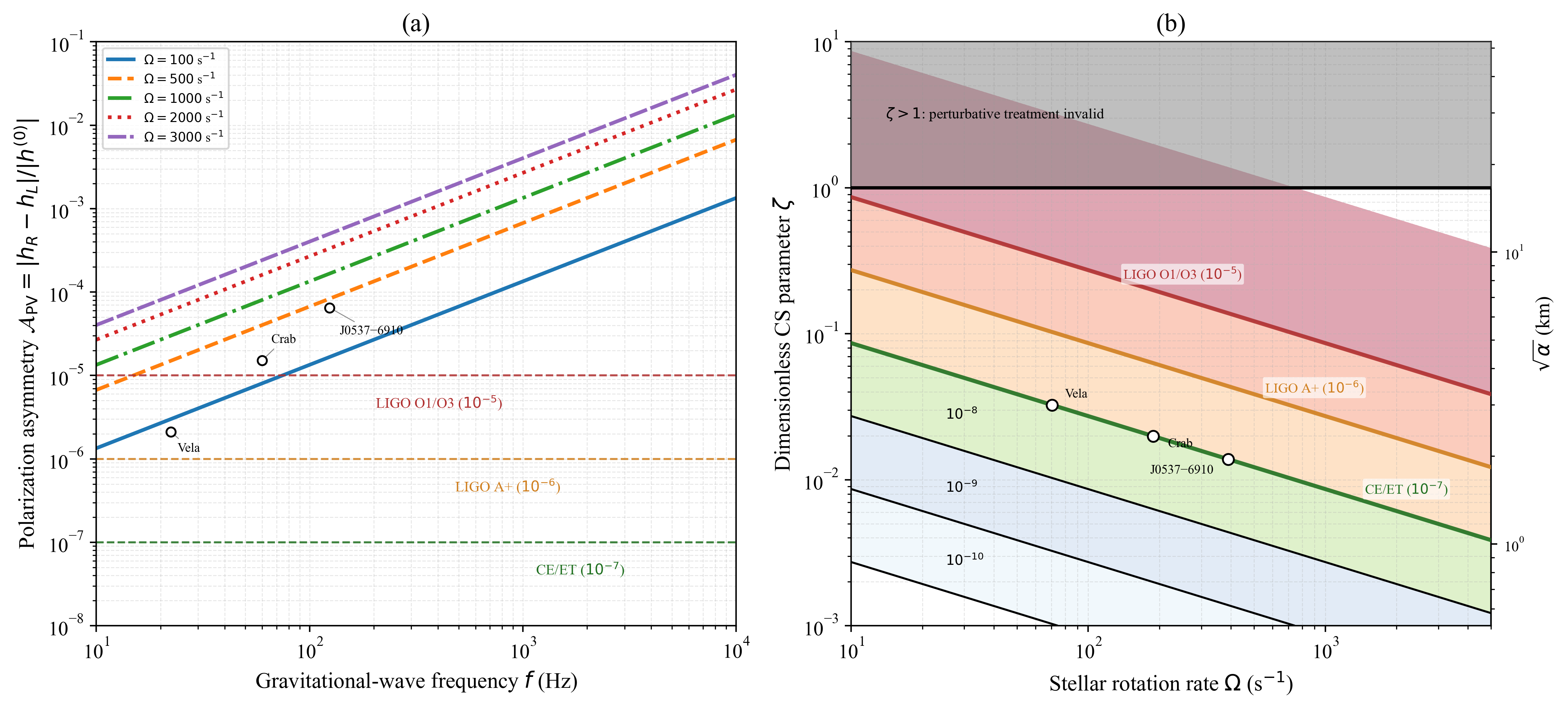}
\caption{(a) Polarization asymmetry $\mathcal{A}_{\rm PV}$ as a function of GW frequency $\nu = \sigma/2\pi$ for different stellar rotation rates $\Omega$, evaluated at the perturbative ceiling $\zeta = 1$ with $M = 1.5\,M_\odot$ and $R_\star = 12$ km; the asymmetry scales as $\zeta^{2}$, so the curves are upper envelopes. The linear scaling $\mathcal{A}_{\rm PV} \propto f$ reflects the cumulative propagation phase acquired through the CS interaction. Horizontal dashed lines mark the strain-asymmetry thresholds of LIGO O1/O3 ($10^{-5}$), LIGO A+ ($10^{-6}$) and CE/ET ($10^{-7}$). Markers indicate known glitching pulsars, each placed at its own rotation rate $\Omega$ and at the quadrupolar emission frequency $\nu=\Omega/\pi$. (b) Contours of constant $\mathcal{A}_{\rm PV}$ in the $\Omega$--$\zeta$ plane at fixed frequency $f = 1$ kHz. Colored regions indicate the asymmetry level, from $10^{-10}$ (pale blue) to $10^{-5}$ (dark red); the coloured curves mark the values of $\zeta$ that each detector generation could probe at a given rotation rate, and the open circles show where the three labelled pulsars fall on the CE/ET curve. The right-hand axis gives the equivalent coupling $\sqrt{\alpha}$, with $\zeta = 1$ corresponding to $\sqrt{\alpha} = 16.6$ km for this star. The shaded region above $\zeta = 1$ lies outside the domain of validity of the perturbative treatment adopted here; note that the entire detectable region lies below it.}
\label{fig:asymmetry}
\end{figure}

Figure~\ref{fig:asymmetry} illustrates this behavior. The left panel shows $\mathcal{A}_{\rm PV}$ as a function of frequency for different rotation rates, demonstrating the linear scaling $\mathcal{A}_{\rm PV} \propto \nu = \sigma/2\pi$. The right panel maps the detectability in the $\Omega$--$\zeta$ plane at fixed frequency $\nu = 1$ kHz, showing that millisecond pulsars ($\Omega \sim 10^3$ s$^{-1}$) observed at kilohertz frequencies are the most promising targets. At the perturbative ceiling $\zeta \sim 1$ the predicted asymmetry for such a source is $\mathcal{A}_{\rm PV} \approx 1.3\times10^{-3}$, rising to $\sim 4\times10^{-3}$ for the fastest known rotators---in both cases above the strain-asymmetry threshold not only of third-generation instruments but for current detectors. Conversely, and more usefully, a third-generation detector reaching $\mathcal{A}_{\rm PV} \sim 10^{-7}$ would constrain $\zeta \lesssim 8.6\times10^{-3}$ for a millisecond pulsar, corresponding to $\sqrt{\alpha} \lesssim 1.5$ km, and $\zeta \lesssim 5.0\times10^{-3}$ ($\sqrt{\alpha} \lesssim 1.2$ km) for the fastest rotators. The essential point is that the entire accessible region of Fig.~\ref{fig:asymmetry} lies two orders of magnitude below the perturbative ceiling, so that such a measurement would constrain the theory in a regime where it admits a controlled description---which the existing bound, at $\zeta_{\rm NS} \sim 10^{13}$, does not.

\begin{figure}[htbp]
\centering
\includegraphics[width=0.9\columnwidth]{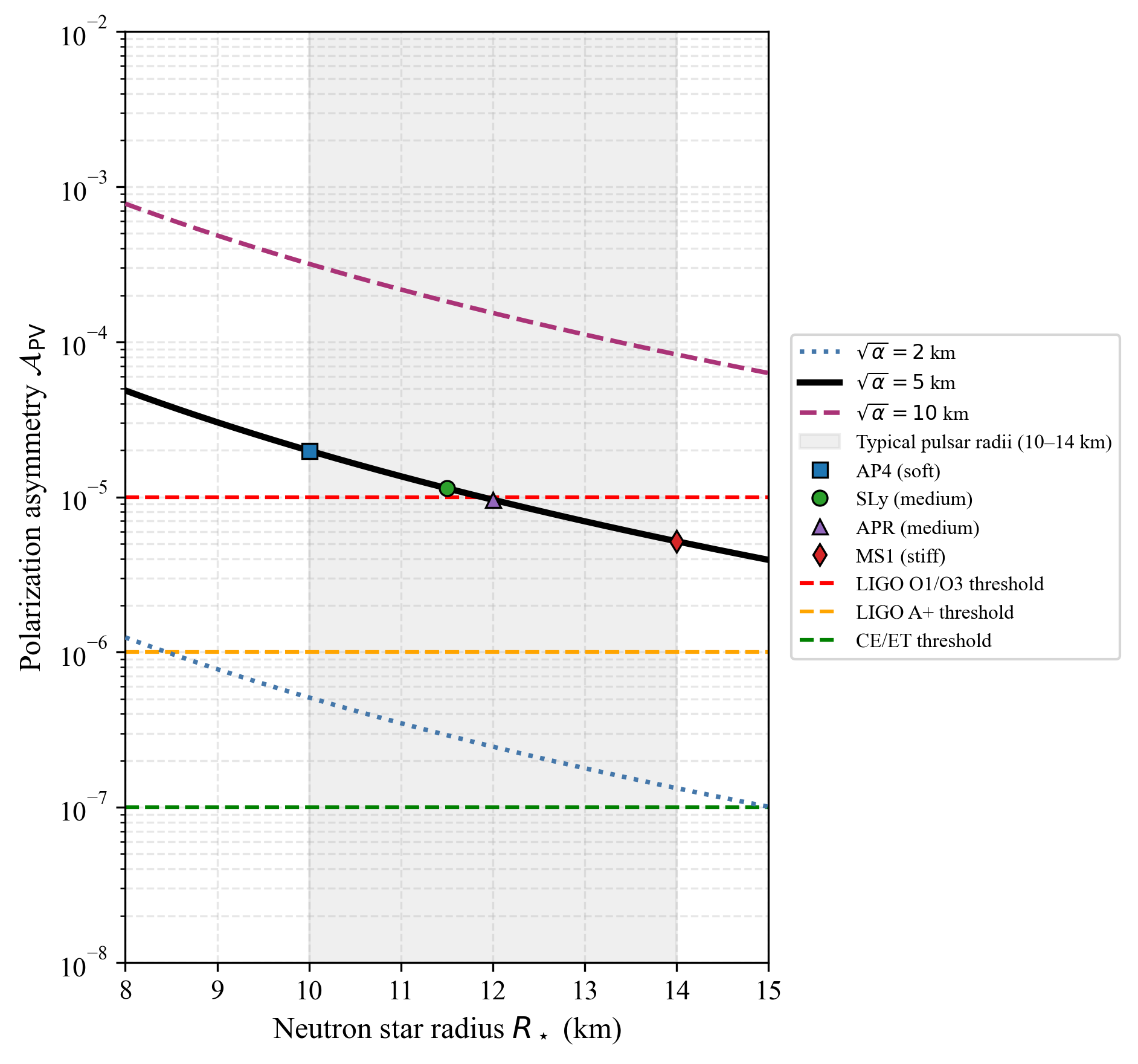}
\caption{Polarization asymmetry \(\mathcal{A}_{\rm PV}\) as a function of neutron star radius \(R_\star\) for a millisecond pulsar (\(\Omega = 1000\) s\(^{-1}\)) at GW frequency \(\nu = 1\) kHz and canonical mass \(M = 1.4\,M_\odot\). Curves are shown at \emph{fixed} CS coupling \(\sqrt{\alpha} = 2\), \(5\), and \(10\) km, so that they exhibit the compactness scaling \(\mathcal{A}_{\rm PV} \propto \chi^4 = (M/R_\star)^4\) derived from the constant-density model, for which the scalar dipole is sourced entirely in the near-zone exterior. Colored markers indicate predictions from selected EoSs, placed on the \(\sqrt{\alpha} = 5\) km curve: AP4 (soft, \(R_\star \sim 10\) km), SLy (medium, \(R_\star \sim 11.5\) km), APR (medium, \(R_\star \sim 12\) km), and MS1 (stiff, \(R_\star \sim 14\) km). Horizontal dashed lines mark detector thresholds (LIGO O1/O3, LIGO A+, and CE/ET). The dark shaded region at small radii indicates where the dimensionless parameter exceeds unity, \(\zeta > 1\), for the largest coupling shown and the perturbative treatment ceases to apply; for \(\sqrt{\alpha} \leq 5\) km this region lies below the plotted $R_{\star}$ range entirely. The predicted asymmetries span a factor of \(3.8\) across the plausible radius range, compared with a factor of \(2.0\) for a \(1/R_\star^2\) dependence; a realistic equation of state adds a positive interior contribution and would raise these values.}
\label{fig:radius}
\end{figure}

The sensitivity of the parity-violating signal to the stellar EoS and the corresponding compactness is illustrated in figure~\ref{fig:radius}. The steep compactness dependence is a distinctive feature of the present result. Because the scalar dipole is sourced predominantly in the near-zone exterior, where $^{\ast}\!RR \propto r^{-7}$, the asymmetry grows as $R_\star^{-4}$ at fixed coupling rather than the shallower $R_\star^{-2}$ that would follow from an interior-dominated source. As discussed in section~\ref{sec:interior_models}, this substantially enhances the sensitivity of the signal to the neutron-star equation of state.
A realistic, centrally condensed EoS reinstates a positive interior contribution through $\mathcal{F}_0$ and yields an intermediate scaling, so that figure~\ref{fig:radius} should be read as a conservative estimate of both the amplitude and the steepness.

For realistic EoSs, the stellar radius typically varies between $10$ and $14$\,km for a $1.4\,M_\odot$ pulsar, leading to a factor of $\sim 4$ variation in the predicted asymmetry [since $(14/10)^4 \approx 3.84$]. Figure~\ref{fig:radius} demonstrates this hierarchy: the ``soft'' and ``medium'' EoSs, such as AP4 ($R_\star \sim 10$\,km), SLy ($R_\star \sim 11.5$\,km)~\cite{DouchinHaensel2001}, and APR ($R_\star \sim 12$\,km)~\cite{Akmal1998}, exhibit the largest asymmetries, with
\begin{equation}
\mathcal{A}_{\rm PV} = \left\{2.0,\;1.1,\;0.96\right\}\times10^{-5}
\left(\frac{\sqrt{\alpha}}{5\ \mathrm{km}}\right)^{\!4}
\end{equation}
respectively, while the ``stiff'' MS1 model ($R_\star \sim 14$\,km)~\cite{MullerSerot1996} produces the weakest signal, $5.2\times10^{-6}(\sqrt{\alpha}/5\,\mathrm{km})^{4}$, on account of its lower compactness. At the reference coupling $\sqrt{\alpha} = 5$ km all four models correspond to $\zeta$ between $0.05$ and $0.15$, comfortably within the perturbative regime, and all lie above the projected sensitivity of third-generation detectors; the three most compact also exceed the strain-asymmetry threshold of current instruments.

Expressed as a constraint rather than a prediction, the same hierarchy implies that a CE/ET measurement at $\mathcal{A}_{\rm PV} = 10^{-7}$ would reach $\sqrt{\alpha} \lesssim 1.33$ km for AP4 and $\sqrt{\alpha} \lesssim 1.86$ km for MS1. Because $\mathcal{A}_{\rm PV} \propto \alpha^2$, the factor of $3.8$ in signal amplitude corresponds to only a factor of $1.4$ in the coupling probed---the ratio of the radii themselves, since the reach scales as $\sqrt{\alpha} \propto R_\star$. These results suggest that a precise measurement of the birefringent phase shift from a pulsar glitch could not only test dynamical CS gravity but also provide a novel, independent constraint on the neutron star EoS.

One caveat should be noted. Because $\zeta \propto \alpha M/R_\star^{3}$, the perturbative ceiling is itself system
dependent; at fixed coupling, more compact stars have larger $\zeta$ and approach the boundary of validity first,
with $\zeta = 1$ reached at $R_\star = \sqrt{2}(\alpha M)^{1/3}$. For $\sqrt{\alpha} = 10$ km this occurs at
$R_\star \approx 8.4$ km, outside the range spanned by realistic EoSs but close enough that the constraint is worth keeping in view when extrapolating to more compact configurations.

\subsection{Detectability of Known Pulsars}
\label{sec:pulsar_detectability}

\begin{figure}[htbp]
\centering
\includegraphics[width=0.9\columnwidth]{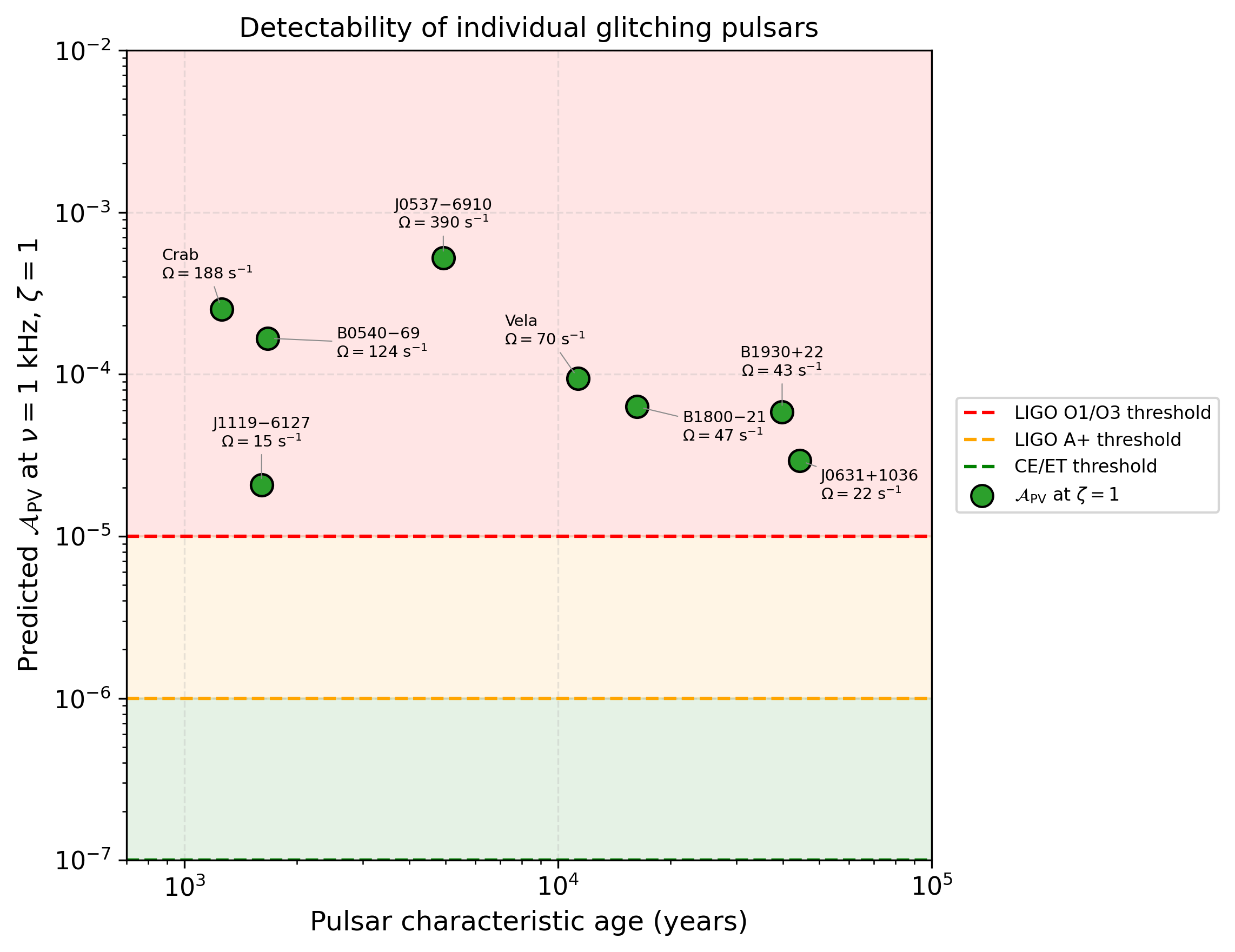}
\caption{Predicted polarization asymmetry \(\mathcal{A}_{\rm PV} = |h_R - h_L|/|h^{(0)}|\) at \(f = 1000\) Hz for known glitching pulsars, evaluated at the perturbative ceiling \(\zeta = 1\) with \(M = 1.5\,M_\odot\) and \(R_\star = 12\) km; the asymmetry scales as \(\zeta^{2}\), so the plotted values are upper envelopes. Shaded horizontal bands indicate detectability regions: LIGO O1/O3 (red, \(\mathcal{A}_{\rm PV} > 10^{-5}\)), LIGO A+ (orange, \(>10^{-6}\)), and CE/ET (green, \(>10^{-7}\)). Each point is labelled with its angular velocity \(\Omega = 2\pi F_0\), which together with the frequency sets the predicted asymmetry; the characteristic age is plotted only to locate each object within the glitching population. At the perturbative ceiling all eight objects lie within reach of third-generation detectors and the fastest rotators exceed current-instrument thresholds; the corresponding constraints on the coupling are listed in Table~\ref{tab:pulsar_parameters}.}
\label{fig:pulsar_detectability}
\end{figure}

Figure~\ref{fig:pulsar_detectability} places known glitching pulsars on the age--asymmetry plane. Because $\mathcal{A}_{\rm PV} \propto \Omega$ for fixed GW frequency and stellar structure, the ordering of the sample is set entirely by rotation rate; the characteristic age is shown to indicate where each object sits in the glitching population rather than because it enters the prediction. The fastest rotator in the sample, J0537$-$6910 ($\Omega = 390$ s$^{-1}$)~\cite{Antonopoulou2018}, produces the largest asymmetry, $\mathcal{A}_{\rm PV} = 5.2\times10^{-4}\,\zeta^{2}$, followed by the Crab ($\Omega = 188$ s$^{-1}$) and B0540$-$69 ($\Omega = 124$ s$^{-1}$) at $2.5\times10^{-4}\,\zeta^{2}$ and $1.7\times10^{-4}\,\zeta^{2}$. Slower objects such as Vela ($\Omega = 70$ s$^{-1}$) and J0631+1036 ($\Omega = 22$ s$^{-1}$) yield correspondingly smaller signals, but at the perturbative ceiling $\zeta \sim 1$ every object in the sample lies above the projected CE/ET threshold, and the three fastest exceed the strain-asymmetry threshold of current instruments.

Since $\mathcal{A}_{\rm PV} \propto \zeta^2$, the more useful statement is the inverse one: each pulsar defines the value of the coupling that a given detector generation could constrain. A CE/ET measurement at $\mathcal{A}_{\rm PV} = 10^{-7}$ would reach $\zeta \lesssim 1.4\times10^{-2}$ for J0537$-$6910, corresponding to $\sqrt{\alpha} \lesssim 2.0$ km, and $\zeta \lesssim 7.0\times10^{-2}$ ($\sqrt{\alpha} \lesssim 4.4$ km) even for the slowest object considered, J1119$-$6127. Because the reach scales only as the fourth root of the signal, $\sqrt{\alpha} \propto \Omega^{-1/2}$, the sample spans a factor of $25$ in predicted asymmetry but merely a factor of $2.2$ in the coupling it can constrain. Target selection is therefore far less critical than the linear $\Omega$ dependence might suggest: any of these objects would improve on the existing bound $\sqrt{\alpha} \lesssim 10^{8}$ km by some seven orders of magnitude, and---crucially---every entry in Table~\ref{tab:pulsar_parameters} lies two orders of magnitude below the perturbative ceiling, so that all of these constraints would be obtained in a regime where the theory admits a controlled description.

We emphasise that the rotation rates quoted here are angular velocities, $\Omega = 2\pi F_0$, computed from catalogue spin frequencies $F_0$; confusing these with $F_0$ itself shifts $\Omega$---and hence $\mathcal{A}_{\rm PV}$---by a factor of $2\pi$. Characteristic ages are obtained as $\tau_c = -F_0/(2\dot{F_0})$.

\begin{table}[h!]
\caption{Parameters and constraints for targeted pulsar searches.}
\label{tab:pulsar_parameters}
\centering
\resizebox{\columnwidth}{!}{%
\begin{tabular}{l c c c c c c}
\hline
\hline
Pulsar & \(P\) (ms) & \(\Omega\) (s\(^{-1}\)) & Age (yr) & \(\mathcal{A}_{\rm PV}/\zeta^{2}\) & \(\zeta\) for CE/ET & \(\sqrt{\alpha}\) (km) \\
\hline
J0537$-$6910 & \(16.1\)  & 390 & \(4.9\times10^{3}\) & \(5.2\times10^{-4}\) & \(0.014\) & \(2.0\) \\
Crab       & \(33.4\)  & 188 & \(1.3\times10^{3}\) & \(2.5\times10^{-4}\) & \(0.020\) & \(2.3\) \\
B0540$-$69   & \(50.6\)  & 124 & \(1.7\times10^{3}\) & \(1.7\times10^{-4}\) & \(0.024\) & \(2.6\) \\
Vela       & \(89.3\)  & 70  & \(1.1\times10^{4}\) & \(9.4\times10^{-5}\) & \(0.033\) & \(3.0\) \\
B1800$-$21   & \(133.7\) & 47  & \(1.6\times10^{4}\) & \(6.3\times10^{-5}\) & \(0.040\) & \(3.3\) \\
B1930+22   & \(144.5\) & 43  & \(4.0\times10^{4}\) & \(5.8\times10^{-5}\) & \(0.041\) & \(3.4\) \\
J0631+1036 & \(287.8\) & 22  & \(4.4\times10^{4}\) & \(2.9\times10^{-5}\) & \(0.058\) & \(4.0\) \\
J1119$-$6127 & \(408.0\) & 15  & \(1.6\times10^{3}\) & \(2.1\times10^{-5}\) & \(0.070\) & \(4.4\) \\
\hline
\hline
\end{tabular}%
}
\end{table}

\subsection{Time-Domain Signature}

An important qualitative feature of the CS correction is that it introduces a time-derivative term into the waveform: \(h_{R,L}(t) \approx h^{(0)}(t) \mp (\alpha \mu_{\mathrm{CS}}/6R_\star^3)\, \dot{h}^{(0)}(t)\). For the constant-density model, \(\mu_{\mathrm{CS}} \approx \alpha M^2\Omega / R_\star\), so the correction is proportional to \(\alpha^2\). This means the polarization asymmetry is transient, appearing only when the waveform is changing rapidly.

\begin{figure}[htbp]
\centering
\includegraphics[width=0.9\columnwidth]{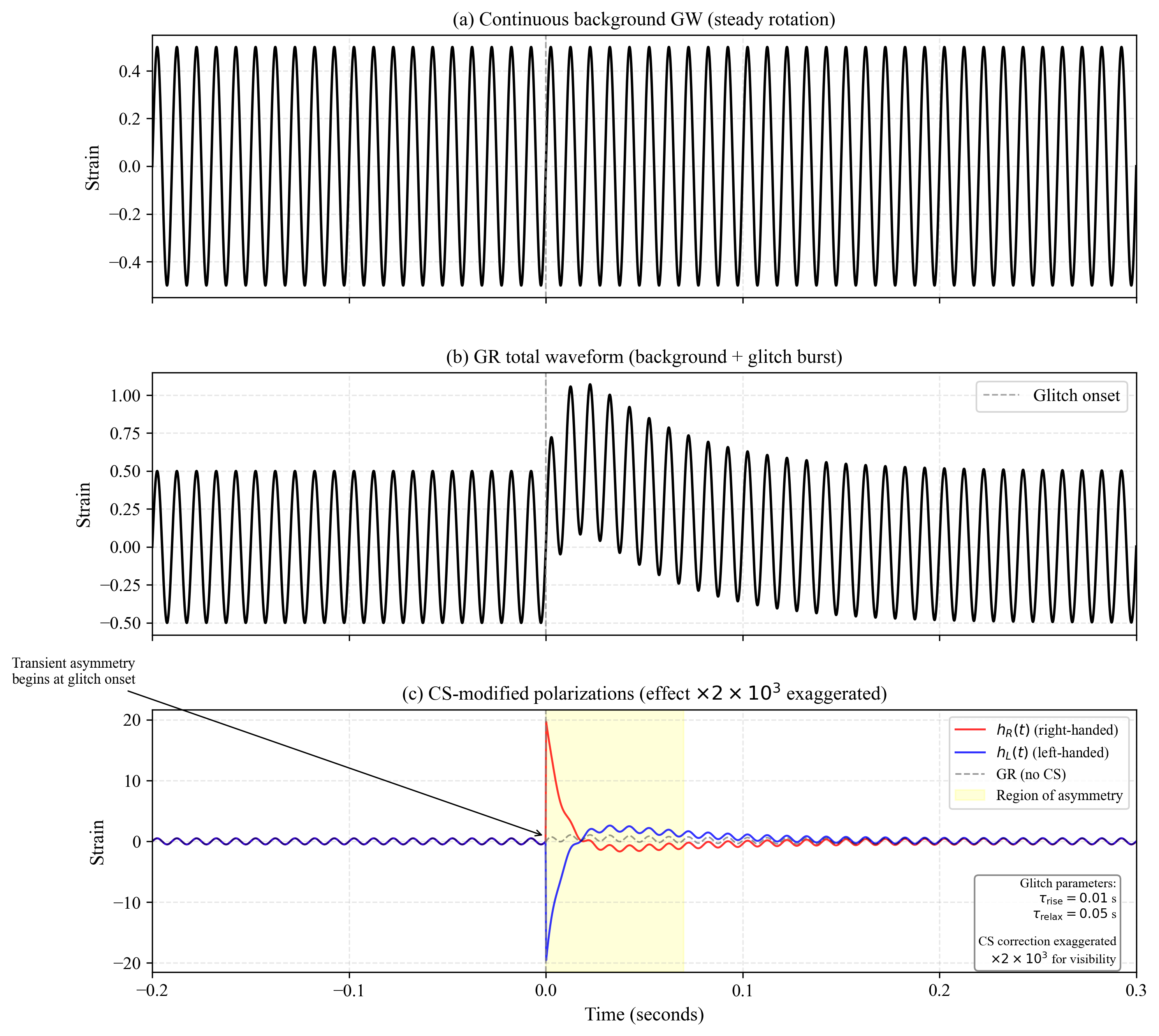}
\caption{Time-domain GW strain from a pulsar glitch in CS gravity, showing the transient polarization asymmetry. (a) Continuous background GW from a steadily rotating neutron star (e.g., from ellipticity or r-mode oscillations). (b) Total GR waveform including the glitch burst, which appears as a small bump at \(t = 0\). (c) Right-handed (\(h_R\), red) and left-handed (\(h_L\), blue) circular polarizations in CS gravity. The glitch triggers a temporary polarization asymmetry: the two polarizations separate during the rise and relaxation phases, then return to being identical after the glitch subsides. The effect is exaggerated for visualization; the actual asymmetry is $\mathcal{A}_{\mathrm{PV}} \approx 1.3\times10^{-3}\,\zeta^2 $ times the GR amplitude at kilohertz frequencies.}
\label{fig:transient}
\end{figure}

Figure~\ref{fig:transient} illustrates this effect in the context of a continuous GW background (e.g., from a rotating neutron star mountain or r-mode oscillations). Before the glitch, the two circular polarizations are identical. At the glitch onset (\(t = 0\)), the CS correction induces a temporary polarization asymmetry that peaks during the rapid rise, reverses sign during the relaxation, and then disappears once the star returns to equilibrium. This transient asymmetry is a distinctive signature of gravitational parity violation.

\subsection{Comparison of Interior Models}
\label{sec:model_comparison}

The two-fluid model, which incorporates differential rotation between the neutron superfluid and the charged component, introduces an additional contribution to the scalar dipole moment. From the matching procedure in appendix~\ref{app:matching}, the scalar dipole moment takes the form \(\mu_{\mathrm{CS}} = \mu_{\mathrm{CS}}^{\rm corot\,int} + \delta\mu_{\mathrm{CS}}^{\rm lag}\), where the corotating part dominates for all realistic glitch parameters.

According to eq.~\eqref{eq:lag_ratio_simplified} the lag alters the effective rotation rate by \(f\,\Delta_2\Omega\), and every quantity in the problem scales with \(\Omega_{\rm eff}\). Because \(f_0\) is a ratio of inertia densities it satisfies \(0 < f_0 < 1\), so the fractional lag itself provides a hard upper bound on the correction; including entrainment drives \(f_0\) toward \(1/2\) (see appendix~\ref{app:entrainment}) and can only lower it further.

\begin{figure}[htbp]
\centering
\includegraphics[width=0.9\columnwidth]{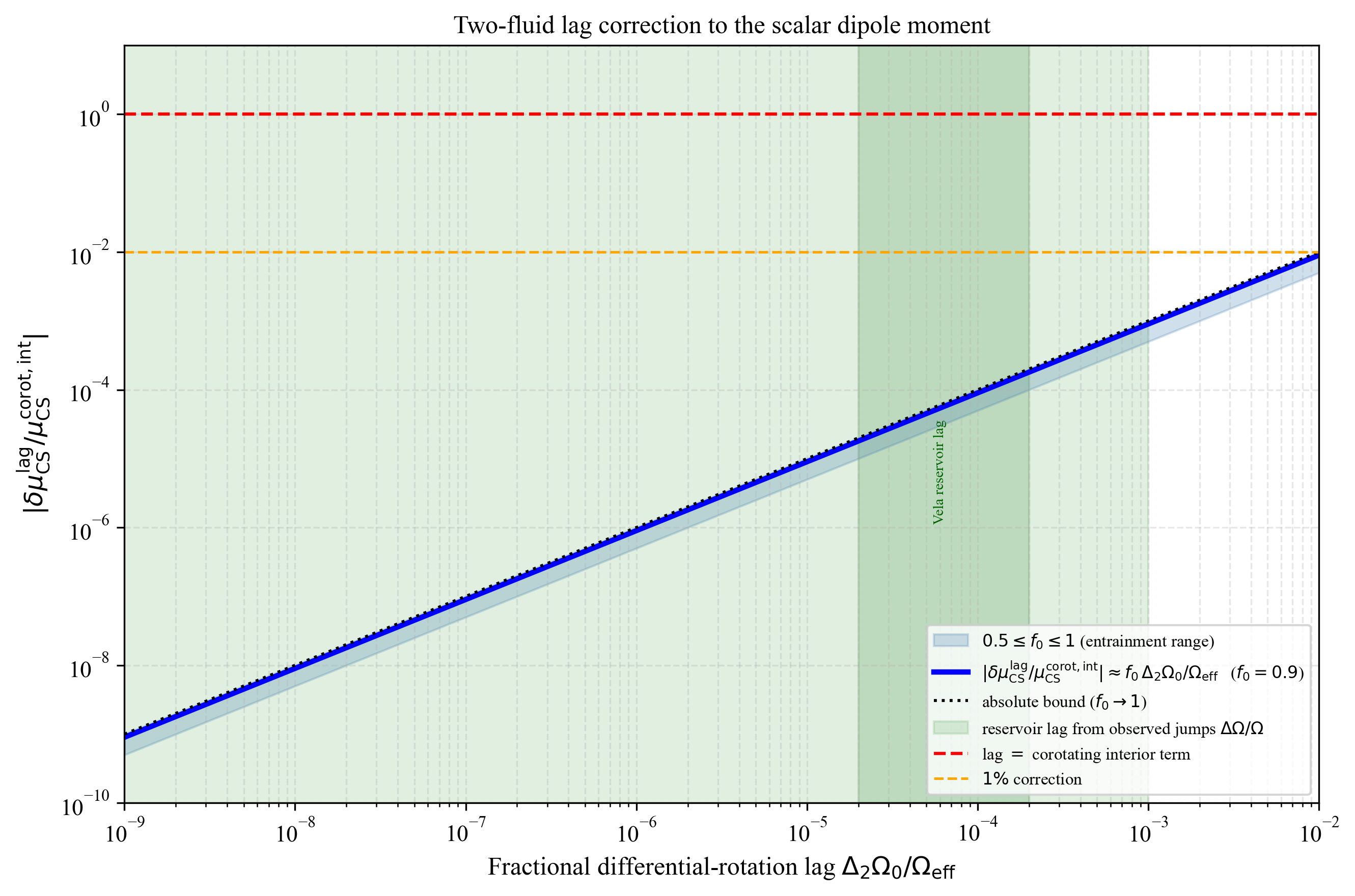}
\caption{Two-fluid lag correction to the scalar dipole moment, relative to the corotating interior contribution, as a function of the fractional differential rotation lag \(\Delta_2\Omega_0/\Omega_{\rm eff}\). The blue curve uses eq.~\eqref{eq:lag_ratio_simplified} with a central neutron inertia fraction \(f_0 = 0.9\); the shaded band spans \(0.5 \leq f_0 \leq 1\), the range accessible once entrainment is included. The dotted line shows the absolute bound \(f_0 \to 1\), which cannot be exceeded since \(f_0\) is a ratio of inertia densities. The ratio is independent of the CS coupling \(\alpha\) and of the equation of state, because both terms carry the same factor \(\alpha\mathcal{F}_0\). The green band marks the range of observed glitch amplitudes, \(\Delta\nu/\nu \sim 10^{-9}\)--\(10^{-5}\); the red dashed line indicates where the lag correction would equal the corotating interior term. Even for the largest recorded glitches the correction remains below \(10^{-5}\), so the corotating contribution determines the birefringent signal entirely.}
\label{fig:model_comparison2}
\end{figure}

Figure~\ref{fig:model_comparison2} shows this ratio as a function of the fractional lag. For observed glitch amplitudes, \(\Delta_2\Omega_0/\Omega_{\rm eff} \sim 10^{-9}\)--\(10^{-5}\), the lag correction lies between \(10^{-9}\) and \(10^{-5}\) of the corotating interior term---and smaller still relative to the total \(\mu_{\mathrm{CS}}\), which is dominated by the exterior contribution. A Vela-sized glitch (\(\Delta\nu/\nu \approx 2\times10^{-6}\)) yields a correction of \(2\times10^{-6}\). Reaching even the percent level would require a fractional lag of order \(10^{-2}\), three orders of magnitude larger than the largest event on record, and a correction of order unity is excluded outright by \(f_0 < 1\).

We therefore conclude that glitch microphysics does not modify the parity-violating GW signal at any observationally relevant level. The birefringent phase shift is set by the star's total angular momentum and time-averaged interior structure, not by the transient redistribution of angular momentum between components. The transient character of the observed asymmetry (figure~\ref{fig:transient}) arises from the glitch-excited GW burst \(h^{(0)}(t)\) itself rather than from any time dependence of \(\mu_{\mathrm{CS}}\). Because eq.~\eqref{eq:lag_ratio_simplified} is independent of the equation of state, this conclusion holds for any interior model, and the constant-density benchmark---supplemented by a realistic \(\mathcal{F}_0\)---provides a reliable estimate of the expected signal strength.

\subsection{Projected Constraints on the Chern-Simons Coupling}
\label{sec:existing_bounds}

The measurement proposed here would be substantially more constraining than any existing or presently projected test of dynamical CS gravity. A third-generation detector resolving $\mathcal{A}_{\rm PV} \sim 10^{-7}$ from a glitch in J0537$-$6910 would reach $\zeta \lesssim 1.4\times10^{-2}$ (Table~\ref{tab:pulsar_parameters}), corresponding to
\begin{equation}
\sqrt{\alpha} \lesssim 2\ \mathrm{km},
\label{eq:projected_bound}
\end{equation}
and even the slowest glitching pulsar in the sample would reach $\sqrt{\alpha} \lesssim 4.4$ km. The significance of these figures is best appreciated by comparison with the existing limits.

The only current astrophysical constraint on \emph{dynamical} CS gravity comes from measurements of frame-dragging around the Earth by Gravity Probe B and the LAGEOS satellites, which give $\xi^{1/4} \lesssim 10^{8}$ km~\cite{Ali-Haimoud2011}. In the normalisation of eq.~\eqref{eq:action} the CS length scale is related to our coupling by $\xi = \alpha^{2}$, so this is $\sqrt{\alpha} \lesssim 10^{8}$ km. We stress that the much tighter double-binary-pulsar bound of Ref.~\cite{Yunes2009b}, and the LAGEOS constraint of Ref.~\cite{Smith2008}, both apply to the \emph{non-dynamical} theory, in which the pseudoscalar is externally prescribed and the bound is placed on $\dot{\vartheta}$ rather than on $\alpha$; as emphasised in Ref.~\cite{Yunes2009b} itself, they do not carry over to the dynamical formulation considered here, where the Pontryagin density vanishes to leading post-Newtonian order.

Expressed through the dimensionless parameter $\zeta$ from eq.~\eqref{eq:zeta_def}, the terrestrial bound is extremely weak when applied to a neutron star. Because $\zeta$ scales with the mean density of the source, $\zeta \propto \alpha M/R_\star^{3}$, the same value of $\alpha$ that is marginally constrained for the Earth corresponds to
\begin{equation}
\zeta_{\rm NS} \sim 4\times10^{13}
\label{eq:zeta_bound}
\end{equation}
for a neutron star, in agreement with the estimate of Ref.~\cite{Ali-Haimoud2011}. This lies thirteen orders of magnitude above the regime $\zeta \lesssim 1$ in which the perturbative treatment---adopted here, and equally in the derivation of that bound---is under control. Existing constraints therefore do not probe dynamical CS gravity in the regime where the effective field theory is valid on stellar scales; they merely exclude couplings so large that the theory has no controlled description there. The perturbative ceiling $\zeta \lesssim 1$, corresponding to $\sqrt{\alpha} \lesssim 17$ km for our fiducial star, is thus a stronger requirement than the observational bound itself.

Against this background, eq.~\eqref{eq:projected_bound} would represent an improvement of some seven to eight orders of magnitude in $\sqrt{\alpha}$---equivalently fifteen orders in $\zeta$, since $\zeta \propto \alpha$---and, crucially, would be obtained \emph{within} the domain of validity of the perturbative expansion rather than far outside it. It also compares favourably with the best presently projected alternative: Ref.~\cite{Ali-Haimoud2011} estimates that a $10\%$ determination of the moment of inertia of pulsar A in the binary PSR J0737$-$3039, should it prove achievable after a further two decades of timing, would yield $\zeta_{\rm NS} \lesssim 1$ and hence $\xi^{1/4} \lesssim 25$ km. Equation~\eqref{eq:projected_bound} is more than an order of magnitude tighter in $\sqrt{\alpha}$, and does not require the moment of inertia to be disentangled from equation-of-state uncertainty.

Finally, GW birefringence from a glitching pulsar probes a different aspect of the theory than either the existing bound or the cosmological birefringence considered elsewhere in the literature. The terrestrial constraint is a \emph{static, near-zone} effect; CS gravity suppresses the gravitomagnetic potential close to the source, and satellite gyroscopes measure the resulting deficit in the Lense-Thirring precession. Searches for birefringence in compact-binary waveforms~\cite{Yagi2018} instead constrain the time evolution of a homogeneous cosmological pseudoscalar, $\dot{\vartheta}_0$ and $\ddot{\vartheta}_0/H_0$, through phase accumulated over cosmological propagation distances. The effect computed here is distinct from both; it is a \emph{near-zone propagation} effect, in which GWs emitted by the star traverse the star's own dipolar scalar field, and it is therefore sensitive directly to $\mu_{\mathrm{CS}}$---that is, to $\alpha$ together with the interior structure and rotation rate of the source. A measurement would thus constrain the local scalar-field configuration around a single compact object rather than a cosmological background or a terrestrial one, and is complementary to existing bounds as well as potentially considerably more stringent.

\subsection{Detectability with Current and Future Detectors}
\label{sec:detectability}

Section~\ref{sec:pulsar_detectability} quoted detectability in terms of a strain-asymmetry threshold. It is important to be explicit about what such a threshold requires. The polarization asymmetry is a \emph{ratio} of strains, so resolving a value $\mathcal{A}_{\rm PV}$ demands a SNR on the glitch waveform itself of
\begin{equation}
\mathrm{SNR} \;\gtrsim\; \frac{1}{\mathcal{A}_{\rm PV}} \;=\; \frac{15}{2\,\zeta^{2}\,(\sigma R_\star)(\Omega R_\star)} .
\label{eq:snr_requirement}
\end{equation}
This, rather than the strain sensitivity of the instrument, is the binding constraint on the measurement. At the perturbative ceiling $\zeta = 1$ and $\nu = 1$ kHz, eq.~\eqref{eq:snr_requirement} requires a SNR $\gtrsim 745$ for a millisecond rotator ($\Omega = 10^3$ s$^{-1}$), $\gtrsim 1.9\times10^{3}$ for J0537$-$6910, and $\gtrsim 1.1\times10^{4}$ for the Vela pulsar. Since $\mathcal{A}_{\rm PV} \propto \zeta^2$, probing smaller values of $\zeta$ is expensive; reaching $\zeta = 0.1$ requires a further factor of $10^{2}$ in signal to noise.

\begin{figure}[t]
\centering
\includegraphics[width=0.9\columnwidth]{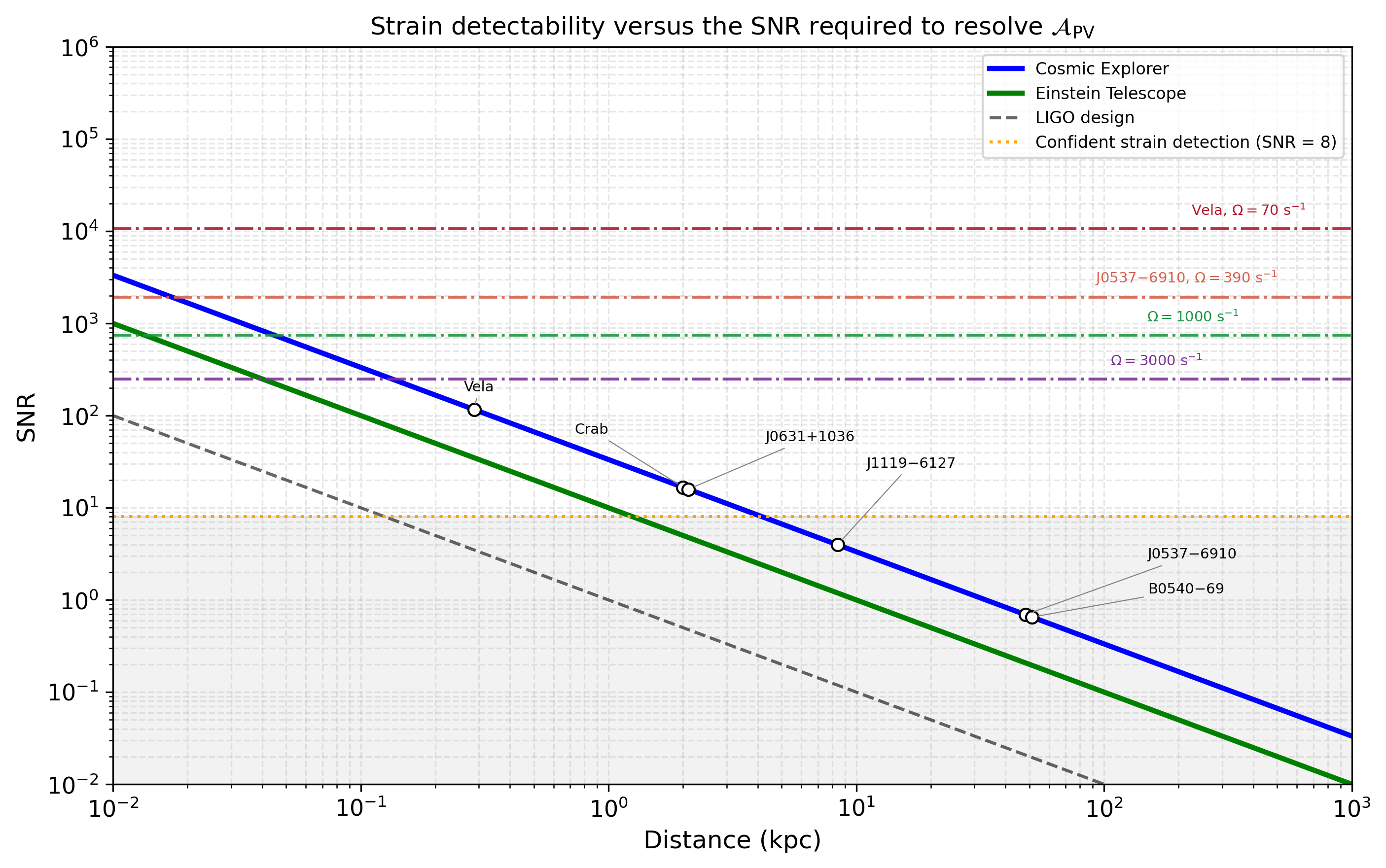}
\caption{Strain SNR for a pulsar glitch as a function of distance, for LIGO design sensitivity (dashed), ET (green) and CE (blue), assuming a glitch strain $h_{\rm ref} = 10^{-24}$ for a source located at a distance of 10 kpc. The dotted horizontal line marks the threshold for confident detection of the burst itself (SNR $=8$). The dash-dotted horizontal lines show the far more demanding SNR required to \emph{resolve} the polarization asymmetry down to the perturbative ceiling $\zeta = 1$, from eq.~\eqref{eq:snr_requirement}, for several rotation rates. Open circles mark known glitching pulsars at their catalogue distances, evaluated at CE sensitivity. The burst itself is comfortably detectable throughout the Milky Way, but no known object simultaneously combines the rotation rate and the proximity needed to resolve the asymmetry from a single glitch of this amplitude.}
\label{fig:detector_horizons}
\end{figure}

Figure~\ref{fig:detector_horizons} compares these requirements with the strain SNR achievable for a glitch of assumed amplitude $h_{\rm ref} = 10^{-24}$ at 10 kpc. CE would detect such an event out to $\sim 4$ kpc at a SNR of $\sim8$, comfortably covering the Galactic population, and would reach $\approx116$ for a Vela glitch at 287 pc. That is ample for detecting the burst, but falls short of eq.~\eqref{eq:snr_requirement} by a factor of $\sim 90$ for Vela and $\sim 240$ for the Crab. Overall, proximity and rotation rate pull in opposite directions across the known sample. J0537$-$6910, although the fastest known glitching rotator and hence intrinsically the most favourable source, lies in the Large Magellanic Cloud at $\sim 50$ kpc, where the strain SNR falls below unity.

Two considerations temper this conclusion. First, the assumed glitch amplitude is uncertain by orders of magnitude; estimates of the fraction of glitch energy channelled into the fundamental quadrupole mode span several decades, and the shortfalls quoted above correspond to required amplitudes $h_{\rm ref} \sim 10^{-22}$ at 10 kpc for Vela and the Crab---optimistic, but within the range considered in the literature for the largest events~\cite{Warszawski2012}. Second, the birefringent phase shift acts on \emph{any} radiation propagating through the star's near zone, not only on the glitch burst. Continuous emission from the same object---from a crustal deformation or a r-mode---can be integrated coherently over months, and the effective SNR then exceeds that of a single burst by orders of magnitude. A search for a persistent circular polarization asymmetry in continuous-wave emission from a rapidly rotating neutron star may therefore be a more promising route to the same physics than the transient signature considered here; we note this as a natural extension of the present work.

Several conclusions follow. First, current-generation detectors will not resolve the parity-violating asymmetry, and neither will third-generation instruments from a single glitch of typical amplitude. Second, the obstacle is the SNR requirement of eq.~\eqref{eq:snr_requirement} and not the predicted signal, which at the perturbative ceiling reaches $\mathcal{A}_{\rm PV} \sim 10^{-3}$---four orders of magnitude larger than the sensitivity of existing instruments to strain asymmetries. Third, the most favourable configuration combines a high rotation rate, a small radius, kilohertz emission frequency and---critically---proximity, since the SNR requirement scales as $\Omega^{-1}$ while the achievable SNR scales as $d^{-1}$. A rapidly rotating glitching pulsar within a few hundred parsecs, were one to be identified, would bring the measurement within reach of CE.

A non-detection at the level accessible to CE/ET would nonetheless improve substantially on existing limits. As noted in section~\ref{sec:existing_bounds}, the current bound on dynamical CS gravity corresponds to $\zeta_{\rm NS} \sim 4\times10^{13}$, thirteen orders of magnitude above the perturbative ceiling; any measurement sensitive to $\zeta \lesssim 1$---that is, $\sqrt{\alpha} \lesssim 17$ km---would be the first constraint on the theory obtained in a regime where it admits this controlled description.

\subsection{Detectability via gravitational-wave polarimetry}
\label{sec:polarimetry}

\subsubsection{Single-source Stokes polarimetry}

The parity-violating amplitude $\mathcal{A}_{\rm PV}$ of qq.~(\ref{eq:parity_general})
is defined as a circular-polarization asymmetry,
$\mathcal{A}_{\rm PV}=|h_R-h_L|/|h^{(0)}|$. A single interferometer, however,
records only one real strain projection at a time,
and cannot isolate $h_R$ from $h_L$. The physically meaningful observable must therefore be
reconstructed, not simply read off. For a continuous-wave
source of known sky location and independently constrained orientation
it may be recovered as a Stokes parameter.

Writing the two polarizations of the emitted continuous-wave signal as
$h_+=A_+\exp i\Phi_{+}(t)$ and $h_\times=A_\times\exp i[\Phi(t)_{+}-\delta]$,
the (time-averaged) GW Stokes parameters are
\begin{align}
I &= A_+^2+A_\times^2, & Q &= A_+^2-A_\times^2,\nonumber\\
U &= 2A_+A_\times\cos\delta, & V &= 2A_+A_\times\sin\delta .
\label{eq:stokes}
\end{align}
The circular Stokes parameter $V\propto\mathrm{Im}(h_+h_\times^{*})$ is the
parity-odd component and is precisely the quantity targeted in
correlation searches for a circularly polarized background
\cite{SetoTaruya2007,SetoTaruya2008}. In standard general relativity the mass-quadrupole
continuous-wave signal at inclination $\iota$ has $A_+=h_0(1+\cos^2\iota)/2$, $A_\times=h_0\cos\iota$, and
$\delta=\pi/2$, so that $U_{\rm GR}=0$ and
\begin{equation}
\left(\frac{V_{\mathrm{GR}}}{I_{\mathrm{GR}}}\right)
=\frac{(1+\cos^2\iota)\cos\iota}
{\tfrac14(1+\cos^2\iota)^2+\cos^2\iota}\equiv v_{\rm GR}(\iota).
\label{eq:vgr}
\end{equation}
Equation~(\ref{eq:vgr}) shows that the polarization state is a
\emph{one-parameter family} in $\cos\iota$. At fixed inclination the
circular content is fully determined. A blind search that marginalizes
over $\iota$ slides along this family and reabsorbs any parity-violating
displacement into an effective inclination; from a single source,
$\mathcal{A}_{\rm PV}$ is then exactly degenerate with $\iota$ and
formally unmeasurable. The degeneracy is broken only by an
\emph{external} determination of the orientation.

For PSR~J0537$-$6910 this determination exists. The pulsar-wind-nebula torus of
N157B has been fit for its symmetry axis \cite{NgRomani2008}, which we
take to coincide with the spin axis, fixing both the inclination $\iota$
and polarization angle $\psi$; this is the same geometric prior adopted
in the LIGO-Virgo-KAGRA directed searches for this pulsar~\cite{LVKJ0537r-mode}. With
$\iota$ fixed, $v_{\rm GR}(\iota)$ is a known number and the
parity-violating signal is the residual
\begin{equation}
\frac{V}{I}\bigg|_{\rm obs}-v_{\rm GR}(\iota)\propto\mathcal{A}_{\rm PV},
\label{eq:residual}
\end{equation}
where the constant of proportionality is $\mathcal{O}(1)$.
There is also an independent, corroborating signature in the form of a nonzero $U$
(equivalently $\delta\neq\pi/2$), which no continuous-wave signal in conventional general relativity can produce at any
inclination. Operationally, the four Jaranowski-Kr\'olak-Schutz
amplitude parameters~\cite{JKS1998} become overdetermined once
$(\iota,\psi)$ are externally fixed, and $\mathcal{A}_{\rm PV}$ enters as
the amplitude of the polarization component that is forbidden in general relativity and which may be recovered using the
$\mathcal{F}$-statistic over the inter-glitch segments.

Because Eq.~(\ref{eq:residual}) is a fractional polarization measurement,
its uncertainty scales as $\sigma_{V/I}\sim\rho^{-1}$ in terms of the coherent
SNR $\rho$, so the smallest resolvable asymmetry is
$\mathcal{A}_{\rm PV}^{\rm min}\sim\rho^{-1}$ because the fitting
prior prevents the available SNR from being spent breaking the
inclination degeneracy. The multi-year coherent SNR achievable on
PSR~J0537$-$6910 with CE and ET then yields the projected bound
$\zeta\lesssim 0.014$ (or $\alpha\lesssim 2~{\rm km}$) quoted in table~\ref{tab:pulsar_parameters}.

\subsubsection{Population spectropolarimetry}

A single pulsar contributes one frequency and hence one value of
$\mathcal{A}_{\rm PV}$, giving no spectral leverage. Across a population
of young pulsars with resolved pulsar-wind-nebula geometries (PSR~J0537$-$6910, Crab, Vela,
B1509$-$58), the prediction
$\mathcal{A}_{\rm PV}=\tfrac{2}{15}\zeta^2(\sigma R_\star)(\Omega R_\star)$
carries an explicit $\mathcal{A}_{\rm PV}\propto\Omega\propto \nu$ scaling
at fixed stellar structure. Treating $\alpha$ as
universal and fitting the ensemble of residuals
(eq.~\ref{eq:residual}) jointly would then serve two useful ends: it would stack the
individual source constraints into a tighter bound on $\alpha$, and---more
importantly---would provide a consistency test, since a genuine
CS signal must return the same $\alpha$ across sources
spanning a decade in $\nu$, whereas an instrumental polarization
systematic is source-independent and will not
track the predicted $\zeta^2\sigma\,\Omega R_\star^2$ weighting. The
sample is limited to pulsars with understood orientations, but
this is precisely the set for which eq.~\eqref{eq:residual} is
degeneracy free.

Moreover, when a sufficiently loud coherent transient accompanies the
glitch, or as a cross check of the systematics of a single-detector
amplitude modulation measurement, a network of detectors with distinct
antenna responses to the $+$ and $\times$ poarlizations may be used to reconstruct the complex
pair $(h_+,h_\times)$---and hence $(I,Q,U,V)$---directly. This would be the resolved-source analogue of the
background $V$-mode measurement of Refs.~\cite{SetoTaruya2007,SetoTaruya2008};
the network response algebra follows the antenna-pattern formalism
of~\cite{SchumacherTalbotHolzYunes2025}. The inclination prior would remain
necessary, however: the network measures the polarization state, but
eq.~(\ref{eq:vgr}) is still required to locate the general relativity reference point for comparison.

\subsection{Limitations and Caveats}
\label{sec:limitations}

Several simplifying assumptions underlie our analysis.
First, as repeatedly noted, we have been working within the weak CS approximation, requiring that CS corrections remain small compared to terms originating in standard general relativity throughout GW propagation. This is the requirement $\zeta \lesssim 1$ of eq.~\eqref{eq:zeta_def}, and it is satisfied throughout the parameter space we consider. However, we emphasise that it is \emph{not} satisfied in the derivation existing observational bound, which corresponds to $\zeta_{\rm NS} \sim 4\times10^{13}$. All quoted signal amplitudes should therefore be read as predictions conditional on the theory being weakly coupled on stellar scales, rather than as consequences of current constraints.

Second, we are neglecting dynamical scalar perturbations \(\delta\vartheta\), focusing on the leading birefringent effect from the background scalar field. A full treatment would require solving the coupled metric-scalar perturbation equations, which we leave for future work.

Third, our glitch model is purely phenomenalistic. A more detailed treatment of the angular momentum transfer during the glitch could affect the GW amplitude, though the polarization asymmetry (being a ratio) is more robust. We note, however, that the glitch amplitude nonetheless enters the observability of the effect through the signal-to-noise requirement of eq.~\eqref{eq:snr_requirement}, and is the dominant uncertainty in the detectability estimates of section~\ref{sec:detectability}.

Fourth, we have neglected entrainment in the two-fluid model. Entrainment does \emph{not} cancel from the lag-induced source; instead it adds directly to the neutron inertia density, $\mathcal{W} \to \rho_n + p_n + \varepsilon$, and modifies the neutron inertia fraction $f$ at the level of tens of percent. The parametric scalings are nevertheless unaffected, because the lag correction enters observables only through $f$, which is a ratio of inertia densities and therefore satisfies $0 < f < 1$; entrainment drives $f$ toward $1/2$ and can only suppress the effect further. (See appendix~\ref{app:entrainment}.)

Finally, our benchmark model assumes a constant-density interior. This choice is not merely a simplification of the numerical coefficients; the interior TOV solution is conformally flat, so its Weyl tensor---and hence the curvature factor $\mathcal{F}_0$---vanishes identically, and the scalar dipole is sourced entirely in the vacuum exterior. The constant-density model therefore represents a degenerate case, and it provides a lower bound on the interior contribution. A realistic, centrally condensed EoS would have $\mathcal{F}_0 > 0$ and would increase the predicted asymmetry above the values quoted here; evaluating $\mathcal{F}_0$ for tabulated equations of state would be the most immediate refinement of this work.

Throughout this work we have set the electromagnetic coupling \(\beta = 0\) in eq.~\eqref{eq:action}, thereby isolating parity violation in the gravitational sector. For completeness, we briefly comment on the possible effects of a nonzero \(\beta\). The scalar field equation \eqref{eq:scalar} would then acquire an additional source term \(\propto \beta F_{\mu\nu}{}^{\ast}{F}^{\mu\nu}\) from the electromagnetic Pontryagin density. For a rotating magnetized neutron star, $F_{\mu\nu}{}^{\ast}{F}^{\mu\nu}$ is nonvanishing even for an aligned dipole configuration, scaling as $F_{\mu\nu}{}^{\ast}{F}^{\mu\nu} \propto (\mu B_0 \Omega / r^6) \cos\theta \sin^2\theta$ in the exterior, where \(\mu\) and \(B_0\) are the magnetic dipole moment and surface field, respectively. This term sources the same dipolar (\(\ell = 1\)) scalar mode as the gravitational Pontryagin density, contributing an additional piece to the scalar dipole moment \(\mu_{\mathrm{CS}}\) with parametric scaling \(\delta\mu_{\mathrm{CS}}^{(\beta)} \propto \beta \mu B_0 \Omega / R_\star^2\).

For typical rotation-powered pulsars (\(B_0 \sim 10^{12}\) G), the ratio \(\delta\mu_{\mathrm{CS}}^{(\beta)}/\mu_{\mathrm{CS}}^{(\alpha)} \sim 10^{-4} (\beta/\alpha)\) is highly suppressed relative to the gravitational source. For magnetars (\(B_0 \sim 10^{15}\) G), however, this ratio can become of order unity or larger if \(\beta \sim \alpha\). Since the glitching pulsars considered in this work (Crab, Vela, and millisecond pulsars) have surface fields well below the magnetar scale, the electromagnetic contribution to the scalar field is negligible. We therefore leave a detailed analysis of the combined \(\alpha\)-\(\beta\) system for magnetar glitches to future work.

\section{Conclusion}
\label{sec:conclusion}

We have investigated GW birefringence from pulsar glitches in dynamical CS gravity, a parity-violating extension of general relativity motivated by string theory and quantum gravity. Using the Hartle-Thorne slow-rotation formalism to describe the stellar background and a perturbative treatment of the CS coupling, we derived the modified Regge-Wheeler equation governing axial gravitational perturbations and computed the resulting phase shift between right- and left-handed circular polarizations.

Our main result is the fractional polarization asymmetry
\begin{equation}
\frac{|h_R - h_L|}{|h^{(0)}|} = \frac{\alpha\sigma \mu_{\mathrm{CS}}}{3R_\star^3},
\end{equation}
where \(\mu_{\mathrm{CS}}\) is the dipole moment of the CS scalar field, determined by matching the interior and exterior solutions. For a constant-density interior model the curvature factor \(\mathcal{F}_0\) vanishes identically---the interior Schwarzschild solution being conformally flat---so that the scalar dipole is sourced entirely in the vacuum exterior and takes the closed form \(\mu_{\mathrm{CS}} = 8\alpha M J/R_\star^3 = \tfrac{16}{5}\alpha M^2\Omega/R_\star\). This leads to the explicit expression
\begin{equation}
\frac{|h_R - h_L|}{|h^{(0)}|} = \frac{16\,\alpha^2 \sigma \Omega M^{2}}{15 R_\star^{4}}
= \frac{2}{15}\,\zeta^{2}\,(\sigma R_\star)(\Omega R_\star),
\end{equation}
which scales as \(\alpha^2\), increases linearly with GW frequency \(\sigma\) and stellar rotation rate \(\Omega\), and decreases as the \emph{fourth} power of the stellar radius \(R_\star\) at fixed mass. Here \(\zeta\) is the dimensionless CS parameter from eq.~\eqref{eq:zeta_def}, for which the perturbative treatment requires \(\zeta \lesssim 1\). The steep compactness dependence follows directly from the exterior source, \(^{\ast}\!RR \propto MJ/r^{7}\), and implies a factor of four variation in the predicted signal across the plausible range of neutron-star radii---twice the dynamic range of a shallower \(R_\star^{-2}\) dependence, and correspondingly greater leverage for probing the equation of state.

We extended the analysis to a two-fluid model incorporating differential rotation between the neutron superfluid and the charged component. In this model the frame-dragging constant \(c_2\) receives a lag correction \(\delta c_2^{\rm lag} \sim \mathcal{W}_0 \Delta_{2}\Omega_0 R_\star^2\) (independent of \(\alpha\)), while the scalar dipole moment acquires an additional contribution \(\delta\mu_{\mathrm{CS}}^{\rm lag} \sim \alpha \mathcal{W}_0 \Delta_{2}\Omega_0 \mathcal{F}_0 R_\star^5\). Because this term and the corotating interior contribution both carry the factor \(\alpha\mathcal{F}_0\), their ratio reduces to \(f_0\,\Delta_{2}\Omega_0/\Omega_{\rm eff}\), the fractional lag weighted by the neutron inertia fraction. This is bounded above by the fractional glitch amplitude itself and is therefore of order \(10^{-9}\)--\(10^{-5}\) for observed glitches, independently of the equation of state and the entrainment model. The constant-density benchmark thus provides a reliable estimate of the expected signal strength.

For a millisecond rotator at kilohertz frequencies the predicted asymmetry reaches \(\mathcal{A}_{\rm PV} \approx 1.3\times10^{-3}\,\zeta^{2}\), attaining \(\sim 10^{-3}\) at the perturbative ceiling---four orders of magnitude above the level to which current instruments are sensitive to strain asymmetries. The obstacle to a measurement is therefore not the size of the effect but the SNR needed to resolve a \emph{ratio} of strains, which for a single glitch of typical amplitude exceeds what even third-generation instruments will achieve for known sources. Nonetheless, the constraint accessible in this way would represent a substantial advance; the only existing bound on dynamical CS gravity, from terrestrial frame-dragging measurements~\cite{Ali-Haimoud2011}, corresponds to \(\zeta_{\rm NS} \sim 4\times10^{13}\) for a neutron star, thirteen orders of magnitude above the regime in which the effective theory is under control. Any measurement sensitive to \(\zeta \lesssim 1\)---that is, \(\sqrt{\alpha} \lesssim 17\) km---would improve on it by some seven orders of magnitude in \(\sqrt{\alpha}\) and would be the first constraint on the theory obtained where it admits a fully controlled description. A detection would provide direct evidence of gravitational parity violation.

Several avenues for future work remain. The most immediate is the evaluation of the interior curvature factor \(\mathcal{F}_0\) for realistic, centrally condensed equations of state. Because \(\mathcal{F}_0\) vanishes for the constant-density benchmark, the estimates given here are conservative, and a realistic interior can only increase the predicted asymmetry. Inclusion of dynamical scalar perturbations could modify the birefringent effect at higher order in the CS coupling. Finally, since the birefringent phase shift acts on any radiation traversing the star's near zone, a search for a persistent circular-polarization asymmetry in continuous-wave emission---which can be integrated coherently over months and so reach far higher effective SNR than a single burst---may prove a more promising route than the transient signature considered here.

In summary, rotating neutron stars can source a parity-odd scalar field whose imprint on gravitational-wave polarization is far larger than existing constraints on dynamical CS gravity would suggest, and GW emissions from these sources probe the theory in a regime where those constraints do not apply. We encourage the community to include pulsar glitches, and the associated continuous-wave polarization asymmetry, in the target lists for the future CE and ET observatories.

\section*{Data Availability Statement}

No new data experimental were generated in this study. The computational data are available from the authors upon request.

\bibliography{references}

\appendix

\section{Hartle-Thorne Metric and Curvature Components}
\label{app:metric}

We present the explicit components of the Hartle-Thorne metric and the associated curvature tensors used in the main text. The metric to first order in rotation is
\begin{equation}
\label{eq:app_metric}
\begin{aligned}
ds^2 = -A(r)\,dt^2 &+ B(r)\,dr^2 + r^2\bigl(d\theta^2 + \sin^2\theta\,d\phi^2\bigr) \\
&- 2\omega(r)\,r^2\sin^2\theta\,dt\,d\phi + \mathcal{O}(\Omega^2),
\end{aligned}
\end{equation}
with \(A(r) = e^{2\Phi(r)}\), \(B(r) = e^{2\Lambda(r)}\), and \(\omega(r)\) the frame-dragging function.

\subsection{Nonzero Christoffel Symbols}
\label{app:nonzero_christoffel}

To linear order in \(\omega\), the nonzero affine connections are
\begin{align}
\Gamma^{t}_{tr} &= \frac{A'}{2A}, &
\Gamma^{r}_{rr} &= \frac{B'}{2B}, &
\Gamma^{t}_{r\phi} &= \frac{\omega' r^2\sin^2\theta}{2A}, \nonumber\\
\Gamma^{r}_{tt} &= \frac{A'}{2B}, &
\Gamma^{r}_{\theta\theta} &= -\frac{r}{B}, &
\Gamma^{r}_{r\phi} &= -\frac{r\sin^2\theta}{B}, \nonumber\\
\Gamma^{\theta}_{r\theta} &= \frac{1}{r}, &
\Gamma^{\phi}_{r\phi} &= \frac{1}{r}, &
\Gamma^{\theta}_{\phi\phi} &= -\sin\theta\cos\theta, \nonumber\\
\Gamma^{\phi}_{\theta\phi} &= \cot\theta, &
\Gamma^{r}_{t\theta} &= -\omega\cot\theta, &
\Gamma^{\phi}_{t\phi} &= \omega\sin\theta\cos\theta,
\end{align}
and additionally
\begin{align}
\Gamma^{r}_{t\phi} &= \frac{r(r\omega' + 2\omega)\sin^2\theta}{2B}, \label{eq:app_gamma_trphi} \\
\Gamma^{\theta}_{tr} &= \frac{\omega A'}{2A} - \frac{\omega}{r} - \frac{\omega'}{2},
\label{eq:app_gamma_ttheta}
\end{align}
the prime denoting differentiation with respect to the radial coordinate $r$.

\subsection{Riemann Tensor Components (Static Sector)}

The static (non-rotating) Riemann tensor components are
\begin{align}
R^{r}_{trt} &= \frac{A''}{2B} - \frac{A'B'}{4B^2} - \frac{(A')^2}{4AB}, \label{eq:app_riemann_rtrt} \\
R^{\theta}_{t\theta t} &= \frac{A'}{2Br}, \qquad
R^{\phi}_{t\phi t} = \frac{A'}{2Br}, \label{eq:app_riemann_thetat} \\
R^{\theta}_{r\theta r} &= \frac{B'}{2Br}, \qquad
R^{\phi}_{r\phi r} = \frac{B'}{2Br}, \label{eq:app_riemann_thetar} \\
R^{t}_{\theta t\theta} &= -\frac{rA'}{2AB}, \qquad
R^{r}_{\theta r\theta} = \frac{rB'}{2B^2}, \label{eq:app_riemann_theta} \\
R^{\phi}_{\theta\phi\theta} &= 1 - \frac{1}{B}. \label{eq:app_riemann_phitheta}
\end{align}

\subsection{Ricci Tensor, Ricci Scalar, and Einstein Tensor}

The Ricci tensor components in the static background are
\begin{align}
R_{tt} &= \frac{A''}{2B} - \frac{A'B'}{4B^2} - \frac{(A')^2}{4AB} + \frac{A'}{Br}, \label{eq:app_ricci_tt} \\
R_{rr} &= -\frac{A''}{2A} + \frac{(A')^2}{4A^2} - \frac{A'B'}{4AB} + \frac{B'}{Br}, \label{eq:app_ricci_rr} \\
R_{\theta\theta} &= 1 - \frac{1}{B} - \frac{rA'}{2AB} + \frac{rB'}{2B^2}, \label{eq:app_ricci_thetatheta} \\
R_{\phi\phi} &= R_{\theta\theta}\sin^2\theta, \label{eq:app_ricci_phiphi}
\end{align}
and the Ricci scalar is
\begin{equation}
R = -\frac{A''}{AB} + \frac{A'B'}{2AB^2} + \frac{2B'}{rB^2} + \frac{2}{r^2} - \frac{2}{Br^2}. \label{eq:app_ricci_scalar}
\end{equation}

The $t\phi$-component receives a rotational correction
\begin{equation}
\label{eq:app_ricci_tphi}
\begin{aligned}
R_{t\phi} = \Bigg[ &\frac{rA'\omega}{2AB} - \frac{r^2A'B'\omega'}{4AB^2} - \frac{r^2\omega'B'}{4B^2} - \frac{r\omega B'}{2B^2} \\
&+ \frac{r^2\omega''}{2B} + \frac{2r\omega'}{B} - \omega + \frac{\omega}{B} \Bigg]\sin^2\theta.
\end{aligned}
\end{equation}
The Einstein tensor components \(G_{\mu\nu} = R_{\mu\nu} - \frac{1}{2}g_{\mu\nu}R\) are therefore
\begin{align}
G_{tt} &= -\frac{(A')^2}{4AB} + \frac{A'}{Br} + \frac{AB'}{rB^2} + \frac{A}{r^2} - \frac{A}{Br^2}, \label{eq:app_einstein_tt} \\
G_{rr} &= -\frac{(A')^2}{4A^2} - \frac{B}{r^2} + \frac{1}{r^2}, \label{eq:app_einstein_rr} \\
G_{\theta\theta} &= -\frac{rA'}{2AB} - \frac{rB'}{2B^2} + \frac{A''r^2}{2AB} - \frac{A'B'r^2}{4AB^2}, \label{eq:app_einstein_thetatheta} \\
G_{\phi\phi} &= G_{\theta\theta}\sin^2\theta, \label{eq:app_einstein_phiphi}
\end{align}
along, obviously, with $G_{t\phi}=R_{t\phi}$.

\section{Pontryagin Density Calculation}
\label{app:pontryagin}

We present a detailed, step-by-step derivation of the Pontryagin density for a slowly rotating neutron star in the Hartle-Thorne formalism. The calculation proceeds in three stages: (i) definition and expansion of the Pontryagin density in terms of Riemann tensor components, (ii) demonstration that the Pontryagin density vanishes for static spherically symmetric spacetimes, and (iii) computation of the leading-order contribution arising from rotation.

\subsection{Definition and General Expansion}

The Pontryagin density is defined as the contraction of the Riemann tensor with its dual,
\begin{equation}
{}^{\ast}\!RR = {}^{\ast}R^{\sigma}_{\ \mu\nu\tau} R^{\tau\ \mu\nu}_{\ \sigma},
\label{eq:app_pontryagin_def}
\end{equation}
where the dual Riemann tensor is given by
\begin{equation}
{}^{\ast}R^{\sigma}_{\ \mu\nu\tau} = \frac{1}{2} \epsilon_{\mu\nu\alpha\beta} R^{\sigma\ \alpha\beta}_{\ \tau}.
\label{eq:app_dual_def}
\end{equation}
Here, \(\epsilon_{\mu\nu\alpha\beta}\) is the Levi-Civita tensor, related to the totally
antisymmetric symbol by \(\epsilon^{\mu\nu\alpha\beta} = \tilde{\epsilon}^{\mu\nu\alpha\beta}/\sqrt{-g}\)
with \(\tilde{\epsilon}^{tr\theta\phi} = +1\). Substituting eq.~\eqref{eq:app_dual_def} into
eq.~\eqref{eq:app_pontryagin_def} and raising the index pairs carried by the Levi-Civita tensor
yields the explicit form

\begin{equation}
{}^{\ast}\!RR = \frac{1}{2}\, \epsilon^{\mu\nu\alpha\beta}\, R^{\sigma}_{\ \tau\alpha\beta}\, R^{\tau}_{\ \sigma\mu\nu}.
\label{eq:app_pontryagin_explicit}
\end{equation}
It is essential to note that the index pairs contracted with \(\epsilon\) occupy the
\emph{final two slots} of each Riemann tensor; contracting the second and third slots instead
pairs different curvature components and does not reproduce
eq.~\eqref{eq:app_pontryagin_explicit}.

In spherical coordinates \((t, r, \theta, \phi)\), the Levi-Civita symbol has only one independent nonvanishing component up to permutations. Because \(\epsilon^{\mu\nu\alpha\beta}\) is totally antisymmetric, the sum receives contributions only when \(\{\mu,\nu\}\) and \(\{\alpha,\beta\}\) partition \(\{t,r,\theta,\phi\}\) into two disjoint pairs. There are three such partitions:
\begin{equation}
\{t,r\}\,|\,\{\theta,\phi\}, \qquad
\{t,\theta\}\,|\,\{r,\phi\}, \qquad
\{t,\phi\}\,|\,\{r,\theta\}.
\end{equation}
For a given partition, the two orderings of \(\{\mu,\nu\}\), the two orderings of \(\{\alpha,\beta\}\), and the interchange of the roles of the two pairs generate \(2\times2\times2 = 8\) equal contributions: each sign change from the Levi-Civita symbol is compensated by the antisymmetry of the corresponding Riemann pair, and the interchange of the pairs is undone by relabelling the summed indices \(\sigma \leftrightarrow \tau\). Together with the factor \(1/2\) in eq.~\eqref{eq:app_pontryagin_explicit} this gives an overall factor of \(4\) per partition. The relative signs follow from
\(\tilde{\epsilon}^{t\theta r\phi} = -\tilde{\epsilon}^{tr\theta\phi}\) and
\(\tilde{\epsilon}^{t\phi r\theta} = +\tilde{\epsilon}^{tr\theta\phi}\); absorbing the middle sign into
the index ordering \(\phi r\) rather than \(r\phi\), the expression reduces to

\begin{equation}
{}^{\ast}\!RR = -\frac{4}{\sqrt{-g}}\left( R^{\sigma}_{\,\,\tau\theta\phi} R^{\tau}_{\,\,\sigma tr}
+ R^{\sigma}_{\,\,\tau\phi r} R^{\tau}_{\,\,\sigma t\theta}
+ R^{\sigma}_{\,\,\tau r\theta} R^{\tau}_{\,\,\sigma t\phi} \right),
\label{eq:app_pontryagin_reduced}
\end{equation}
with \(\sigma\) and \(\tau\) summed. The overall sign corresponds to the orientation for which
the Kerr limit obtained below is positive; reversing it flips the sign of \(\vartheta\) and of
\(\mu_{\mathrm{CS}}\), but leaves the observable polarization asymmetry unchanged. This compact
form is the starting point for explicit evaluation in any given spacetime.

\subsection{Vanishing of the Pontryagin Density for Static Spherically Symmetric Spacetimes}

Consider a general static, spherically symmetric metric in Schwarzschild coordinates:

\begin{equation}
ds^{2} = -A(r) dt^{2} + B(r) dr^{2} + r^{2} \left(d\theta^{2} + \sin^{2}\theta\, d\phi^{2}\right).
\label{eq:app_static_metric}
\end{equation}

We now evaluate eq.~\eqref{eq:app_pontryagin_reduced} using the nonvanishing Riemann tensor components of this metric. Every product appearing in eq.~\eqref{eq:app_pontryagin_reduced} pairs a component whose final two indices are \emph{mixed} between the \(tr\)-sector and the angular sector with one whose final two indices are not. For the static metric the Riemann tensor is block diagonal: its nonvanishing components have their final index pair drawn entirely from \(\{t,r\}\) or entirely from \(\{\theta,\phi\}\), or else repeat one index from each block in the pattern \(R^{a}_{\ b a b}\). Components such as \(R^{\sigma}_{\ \tau r\phi}\) and \(R^{\sigma}_{\ \tau t\theta}\) with \(\sigma \neq \tau\) mixing the two blocks vanish identically, because the metric is diagonal and there is no coupling between the \(tr\)-sector and the angular sector. Consequently every term in eq.~\eqref{eq:app_pontryagin_reduced} contains at least one vanishing factor, and

\begin{equation}
{}^{\ast}\!RR = 0 \qquad \text{(static, spherically symmetric)}.
\label{eq:app_pontryagin_static_zero}
\end{equation}

This result is expected on physical grounds: the Pontryagin density is odd under parity, and a nonrotating, spherically symmetric spacetime possesses no preferred direction to generate a parity-odd quantity.

\subsection{Rotating Spacetime: Hartle-Thorne Metric}

To introduce rotation, we adopt the Hartle-Thorne metric to first order in the rotation parameter as given in appendix~\ref{app:metric}. The spacetime becomes axially symmetric, and the metric acquires an off-diagonal component. To linear order in \(\omega\), the additional nonzero Christoffel symbols beyond the static ones are listed in section~\ref{app:nonzero_christoffel}. The static Christoffel symbols remain unchanged to this order.

The rotation introduces new nonzero Riemann tensor components that are linear in \(\omega\) and its derivatives. Those required for the contraction~\eqref{eq:app_pontryagin_reduced} are

\begin{align}
R^{t}_{\ r\theta\phi} &= \frac{\omega' r^{2}}{A} \sin\theta \cos\theta, \label{eq:app_riemann_rot1} \\
R^{r}_{\ t\theta\phi} &= \frac{\omega' r^{2}}{B} \sin\theta \cos\theta, \label{eq:app_riemann_rot2} \\
R^{\theta}_{\ \phi tr} &= -\,\omega' \sin\theta \cos\theta, \label{eq:app_riemann_rot3} \\
R^{\phi}_{\ \theta tr} &= \omega' \cot\theta, \label{eq:app_riemann_rot4} \\
R^{t}_{\ \theta\phi r} &= -\frac{\omega' r^{2}}{2A} \sin\theta\cos\theta,
\qquad
R^{r}_{\ \phi t\theta} = -\frac{\omega' r^{2}}{2B} \sin\theta\cos\theta, \label{eq:app_riemann_rot5} \\
R^{\theta}_{\ t\phi r} &= -\frac{\omega'}{2} \sin\theta\cos\theta,
\qquad\ \ \;
R^{\phi}_{\ r t\theta} = \frac{\omega'}{2}\cot\theta, \label{eq:app_riemann_rot6} \\
R^{t}_{\ \phi r\theta} &= -\frac{\omega' r^{2}}{2A}\sin\theta\cos\theta,
\qquad
R^{\theta}_{\ r t\phi} = -\frac{\omega'}{2}\cot\theta, \label{eq:app_riemann_rot7} \\
R^{r}_{\ \theta t\phi} &= \frac{\omega' r^{2}}{2B}\sin\theta\cos\theta,
\qquad\ \;
R^{\phi}_{\ t r\theta} = -\frac{\omega'}{2}\sin\theta\cos\theta. \label{eq:app_riemann_rot8}
\end{align}
All other components either vanish or are of higher order in \(\omega\).

\subsection{Evaluation of the Pontryagin Density for the Rotating Spacetime}

We now substitute the Riemann components into the reduced expression eq.~\eqref{eq:app_pontryagin_reduced}. Since we are working to linear order in \(\omega\), we only keep terms that are products of a static Riemann component (zeroth order in \(\omega\)) with a rotational Riemann component (first order in \(\omega\)). Products of two rotational components are of \(\mathcal{O}(\omega^{2})\) and are neglected.

Each of the three groups in eq.~\eqref{eq:app_pontryagin_reduced} carries an implicit sum over \(\sigma\) and \(\tau\), and in each group \emph{four} index pairs contribute; the four are equal in pairs. Writing \(\mathrm{s} \equiv \sin\theta\) and \(\mathrm{c} \equiv \cos\theta\), we evaluate the groups in turn.

\paragraph{First group, \(R^{\sigma}_{\ \tau\theta\phi}R^{\tau}_{\ \sigma tr}\).}
The contributing pairs are \((\sigma,\tau) = (t,r), (r,t), (\theta,\phi), (\phi,\theta)\):
\begin{align}
R^{t}_{\ r\theta\phi}R^{r}_{\ ttr}
&= \frac{\omega' r^{2}}{A}\mathrm{s}\mathrm{c}
\left[-\frac{A''}{2B} + \frac{A'B'}{4B^{2}} + \frac{(A')^{2}}{4AB}\right], \nonumber\\
R^{r}_{\ t\theta\phi}R^{t}_{\ rtr}
&= \frac{\omega' r^{2}}{B}\mathrm{s}\mathrm{c}
\left[-\frac{A''}{2A} + \frac{A'B'}{4AB} + \frac{(A')^{2}}{4A^{2}}\right], \nonumber\\
R^{\theta}_{\ \phi\theta\phi}R^{\phi}_{\ \theta tr}
&= \frac{(B-1)}{B}\mathrm{s}^{2}\cdot \omega'\cot\theta
= \frac{(B-1)}{B}\,\omega'\,\mathrm{s}\mathrm{c}, \nonumber\\
R^{\phi}_{\ \theta\theta\phi}R^{\theta}_{\ \phi tr}
&= -\frac{(B-1)}{B}\cdot\left(-\omega'\,\mathrm{s}\mathrm{c}\right)
= \frac{(B-1)}{B}\,\omega'\,\mathrm{s}\mathrm{c}.
\end{align}
The first two products are equal and combine to
\(-\,\bigl[2AA''B - (A')^{2}B - AA'B'\bigr]\,r^{2}\omega'\mathrm{s}\mathrm{c}/(2A^{2}B^{2})\);
the last two are equal and sum to \(-2(1-B)\omega'\mathrm{s}\mathrm{c}/B\). Hence
\begin{equation}
\begin{aligned}
    \text{group 1} = -\,\omega'\,\mathrm{s}\mathrm{c}
&\Bigg\{ \frac{\bigl[2AA''B - (A')^{2}B - AA'B'\bigr] r^{2}}{2A^{2}B^{2}}\\
&+ \frac{2(1-B)}{B} \Bigg\}.
\end{aligned}
\label{eq:app_group1}
\end{equation}

\paragraph{Second and third groups.}
For \(R^{\sigma}_{\ \tau\phi r}R^{\tau}_{\ \sigma t\theta}\) the contributing pairs are \((t,\theta), (r,\phi), (\theta,t), (\phi,r)\):
\begin{align}
R^{t}_{\ \theta\phi r}R^{\theta}_{\ tt\theta} &= \frac{r\omega' A'}{4AB}\mathrm{s}\mathrm{c},
& R^{r}_{\ \phi\phi r}R^{\phi}_{\ rt\theta} &= -\frac{r\omega' B'}{4B^{2}}\mathrm{s}\mathrm{c},
\nonumber\\
R^{\theta}_{\ t\phi r}R^{t}_{\ \theta t\theta} &= \frac{r\omega' A'}{4AB}\mathrm{s}\mathrm{c},
& R^{\phi}_{\ r\phi r}R^{r}_{\ \phi t\theta} &= -\frac{r\omega' B'}{4B^{2}}\mathrm{s}\mathrm{c},
\end{align}
so that
\begin{equation}
\text{group 2} = r\,\omega'\,\mathrm{s}\mathrm{c}
\left[\frac{A'}{2AB} - \frac{B'}{2B^{2}}\right]
= -\,\frac{r\,\omega'\,\mathrm{s}\mathrm{c}}{2AB^{2}}\bigl(AB' - BA'\bigr).
\label{eq:app_group2}
\end{equation}
The third group is obtained from the second by the interchange \(\theta \leftrightarrow \phi\) in the index pattern, which permutes the four products among themselves and leaves the sum unchanged:
\begin{equation}
\text{group 3} = \text{group 2}
= -\,\frac{r\,\omega'\,\mathrm{s}\mathrm{c}}{2AB^{2}}\bigl(AB' - BA'\bigr).
\label{eq:app_group3}
\end{equation}
Groups 2 and 3 are not proportional to group 1: they carry a single radial derivative and a single power of \(r\), a structure absent from eq.~\eqref{eq:app_group1}. They therefore modify the radial profile of the source, not merely its normalization. Only in vacuum, where \(AB = 1\), do they reduce to one half of group 1 and mimic an overall rescaling.

\paragraph{Collecting the terms.}
Inserting eqs.~\eqref{eq:app_group1}--\eqref{eq:app_group3} into eq.~\eqref{eq:app_pontryagin_reduced} and substituting \(\sqrt{-g} = \sqrt{AB}\,r^{2}\sin\theta\), the factor \(\sin\theta\cos\theta/\sin\theta = \cos\theta\) factors out, confirming the purely dipolar angular dependence:
\begin{equation}
\begin{aligned}
{}^{\ast}\!RR = \frac{4\,\omega'\cos\theta}{\sqrt{AB}\,r^{2}} \Bigg\{
&\frac{\bigl[2AA''B - (A')^{2}B - AA'B'\bigr] r^{2}}{2A^{2}B^{2}} \\
&+ \frac{2(1-B)}{B} + \frac{r\bigl(AB'-BA'\bigr)}{AB^{2}} \Bigg\}.
\end{aligned}
\label{eq:app_pontryagin_prefinal}
\end{equation}
Combining over the common denominator \((AB)^{5/2}r^{2}\) yields the final expression
\begin{equation}
{}^{\ast}\!RR = \omega'(r) \cos\theta\; \mathcal{F}(r),
\label{eq:app_pontryagin_final}
\end{equation}
where the radial function \(\mathcal{F}(r)\) is given by
\begin{multline}
\mathcal{F}(r) = \frac{2}{(AB)^{5/2} r^{2}}
\Bigl\{ \bigl[2AA''B - (A')^{2}B - AA'B'\bigr] r^{2} \\
+ 4A^{2}B(1 - B) + 2rA\bigl(AB' - BA'\bigr) \Bigr\}.
\label{eq:app_Fr_final}
\end{multline}

Equation~\eqref{eq:app_pontryagin_final} is the general expression for the Pontryagin density in a slowly rotating, axially symmetric spacetime to linear order in the frame-dragging function \(\omega(r)\). Two structural features are worth recording: the result is linear in \(\omega'\), with no \(\omega''\) and no undifferentiated \(\omega\) surviving the contraction; and the separation of the angular dependence into \(\cos\theta\) is exact at this order, so that the ansatz \(\vartheta = \hat{\vartheta}(r)\cos\theta\) used in section~\ref{sec:scalar} involves no multipole truncation. This result was used in section~\ref{sec:pontryagin} to compute the explicit form of the source term for the CS scalar field.

\subsection{Verification: Schwarzschild Exterior and the Slow-Rotation Kerr Limit}

To verify eq.~\eqref{eq:app_pontryagin_final}, we evaluate it in the exterior region of a neutron star, where the metric functions reduce to the Schwarzschild form:

\begin{equation}
A(r) = 1 - \frac{2M}{r}, \qquad B(r) = \frac{1}{A(r)}.
\label{eq:app_schwarzschild}
\end{equation}

In this region, the frame-dragging function is given by the exterior solution
\begin{equation}
\omega(r) = \frac{2J}{r^{3}},
\label{eq:app_omega_exterior}
\end{equation}
where \(J = aM\) is the angular momentum and \(a = J/M\) is the Kerr parameter.

We first compute the necessary derivatives. With \(AB = 1\), we have \(B = 1/A\), \(B' = -A'/A^{2}\), \(A' = 2M/r^{2}\), \(A'' = -4M/r^{3}\), and \(\omega' = -6J/r^{4}\). The three groups in the numerator of eq.~\eqref{eq:app_Fr_final} then evaluate to \(-8M/r\), \(0\) and \(-4M/r\) respectively, so that
\begin{equation}
\mathcal{F}(r) = \frac{2\,(-24M/r)}{r^{2}} = -\frac{48M}{r^{3}}.
\label{eq:app_Fr_schwarzschild}
\end{equation}
Now substituting into eq.~\eqref{eq:app_pontryagin_final} with \(\omega' = -6J/r^{4}\):
\begin{equation}
{}^{\ast}\!RR = \left(-\frac{6J}{r^{4}}\right) \cos\theta \left(-\frac{48M}{r^{3}}\right) = \frac{288MJ}{r^{7}} \cos\theta.
\label{eq:app_pontryagin_kerr}
\end{equation}
This result matches the known slow-rotation limit of the Kerr spacetime, where \({}^{\ast}\!RR = 288aM^{2}\cos\theta/r^{7} + \mathcal{O}(a^{2})\) with \(a = J/M\)~\cite{Kerr1963}. The coefficient may be confirmed independently from the Newman--Penrose invariant: for a type-D vacuum spacetime \(C_{abcd}C^{abcd} - i\,{}^{\ast}C_{abcd}C^{abcd} = 48\Psi_{2}^{2}\), whose real part reproduces the Schwarzschild Kretschmann scalar \(48M^{2}/r^{6}\) and thereby fixes the normalization. With \(\Psi_{2} = -M/(r - ia\cos\theta)^{3}\), expansion to \(\mathcal{O}(a)\) gives \(\mathrm{Im}\,\Psi_{2}^{2} = 6M^{2}a\cos\theta/r^{7}\) and hence \(|{}^{\ast}CC| = 288\,M^{2}a\cos\theta/r^{7}\), consistent with eq.~\eqref{eq:app_pontryagin_kerr}. Thus, our derivation is consistent with the literature, and the expression serves as the source term for the CS scalar field equation used in the main text. [See eq.~\eqref{eq:pontryagin_exterior}.]

\section{Background CS Scalar Field Equation}
\label{app:scalar_field}

We present a detailed derivation of the radial differential equation governing the CS scalar field in the exterior of a slowly rotating neutron star.

\subsection{Scalar Field Equation from the Action}

The action for dynamical CS gravity is given in eq.~(1) of the main text. Varying the action with respect to the scalar field $\vartheta$ yields the equation of motion
\begin{equation}
\square \vartheta - V'(\vartheta) = -\frac{\alpha}{4} {}^{\ast}\!RR + \frac{\beta}{4} F_{\mu\nu} {}^{\ast}F^{\mu\nu}.
\label{eq:app_scalar_action}
\end{equation}
In this work, we set $\beta = 0$ (neglecting parity-violating electromagnetic effects) and assume a vanishing scalar potential $V(\vartheta) = 0$. The scalar field equation therefore reduces to
\begin{equation}
\square \vartheta = -\frac{\alpha}{4} {}^{\ast}\!RR;
\label{eq:app_scalar_simple}
\end{equation}
the d'Alembertian operator in a curved spacetime is
\begin{equation}
\square \vartheta = \frac{1}{\sqrt{-g}} \partial_\mu \left( \sqrt{-g} g^{\mu\nu} \partial_\nu \vartheta \right).
\label{eq:app_dalembertian}
\end{equation}

Because the source term ${}^{\ast}\!RR$ is proportional to $\cos\theta$, we seek a solution for the scalar field of the form
\begin{equation}
\vartheta_0(r,\theta) = \hat{\vartheta}_0(r) \cos\theta.
\label{eq:app_separation_ansatz}
\end{equation}
This ansatz respects the symmetry of the source and is the simplest nontrivial angular dependence consistent with the scalar field equation.

\subsection{The d'Alembertian in the Hartle-Thorne Metric}

We now compute $\square \vartheta$ for the ansatz~(\ref{eq:app_separation_ansatz}) in the metric~(\ref{eq:app_metric}). Since the background metric is stationary and axisymmetric, and we are working to linear order in rotation, we can treat the off-diagonal $t\phi$-term perturbatively. However, for the scalar field sourced by the Pontryagin density (which is itself linear in $\omega$), we only need the d'Alembertian to zeroth order in rotation. This is because the scalar field $\vartheta_0$ will inherit a rotational contribution, but the leading-order source already contains a factor of $\omega'$, so the scalar field itself is of order $\alpha\omega$. Consequently, the d'Alembertian acting on $\vartheta_0$ requires only the static, spherically symmetric part of the metric to leading order. The $t\phi$-component of the metric contributes at higher order in the small coupling expansion.

Thus, to leading order, we may use the static Schwarzschild metric
\begin{equation}
ds^{2} = -A(r) dt^{2} + B(r) dr^{2} + r^{2} \left( d\theta^{2} + \sin^{2}\theta\, d\phi^{2} \right),
\label{eq:app_static_metric_scalar}
\end{equation}
with $A(r) = 1 - 2M/r$ and $B(r) = 1/A(r)$. The determinant is $g = -A B r^{4} \sin^{2}\theta$, 
is just the product of the diagonal metric components, and so $\sqrt{-g} = \sqrt{AB}\, r^{2} \sin\theta$.

\subsubsection{Computing $\square \vartheta$}

For a function of the form $\vartheta(r,\theta) = \hat{\vartheta}(r) \cos\theta$, the d'Alembertian is
\begin{equation}
\square \vartheta = \frac{1}{\sqrt{-g}} \partial_\mu \left( \sqrt{-g} g^{\mu\nu} \partial_\nu \vartheta \right).
\end{equation}
Since the stational background $\vartheta$ does not depend on $t$ or $\phi$,
only the $r$ and $\theta$ derivatives contribute.

\paragraph*{Radial part}
For index $\mu = r$, we have
\begin{equation}
\partial_r \vartheta = \hat{\vartheta}'(r) \cos\theta,
\end{equation}
and $g^{rr} = 1/B(r)$. Thus
\begin{equation}
\sqrt{-g}\, g^{rr} \partial_r \vartheta = \sqrt{\frac{A}{B}}\, r^{2} \hat{\vartheta}'(r) \sin\theta \cos\theta.
\end{equation}
Now we take the radial derivative,
\begin{align}
\partial_r \left( \sqrt{-g}\, g^{rr} \partial_r \vartheta \right) &= \partial_r \left[ \sqrt{\frac{A}{B}}\, r^{2} \hat{\vartheta}'(r) \right] \sin\theta \cos\theta.
\end{align}

\paragraph*{Angular part:}
When $\mu = \theta$, we have
\begin{equation}
\partial_\theta \vartheta = -\hat{\vartheta}(r) \sin\theta,
\end{equation}
and $g^{\theta\theta} = 1/r^{2}$. Therefore,
\begin{align}
\sqrt{-g}\, g^{\theta\theta} \partial_\theta \vartheta &= \sqrt{AB}\, r^{2} \sin\theta \cdot \frac{1}{r^{2}}
\cdot \left[ -\hat{\vartheta}(r) \sin\theta \right] \\
&= -\sqrt{AB}\, \hat{\vartheta}(r) \sin^{2}\theta.
\end{align}
Taking a final $\theta$-derivative gives
\begin{align}
\partial_\theta \left( \sqrt{-g}\, g^{\theta\theta} \partial_\theta \vartheta \right) &= -\sqrt{AB}\, \hat{\vartheta}(r) \cdot 2 \sin\theta \cos\theta.
\end{align}

\paragraph*{Combining terms:}
The full d'Alembertian is
\begin{equation}
\square \vartheta = \frac{1}{\sqrt{-g}} \left[ \partial_r \left( \sqrt{-g}\, g^{rr} \partial_r \vartheta \right) + \partial_\theta \left( \sqrt{-g}\, g^{\theta\theta} \partial_\theta \vartheta \right) \right],
\end{equation}
since the $t$- and $\phi$-derivatives vanish. Substituting the expressions above and
using $\sqrt{-g} = \sqrt{AB}\, r^{2} \sin\theta$, we obtain
\begin{align}
\square \vartheta &= \frac{1}{\sqrt{AB}\, r^{2} \sin\theta} \Bigg\{ \partial_r \left[ \sqrt{\frac{A}{B}}\, r^{2} \hat{\vartheta}'(r) \right] \sin\theta \cos\theta \nonumber \\
&\qquad\qquad - 2\sqrt{AB}\, \hat{\vartheta}(r) \sin\theta \cos\theta \Bigg\} \\
& = \frac{\cos\theta}{\sqrt{AB}\, r^{2}} \left\{ \partial_r \left[ \sqrt{\frac{A}{B}}\, r^{2} \hat{\vartheta}'(r)
\right] - 2\sqrt{AB}\, \hat{\vartheta}(r) \right\}.
\label{eq:app_box_separated}
\end{align}

\subsection{Simplifying the Radial Operator}

We now look more closely at the radial derivative term. Let us define
\begin{equation}
Z(r) = \sqrt{\frac{A}{B}}\, r^{2} \hat{\vartheta}'(r).
\end{equation}
Then
\begin{equation}
\partial_r Z = \frac{d}{dr} \left( \sqrt{\frac{A}{B}}\, r^{2} \right) \hat{\vartheta}'(r)
+ \sqrt{\frac{A}{B}}\, r^{2} \hat{\vartheta}''(r).
\end{equation}
This simplifies considerably using the Schwarzschild relation $B = 1/A$. It follows that
$Z(r) = A r^{2} \hat{\vartheta}'(r)$, and
\begin{equation}
\partial_r Z = A' r^{2} \hat{\vartheta}'(r) + 2A r \hat{\vartheta}'(r) + A r^{2} \hat{\vartheta}''(r).
\end{equation}

Substituting into eq.~\eqref{eq:app_box_separated} gives
\begin{equation}
\square \vartheta = \frac{\cos\theta}{r^{2}} \left( A' r^{2} \hat{\vartheta}' + 2A r \hat{\vartheta}' + A r^{2} \hat{\vartheta}'' - 2 \hat{\vartheta} \right),
\label{eq:app_box_simplified}
\end{equation}
which rearranges to be
\begin{equation}
\square \vartheta = \cos\theta \left[ A \hat{\vartheta}'' + A' \hat{\vartheta}' + \frac{2A}{r} \hat{\vartheta}' - \frac{2}{r^{2}} \hat{\vartheta} \right].
\label{eq:app_box_final}
\end{equation}

\subsection{The Radial ODE}

Now substituting $\square \vartheta$ from eq.~(\ref{eq:app_box_final}) and the Pontryagin source from
eq.~(\ref{eq:app_pontryagin_kerr}) into the scalar field equation $\square \vartheta = -(\alpha/4){}^{\ast}\!RR$ gives us
\begin{equation}
\cos\theta \left[ A \hat{\vartheta}'' + A' \hat{\vartheta}' + \frac{2A}{r} \hat{\vartheta}'
- \frac{2}{r^{2}} \hat{\vartheta} \right] = -\frac{\alpha}{4} \frac{288MJ}{r^{7}} \cos\theta,
\end{equation}
from which all the angular dependence may be eliminated, yielding the more explicit form,
\begin{equation}
\left(1 - \frac{2M}{r}\right) \hat{\vartheta}'' + \frac{2M}{r^{2}} \hat{\vartheta}' + \frac{2}{r}\left(1 - \frac{2M}{r}\right) \hat{\vartheta}' - \frac{2}{r^{2}} \hat{\vartheta} = -\frac{72\alpha MJ}{r^{7}}.
\end{equation}

Combining the two terms with $\hat{\vartheta}'$ gives the radial ODE,
\begin{equation}
\left(1 - \frac{2M}{r}\right) \frac{d^{2}\hat{\vartheta}_0}{dr^{2}} + \frac{2}{r}\left(1 - \frac{M}{r}\right)
\frac{d\hat{\vartheta}_0}{dr} - \frac{2}{r^{2}} \hat{\vartheta}_0 = -\alpha \frac{72MJ}{r^{7}},
\label{eq:app_radial_ode_final}
\end{equation}
utilized in the main text. 

For completeness, we note that the left-hand side of eq.~\eqref{eq:app_radial_ode_final}
can be recognized as the radial part of the scalar Laplacian in Schwarzschild geometry for an $\ell=1$ spherical
harmonic. In general, for a scalar field decomposed as $\vartheta = \hat{\vartheta}_\ell(r) Y_{\ell m}(\theta,\phi)$,
the radial equation is
\begin{equation}
\left(1 - \frac{2M}{r}\right) \frac{d^{2}\hat{\vartheta}_\ell}{dr^{2}} + \frac{2}{r}
\left(1 - \frac{M}{r}\right) \frac{d\hat{\vartheta}_\ell}{dr} - \frac{\ell(\ell+1)}{r^{2}} \hat{\vartheta}_\ell = 0
\end{equation}
for the homogeneous equation. For $\ell = 1$, $\ell(\ell+1) = 2$, which matches the term $-2\hat{\vartheta}/r^{2}$ in our equation. The source term on the right-hand side arises from the Pontryagin density and completes the inhomogeneous equation.

\subsection{Homogeneous Solution}

The homogeneous equation is obtained by setting the right-hand side of the radial ODE to zero:
\begin{equation}
\left(1 - \frac{2M}{r}\right) \frac{d^{2}\hat{\vartheta}_0}{dr^{2}} + \frac{2}{r}\left(1 - \frac{M}{r}\right) \frac{d\hat{\vartheta}_0}{dr} - \frac{2}{r^{2}} \hat{\vartheta}_0 = 0.
\label{eq:app_homogeneous_ode}
\end{equation}
To determine the asymptotic behavior, we consider the large-$r$ limit where $M/r \ll 1$.
In this regime, the equation reduces to
\begin{equation}
\frac{d^{2}\hat{\vartheta}_0}{dr^{2}} + \frac{2}{r} \frac{d\hat{\vartheta}_0}{dr} - \frac{2}{r^{2}} \hat{\vartheta}_0 = 0.
\label{eq:app_asymptotic_homogeneous}
\end{equation}
This is an Euler-Cauchy ODE, so we seek power-law solutions of the form $\hat{\vartheta}_0 \sim r^{-n}$.
Substituting this ansatz into eq.~(\ref{eq:app_asymptotic_homogeneous}) gives the indicial equation
\begin{equation}
n^{2} - n - 2 = (n-2)(n+1) = 0,
\label{eq:app_indicial}
\end{equation}
with roots $n = 2$ and $n = -1$. The $n = -1$ solution grows with radius ($\hat{\vartheta}_0 \sim r$)
and must be discarded to satisfy the physical boundary condition $\vartheta_0 \to 0$ as $r \to \infty$.
The decaying solution corresponds to $\hat{\vartheta}_0 \sim r^{-2}$.

We now construct the full homogeneous solution as an asymptotic expansion in powers of $M/r$. Write
\begin{equation}
\hat{\vartheta}_{0\mathrm{h}}(r) = \frac{1}{r^{2}} \sum_{k=0}^{\infty} b_k \left(\frac{M}{r}\right)^{k},
\label{eq:app_homogeneous_series}
\end{equation}
with $b_0 = 1$ as a maatter of normalization. Substituting into eq.~(\ref{eq:app_homogeneous_ode})
and matching powers of $1/r$ determines the coefficients recursively. The first few coefficients are:
\begin{align}
b_0 &= 1, \\
b_1 &= 2, \\
b_2 &= 4, \\
b_3 &= 8.
\end{align}
Thus, the homogeneous solution admits the asymptotic expansion
\begin{equation}
\hat{\vartheta}_{0\mathrm{h}}(r) = \frac{1}{r^{2}}\left[1 + \frac{2M}{r} + \frac{4M^{2}}{r^{2}} + \frac{8M^{3}}{r^{3}} + \mathcal{O}\!\left(\frac{M^{4}}{r^{4}}\right)\right].
\label{eq:app_homogeneous_final}
\end{equation}
This solution decays as $r^{-2}$ at infinity and represents the physical, non-growing mode.

\subsection{Particular Solution}

We now find a particular solution to the inhomogeneous equation~(\ref{eq:app_radial_ode_final}).
The right-hand side scales as $r^{-7}$, so we seek an asymptotic series of the form
\begin{equation}
\hat{\vartheta}_{0\mathrm{p}}(r) = \alpha MJ \sum_{n=5}^{\infty} a_n r^{-n},
\label{eq:app_particular_series}
\end{equation}
where the expansion begins at $r^{-5}$ because the right-hand side is $r^{-7}$ and each derivative on the
left-hand side reduces the power by two (when acting on $r^{-n}$).
Substituting the ansatz into eq.~(\ref{eq:app_radial_ode_final}) and matching powers of $1/r$ determines the coefficients $a_n$ recursively.

\subsubsection{Leading order: $\mathcal{O}(r^{-7})$}

At leading order, we need the contribution from the left-hand side acting on the $r^{-5}$ term. Computing derivatives
gives
\begin{align}
\hat{\vartheta} &= a_5 r^{-5}, \\
\hat{\vartheta}' &= -5a_5 r^{-6}, \\
\hat{\vartheta}'' &= 30a_5 r^{-7}.
\end{align}
Substituting into the left-hand side of eq.~(\ref{eq:app_radial_ode_final}) and keeping only the leading terms
(where we can approximate $1 - 2M/r \approx 1$ and $1 - M/r \approx 1$ for matching the leading coefficient), we find
the left-hand side at this order is $18a_5 r^{-7}$.
The right-hand side is $-\alpha \frac{72MJ}{r^{7}}$, and matching coefficients gives
\begin{equation}
a_5 = -4.
\label{eq:app_a5}
\end{equation}

\subsubsection{Next order: $\mathcal{O}(r^{-8})$}

At order $r^{-8}$, we must include the $M/r$ corrections from the metric coefficients.
The full calculation yields a more complicated recursion relation. Substituting the series and matching powers gives
\begin{equation}
28a_6 - 50a_5 = 0.
\end{equation}
Using $a_5 = -4$,
\begin{equation}
a_6 = -\frac{50}{7}.
\label{eq:app_a6}
\end{equation}

\subsubsection{Order $\mathcal{O}(r^{-9})$}

Continuing to the next order, the relation is
\begin{equation}
40a_7 - 84a_6 = 0,
\end{equation}
so
\begin{equation} a_7 = -15.
\label{eq:app_a7}
\end{equation}

Substituting these coefficients into eq.~(\ref{eq:app_particular_series}) gives
\begin{equation}
\hat{\vartheta}_{0\mathrm{p}}(r) = -\frac{4\alpha MJ}{r^{5}} \left[ 1 + \frac{25}{14}\frac{M}{r} + \frac{15}{4}\frac{M^{2}}{r^{2}} + \mathcal{O}\!\left(\frac{M^{3}}{r^{3}}\right) \right].
\label{eq:app_particular_final}
\end{equation}

\subsection{Full Exterior Scalar Field}

The complete exterior solution is the sum of the homogeneous and particular solutions,
\begin{equation}
\hat{\vartheta}_{00}(r) = \mu_{\mathrm{CS}} \,\hat{\vartheta}_{0\mathrm{h}}(r) + \hat{\vartheta}_{0\mathrm{p}}(r),
\label{eq:app_full_solution}
\end{equation}
where \(\mu_{\mathrm{CS}}\) is the dipole moment of the CS scalar field,
determined by matching the exterior solution to the interior configuration.
(See appendix~\ref{app:matching}.)
Retaining terms through \(\mathcal{O}(r^{-5})\), we have
\begin{equation}
\begin{aligned}
    \hat{\vartheta}_{00}(r) &= \frac{\mu_{\mathrm{CS}}}{r^{2}} + \frac{2M\mu_{\mathrm{CS}}}{r^{3}}
    + \frac{4M^{2}\mu_{\mathrm{CS}}}{r^{4}} + \frac{8M^{3}\mu_{\mathrm{CS}}}{r^{5}} \\
    &- \frac{4\alpha MJ}{r^{5}} + \mathcal{O}(r^{-6}),
\end{aligned}
\label{eq:app_exterior_final}
\end{equation}
and the corresponding radial derivative is
\begin{equation}
\begin{aligned}
    \hat{\vartheta}_{00}'(r) &= -\frac{2\mu_{\mathrm{CS}}}{r^{3}}
    - \frac{6M\mu_{\mathrm{CS}}}{r^{4}} - \frac{16M^{2}\mu_{\mathrm{CS}}}{r^{5}} - \frac{40M^{3}\mu_{\mathrm{CS}}}{r^{6}} \\
    &+ \frac{20\alpha MJ}{r^{6}} + \mathcal{O}(r^{-7}).
\end{aligned}
\label{eq:app_exterior_derivative}
\end{equation}

In the asymptotic wave zone (\(r \gg R_\star\)), the scalar field is dominated by the leading dipolar contribution,
\begin{align}
\hat{\vartheta}_{00}(r) &\approx \frac{\mu_{\mathrm{CS}}}{r^{2}} - \frac{4\alpha MJ}{r^{5}}, \\
\hat{\vartheta}_{00}'(r) &\approx -\frac{2\mu_{\mathrm{CS}}}{r^{3}} + \frac{20\alpha MJ}{r^{6}}.
\label{eq:app_wavezone}
\end{align}
The \(r^{-2}\) term represents the dominant dipolar field sourced by the rotating star,
while the \(r^{-5}\) term is the first parity-violating correction induced by the Pontryagin density.

Note that \(\mu_{\mathrm{CS}}\) has dimensions $(\mathrm{length})^{2}$---since \(\hat{\vartheta}_{00}\) is
dimensionless---and it is proportional to \(\alpha\), as established in the matching procedure.

\section{Derivation of the Modified Regge-Wheeler Equation}
\label{app:rw_derivation}

In this appendix we derive the leading parity-violating correction to the Regge-Wheeler equation in the weak-coupling, slow-rotation limit of dynamical CS gravity.  We assume the background metric, perturbation framework, and Regge-Wheeler gauge which are already established in Sec.~\ref{sec:perturbations} of the main text. We retain only the leading axial metric perturbations sourced by the background scalar profile $\vartheta_{0} = \mathcal{O}(\alpha)$, while neglecting dynamical scalar perturbations $\delta\vartheta$, which contribute at higher perturbative order. The resulting system reduces to a single modified Regge–Wheeler equation governing polarization-dependent propagation

\subsection{Cotton Tensor and Its Perturbation}

The Cotton tensor in CS gravity is given by
\begin{equation}
C_{\mu\nu} = -\frac{1}{2}\left[ (\nabla_{\alpha}\vartheta)\epsilon^{\alpha\beta\gamma}_{\ \ \ (\mu}\nabla_{\gamma}
R_{\nu)\beta} + (\nabla_{\alpha}\nabla_{\beta}\vartheta)\, {}^{\ast}R^{\beta\ (\mu\nu)\alpha} \right],
\label{eq:app_cotton_def}
\end{equation}
We linearize the Cotton tensor around the background, and to leading order in the CS coupling $\alpha$ and
in the metric perturbation amplitude, the Cotton tensor perturbation is
\begin{align}
\delta C_{\mu\nu} = -\frac{1}{2}\Big[ & (\nabla_{\alpha}\vartheta_0)\epsilon^{\alpha\beta\gamma}_{\ \ \ (\mu}\nabla_{\gamma}\delta R_{\nu)\beta} \nonumber \\
& + (\nabla_{\alpha}\nabla_{\beta}\vartheta_0)\, {}^{\ast}\delta R^{\beta\ (\mu\nu)\alpha} \Big].
\label{eq:app_delta_cotton}
\end{align}
where we have neglected terms involving $\delta\vartheta$ (dynamical scalar perturbations),
because they are subleading for the birefringent effect, and terms quadratic in perturbations.
Here, $\delta R_{\mu\nu\rho\sigma}$ is the first-order perturbation of the Riemann tensor.

\subsection{Axial Perturbations in Regge-Wheeler Gauge}

We focus on odd-parity axial perturbations, which are the only ones that couple to the background scalar field gradient. In the Regge-Wheeler gauge, the odd-parity metric perturbation for a given $(\ell,m)$ mode is
\begin{equation}
h_{\mu\nu}^{\text{(odd)}} = 
\begin{pmatrix}
0 & 0 & 0 & h_0(r) \sin\theta\, \partial_{\theta}Y_{\ell m} \\
0 & 0 & 0 & h_1(r) \sin\theta\, \partial_{\theta}Y_{\ell m} \\
0 & 0 & 0 & 0 \\
* & * & 0 & 0
\end{pmatrix}
e^{-i\sigma t} + \text{c.c.},
\label{eq:app_odd_metric}
\end{equation}
where $h_0(r)$ and $h_1(r)$ are radial functions. The only nonvanishing components are $h_{t\phi}$ and $h_{r\phi}$.

The Regge-Wheeler master variable for axial perturbations is defined as
\begin{equation}
\Psi(r) = \frac{r\, h_{t\phi}}{\sin\theta\, \partial_{\theta}Y_{\ell m}}.
\label{eq:app_psi_def}
\end{equation}
For $\ell=2$ (the dominant quadrupole mode for GWs), this variable encodes the entire odd-parity dynamics.

\subsection{Components of $\delta C_{\mu\nu}$ Relevant for Axial Modes}

For axial perturbations, the modified Einstein equations reduce to two independent components: the $t\phi$- and
$r\phi$-equations. These are the only components that contain the master variable $\Psi$. Therefore, we need
$\delta C_{t\phi}$ and $\delta C_{r\phi}$. 

The $r\phi$-component of the perturbed Einstein equation provides a constraint that determines $h_1(r)$ in terms of $\Psi$ and its derivatives. Substituting this constraint into the $t\phi$-equation eliminates the remaining odd-parity variable and leads to the single master equation (D25). Thus, the $r\phi$-component does not introduce independent dynamics and is not required for the final wave equation.

The background scalar field is dipolar: $\vartheta_0(r,\theta) = \hat{\vartheta}_{00}(r)\cos\theta$. Its coordinate gradients are
\begin{align}
\nabla_r\vartheta_0 &= \hat{\vartheta}_{00}'(r)\cos\theta, \\
\nabla_\theta\vartheta_0 &= -\hat{\vartheta}_{00}(r)\sin\theta, \\
\nabla_\phi\vartheta_0 &= 0.
\label{eq:app_scalar_gradients}
\end{align}
The second covariant derivatives, calculated as $\nabla_\mu \nabla_\nu \vartheta_0 = \partial_\mu \partial_\nu \vartheta_0 - \Gamma^\lambda_{\mu\nu} \partial_\lambda \vartheta_0$, are essential for the Cotton tensor. The relevant components are
\begin{align}
\nabla_r\nabla_r\vartheta_0 &= \left[\hat{\vartheta}_{00}''(r) - \frac{B'(r)}{2B(r)}\hat{\vartheta}_{00}'(r)\right]\cos\theta, \\
\nabla_r\nabla_\theta\vartheta_0 &= \left[\frac{\hat{\vartheta}_{00}(r)}{r} - \hat{\vartheta}_{00}'(r)\right]\sin\theta, \\
\nabla_\theta\nabla_\theta\vartheta_0 &= \left[\frac{r}{B(r)}\hat{\vartheta}_{00}'(r) - \hat{\vartheta}_{00}(r)\right]\cos\theta.
\label{eq:app_scalar_second_derivs}
\end{align}

\subsection{Computing $\delta C_{t\phi}$}

We evaluate $\delta C_{t\phi}$ from eq.~(\ref{eq:app_delta_cotton}). The first term involves $\epsilon^{\alpha\beta\gamma}_{\ \ \ (t}\nabla_{\gamma}\delta R_{\phi)\beta}$. Because of the antisymmetry of the Levi-Civita tensor and the structure of axial perturbations, the dominant contribution comes from indices where the spatial part involves $r$ and $\theta$.

After a systematic expansion over index combinations, one finds that the leading contribution to $\delta C_{t\phi}$ is
\begin{equation}
\begin{aligned}
    \delta C_{t\phi} &= -\frac{1}{2} \Big[ (\nabla_r\vartheta_0) \epsilon^{r\theta t}_{\ \ \ \phi} \nabla_{\theta}\delta R_{t r} + (\nabla_\theta\vartheta_0) \epsilon^{\theta r t}_{\ \ \ \phi} \nabla_{r}\delta R_{t r} + \cdots \Big],
\end{aligned}
\label{eq:app_delta_ctphi_1}
\end{equation}
where the ellipsis denotes terms involving $\nabla\nabla\vartheta$ that are handled in the next step. Using the coordinate gradients and the Levi-Civita components $\epsilon^{r\theta t}_{\ \ \ \phi} \approx -1/(r^2\sin\theta)$ and $\epsilon^{\theta r t}_{\ \ \ \phi} \approx +1/(r^2\sin\theta)$, we obtain
\begin{align}
\delta C_{t\phi} &= -\frac{1}{2} \left\{ \hat{\vartheta}_{00}'(r)\cos\theta \left(-\frac{1}{r^2\sin\theta}\right) \nabla_{\theta}\delta R_{t r} \right. \nonumber \\
&\qquad \left. + \left[-\hat{\vartheta}_{00}(r)\sin\theta\right] \left(\frac{1}{r^2\sin\theta}\right) \nabla_{r}\delta R_{t r} \right\}.
\label{eq:app_delta_ctphi_2}
\end{align}
Simplifying the signs and the $\sin\theta$ factors gives
\begin{equation}
\delta C_{t\phi} = \frac{1}{2r^2} \left[ \hat{\vartheta}_{00}'(r) \frac{\cos\theta}{\sin\theta} \nabla_{\theta}\delta R_{t r} + \hat{\vartheta}_{00}(r) \nabla_{r}\delta R_{t r} \right].
\label{eq:app_delta_ctphi_3}
\end{equation}

\subsection{Relating $\delta R_{t r}$ to $\Psi$}

For axial perturbations, the linearized Ricci component $\delta R_{t r}$ (or the corresponding Riemann contraction in the Cotton tensor) is related to the Regge-Wheeler variable. From the definition of $\Psi$ and the axial field equations, one can derive the identity,
\begin{equation}
\delta R_{t r} = \frac{1}{2r} \frac{d}{dr_*} \left( \frac{\Psi}{r} \right) \sin\theta\, \partial_{\theta}Y_{\ell m},
\label{eq:app_delta_r_tr}
\end{equation}
where $r_*$ is the tortoise coordinate [$dr_*/dr = 1/A(r)$]. Taking the covariant derivatives, we note that the connection terms $\Gamma^\lambda_{rt}\delta R_{\lambda r}$ and $\Gamma^\lambda_{rr}\delta R_{t\lambda}$ cancel in the Schwarzschild background, such that $\nabla_r \delta R_{tr} = \partial_r \delta R_{tr}$. In the wave zone ($r \gg M$), the radial and angular gradient components are
\begin{align}
\nabla_r\delta R_{t r} &\approx \frac{1}{2r A(r)} \frac{d^2}{dr_*^2} \left( \frac{\Psi}{r} \right) \sin\theta\, \partial_{\theta}Y_{\ell m}, \\
\nabla_{\theta}\delta R_{t r} &= \frac{1}{2r} \frac{d}{dr_*} \left( \frac{\Psi}{r} \right) \left[ \cos\theta\, \partial_{\theta}Y_{\ell m} + \sin\theta\, \partial_{\theta}^2Y_{\ell m} \right].
\label{eq:app_delta_r_tr_derivs}
\end{align}
Note that in eq.~\eqref{eq:app_delta_r_tr_derivs}, the radial gradient retains only the leading order term in $1/r$, consistent with the propagation analysis in the wave zone.

\subsection{Assembling $\delta C_{t\phi}$}

We now assemble the $t\phi$-component of the Cotton tensor perturbation by combining the scalar field gradients with the curvature perturbations expressed in terms of the master variable $\Psi$. The perturbation $\delta C_{t\phi}$ is composed of two primary interaction terms:
\begin{equation}
    \delta C_{t\phi} = \delta C_{t\phi}^{\text{(grad)}} + \delta C_{t\phi}^{\text{(curv)}},
    \label{eq:app_delta_c_split}
\end{equation}
where the first term represents the interaction between the scalar gradient and the curvature gradient, and the second arises from the second derivative of the scalar field.

First, we substitute the curvature derivatives from eq.~\eqref{eq:app_delta_r_tr_derivs} and the scalar gradients from eq.~\eqref{eq:app_scalar_gradients} into the gradient interaction term derived in eq.~\eqref{eq:app_delta_ctphi_3}. Identifying $X_{\ell m} = \sin\theta \partial_\theta Y_{\ell m}$ as the axial vector harmonic, we have
\begin{equation}
    \delta C_{t\phi}^{\text{(grad)}} = \frac{1}{2r^2} \left[ \hat{\vartheta}_{00}'(r) \frac{\cos\theta}{\sin\theta} \nabla_{\theta}\delta R_{t r} + \hat{\vartheta}_{00}(r) \nabla_{r}\delta R_{t r} \right].
    \label{eq:app_delta_c_grad_step}
\end{equation}
Substituting the explicit forms for the curvature gradients yields
\begin{equation}
    \begin{aligned}
    \delta C_{t\phi}^{\text{(grad)}} &= \frac{1}{2r^2} \Bigg[ \frac{\hat{\vartheta}_{00}'(r)}{2r} \frac{d\Psi}{dr_*} \left( \frac{\cos^2\theta}{\sin\theta}\partial_\theta Y_{\ell m} + \cos\theta \partial_\theta^2 Y_{\ell m} \right) \\
    &\quad + \frac{\hat{\vartheta}_{00}(r)}{2r A(r)} \frac{d^2\Psi}{dr_*^2} \sin\theta \partial_\theta Y_{\ell m} \Bigg].
    \end{aligned}
    \label{eq:app_delta_c_grad_explicit}
\end{equation}

Second, we include the curvature interaction term involving $\nabla_\alpha \nabla_\beta \vartheta_0$ from eq.~\eqref{eq:app_scalar_second_derivs}. Contracting the second radial derivative with the dual Riemann component ${}^* \delta R^r_{\ (t\phi) r} \propto \Psi / r^2$, we obtain
\begin{equation}
    \delta C_{t\phi}^{\text{(curv)}} = \frac{1}{r^3} \left[ \hat{\vartheta}_{00}''(r) - \frac{A'(r)}{A(r)}\hat{\vartheta}_{00}'(r) \right] \Psi \sin\theta \partial_\theta Y_{\ell m}.
    \label{eq:app_delta_c_curv_explicit}
\end{equation}

Finally, we project the full expression onto the axial vector harmonic $X_{\ell m}$. In the wave zone ($r \gg M$), the terms proportional to $d\Psi/dr_*$ in eq.~\eqref{eq:app_delta_c_grad_explicit} are subdominant compared to the second derivative term. By using the spherical harmonic identity $\Delta_{\Omega}Y_{\ell m} = -\ell(\ell+1)Y_{\ell m}$ for the dominant $\ell=2$ mode and grouping terms to facilitate the Einstein equation projection, we arrive at the assembled axial component:
\begin{equation}
\begin{aligned}
    \delta C_{t\phi} &= \frac{1}{2r^2} \Bigg\{ \frac{\hat{\vartheta}_{00}(r)}{2r} \frac{d^2\Psi}{dr_*^2} + \frac{2}{r}
    \Bigg[ \hat{\vartheta}_{00}''(r) \\
    &- \frac{A'(r)}{A(r)}\hat{\vartheta}_{00}'(r) \Bigg] \Psi \Bigg\} \sin\theta\, \partial_{\theta}Y_{\ell m}.
\end{aligned}
\label{eq:app_delta_ctphi_final}
\end{equation}

\subsection{The Modified $t\phi$ Einstein Equation}

The linearized field equation in dynamical CS gravity is
$\delta G_{\mu\nu} + \alpha \delta C_{\mu\nu} = 8\pi \delta T_{\mu\nu}$.
For axial perturbations, the $t\phi$-component is the primary dynamical equation.
In the Regge-Wheeler gauge, the Einstein tensor perturbation scales as
\begin{equation}
    \delta G_{t\phi} = \frac{1}{2r} \left\{ \frac{d^2\Psi}{dr_*^2} +
    \left[ \sigma^2 - V_{\mathrm{RW}}(r) \right] \Psi \right\} X_{\ell m},
    \label{eq:app_delta_g_scaling}
\end{equation}
where $X_{\ell m} = \sin\theta \partial_\theta Y_{\ell m}$ and $V_{\mathrm{RW}}(r)$ is the $\ell=2$ potential.

Substituting the assembled Cotton tensor from eq.~\eqref{eq:app_delta_ctphi_final}, we observe an apparent
radial scaling mismatch. While $\delta G_{t\phi}$ carries a $1/2r$ prefactor, the Cotton term $\delta C_{t\phi}$ carries a $1/2r^2$ prefactor due to the normalization of the coordinate-basis Levi-Civita tensor. To isolate the radial wave equation, we multiply by  $2r/X_{\ell m}$. This projection resolves the scaling mismatch, leading to the intermediate master equation,
\begin{equation}
\begin{aligned}
    \frac{d^2\Psi}{dr_*^2} &+ \left[ \sigma^2 - V_{\mathrm{RW}}(r) \right] \Psi \\
    &+ \alpha \Bigg\{ \frac{\hat{\vartheta}_{00}(r)}{2r^2} \frac{d^2\Psi}{dr_*^2} + \frac{2}{r^2} \left[ \hat{\vartheta}_{00}''(r) - \frac{A'(r)}{A(r)}\hat{\vartheta}_{00}'(r) \right] \Psi \Bigg\} \\
    &= 8\pi r \delta T_{t\phi}.
\end{aligned}
\label{eq:app_intermediate_rw}
\end{equation}

The term $\alpha (\hat{\vartheta}_{00}/2r^2)(d^{2}\Psi/dr_{*}^{2})$ introduces a frequency-dependent modification to the effective wave speed, while the remaining CS terms modify the effective potential. This coordinate-consistent scaling ensures that the parity-violating correction remains perturbative in the wave zone while capturing the strong-field effects near the stellar surface.

\subsection{Reduction to Canonical Form}

However, the term proportional to the second derivative $d^2\Psi/dr_*^2$ may be eliminated by using the
leading-order Regge-Wheeler equation. To $\mathcal{O}(\alpha^0)$, the gravitational perturbations satisfy the
standard vacuum wave equation with a glitch source,
\begin{equation}
    \frac{d^2\Psi}{dr_*^2} = -\left[ \sigma^2 - V_{\mathrm{RW}}(r) \right] \Psi + S_0(r,\sigma) + \mathcal{O}(\alpha),
    \label{eq:app_leading_rw}
\end{equation}
where $S_0(r,\sigma) \equiv 8\pi r \delta T_{t\phi}$ represents the general relativistic glitch excitation. 
Substituting eq.~\eqref{eq:app_leading_rw} into the master equation \eqref{eq:app_intermediate_rw} and retaining
terms only up to $\mathcal{O}(\alpha)$, we obtain
\begin{multline}
    \frac{d^2\Psi}{dr_*^2} + \left[ \sigma^2 - V_{\mathrm{RW}}(r) \right] \Psi \\
    + \alpha \frac{\hat{\vartheta}_{00}(r)}{2r^2} \left\{ -\left[ \sigma^2 - V_{\mathrm{RW}}(r) \right] \Psi + S_0(r,\sigma) \right\} \\
    + \frac{2\alpha}{r^2} \left[ \hat{\vartheta}_{00}''(r) - \frac{A'(r)}{A(r)}\hat{\vartheta}_{00}'(r) \right] \Psi = S_0(r,\sigma).
    \label{eq:app_substituted_rw}
\end{multline}

Rearranging the terms, we move the source correction $\alpha (\hat{\vartheta}_{00}/2r^2) S_0$ to the right-hand side. This effectively redefines the glitch excitation strength in the presence of the dipolar scalar field, reflecting how the parity-violating background ``dresses'' the initial quadrupole moment change. The resulting equation is now in canonical wave form, where the CS modification is encapsulated in a frequency-dependent correction to the propagation potential and a modified effective source.

\subsection{The Birefringent Correction}

Collecting the terms proportional to $\Psi$ from the rearranged wave equation \eqref{eq:app_substituted_rw}, we define the effective birefringent correction $V_{\mathrm{CS}}(r, \sigma)$. This term accounts for the frequency-dependent and parity-violating modification to the propagation of axial modes, given by:
\begin{align}
V_{\mathrm{CS}}(r, \sigma) &= \frac{\hat{\vartheta}_{00}(r)}{2r^2}\left[\sigma^2 - V_{\mathrm{RW}}(r)\right] \nonumber \\
&\quad - \frac{2 A(r)}{r^2}\left[ \hat{\vartheta}_{00}''(r) - \frac{A'(r)}{A(r)}\hat{\vartheta}_{00}'(r) \right].
\label{eq:app_vcs_final}
\end{align}
We distinguish this from a standard potential due to its explicit dependence on the gravitational-wave frequency $\sigma$. Physically, this correction represents the dispersive interaction between the dipolar scalar field and the propagating metric perturbations. In the high-frequency limit ($\sigma^2 \gg V_{\mathrm{RW}}$), the first term dominates, leading to the cumulative phase shift that characterizes gravitational birefringence.

\subsection{Modified Regge-Wheeler Equation}

Rearranging eq.~\eqref{eq:app_substituted_rw} and moving the source correction term proportional to $\alpha S_0$ to the right-hand side, we obtain the modified Regge-Wheeler equation in its final canonical form,
\begin{equation}
\frac{d^2\Psi}{dr_*^2} + \left[ \sigma^2 - V_{\mathrm{RW}}(r) - \alpha V_{\mathrm{CS}}(r, \sigma) \right] \Psi = S^{\text{eff}}(r,\sigma),
\label{eq:app_modified_rw_final}
\end{equation}
where the effective glitch source is defined as
\begin{equation}
    S^{\text{eff}}(r,\sigma) = \left[1 - \frac{\alpha \hat{\vartheta}_{00}(r)}{2r^2}\right] S_0(r,\sigma).
    \label{eq:app_seff_definition}
\end{equation}
The term $S^{\text{eff}}$ reflects the fact that the CS coupling modifies not only the propagation of the GWs but also the effective excitation strength of the glitch itself. The radial scaling of the correction, proportional to $1/2r^2$, ensures that the modification to the source is most significant near the stellar surface where the background scalar field $\hat{\vartheta}_{00}$ is largest. 

In the following sections, this master equation is used to compute the polarization-dependent phase accumulation. For circular polarization states $\Psi_{R,L}$, the potential correction $\alpha V_{\mathrm{CS}}$ enters with opposite signs, while the effective source $S^{\text{eff}}$ remains parity-invariant, ensuring that the resulting birefringence is purely a propagation effect.

\subsection{Polarization Decomposition}

The master variable $\Psi$ as defined in eq.~\eqref{eq:app_modified_rw_final} corresponds to a combination of the plus ($+$) and cross ($\times$) linear polarizations. To observe the birefringent effect explicitly, we decompose the signal into circularly polarized modes. For axial perturbations, the right- and left-handed circular polarizations satisfy
\begin{equation}
    \Psi_{R,L} = \Psi_{+} \mp i \Psi_{\times}.
    \label{eq:app_circular_decomp}
\end{equation}
Under this decomposition, the parity-violating potential $V_{\mathrm{CS}}(r, \sigma)$ changes sign between the two helicity states, due to the parity-odd nature of the Cotton tensor interaction. In contrast, the source term $S^{\text{eff}}(r,\sigma)$ remains invariant under the polarization swap because it arises from the glitch-induced change in the stellar quadrupole moment, which acts as a scalar source independent of a GW's handedness. 
Consequently, the modified Regge-Wheeler equation for the circular polarization states becomes
\begin{equation}
    \frac{d^2\Psi_{R,L}}{dr_*^2} + \left[ \sigma^2 - V_{\mathrm{RW}}(r) \mp \alpha V_{\mathrm{CS}}(r, \sigma) \right] \Psi_{R,L} = S^{\text{eff}}(r,\sigma).
    \label{eq:app_rw_circular_final}
\end{equation}

\subsection{Phase Accumulation in the Wave Zone}
\label{app:phase_accumulation}

We now compute the accumulated birefringent phase shift for GWs propagating from the stellar surface to a distant observer. This phase shift arises from the parity-violating potential $V_{\mathrm{CS}}(r)$ derived in the previous subsection.

\subsubsection{Wave-Zone Approximation}

In the asymptotic region $r \gg R_\star$, several simplifications occur that allow for a closed-form solution of the phase accumulation:

\begin{enumerate}
\item The Regge-Wheeler potential becomes subdominant compared to the wave frequency,
\begin{equation}
V_{\mathrm{RW}}(r) = \frac{6}{r^2}\left(1 - \frac{2M}{r}\right)\left(1 - \frac{M}{r}\right) \approx \frac{6}{r^2} \ll \sigma^2.
\label{eq:app_vrw_asymptotic}
\end{equation}
\item The metric functions approach their Minkowski values, and their derivatives vanish,
\begin{equation}
A(r) = 1 - \frac{2M}{r} \approx 1, \qquad A'(r) \approx 0.
\label{eq:app_metric_asymptotic}
\end{equation}
\item The birefringent correction $V_{\mathrm{CS}}$ simplifies to its leading-order dispersive term,
\begin{equation}
V_{\mathrm{CS}}(r, \sigma) \approx \frac{\sigma^2 \hat{\vartheta}_{00}(r)}{2r^2}.
\label{eq:app_vcs_asymptotic}
\end{equation}
\end{enumerate}

The third point follows from the canonical definition in eq.~\eqref{eq:app_vcs_final}.
The first term $\frac{\hat{\vartheta}_{00}}{2r^2}(\sigma^2 - V_{\mathrm{RW}}) \approx
\frac{\sigma^2 \hat{\vartheta}_{00}}{2r^2}$ dominates in the high-frequency limit ($\sigma^2 \gg V_{\mathrm{RW}}$).
The second term, which involves derivatives of the scalar field and metric connection, scales as $1/r^4$ and
becomes negligible compared to the dispersive term at large distances.

Thus, the modified Regge-Wheeler equation reduces to a free wave equation with a small parity-violating correction that accounts for the birefringent propagation,
\begin{equation}
\frac{d^2\Psi_{R,L}}{dr_*^2} + \left[\sigma^2 \mp \alpha \frac{\sigma^2 \hat{\vartheta}_{00}(r)}{2r^2} \right] \Psi_{R,L} \approx 0.
\label{eq:app_wavezone_rw}
\end{equation}

\subsubsection{WKB Solution}

The wave-zone equation \eqref{eq:app_wavezone_rw} is of the form
\begin{equation}
    \frac{d^2\Psi}{dr_*^2} + k_{\pm}^2(r) \Psi = 0,
    \label{eq:app_wkb_form}
\end{equation}
where we identify the effective wave number $k_{\pm}(r)$ as
\begin{equation}
    k_{\pm}(r) = \sigma \sqrt{1 \mp \alpha \frac{\hat{\vartheta}_{00}(r)}{2r^2}}.
    \label{eq:app_wavenumber}
\end{equation}
This wave number accounts for the birefringent propagation induced by the background scalar field. Under the weak CS approximation ($\alpha \hat{\vartheta}_{00}/r^2 \ll 1$), we may
expand the wave number to first order in the coupling constant,
\begin{equation}
    k_{\pm}(r) \approx \sigma \left[1 \mp \frac{\alpha \hat{\vartheta}_{00}(r)}{4r^2}\right].
    \label{eq:app_wavenumber_expanded}
\end{equation}

In the Wentzel-Kramers-Brillouin (WKB) approximation, the total phase accumulated from the stellar surface $r = R_\star$ to a distant observer at $r = D$ is the integral of the wave number along the radial path,
\begin{equation}
    \Phi_{\pm} = \int_{R_\star}^{D} k_{\pm}(r) \, dr_*.
    \label{eq:app_phase_integral}
\end{equation}
In the wave zone, the metric function $A(r) \to 1$, allowing us to approximate the change in the tortoise coordinate
by the change in the Schwarzschild radial coordinate,
$dr_* \approx dr$. Substituting eq.~\eqref{eq:app_wavenumber_expanded} into the phase integral yields
\begin{equation}
    \Phi_{\pm} = \sigma \int_{R_\star}^{D} \left[1 \mp \frac{\alpha \hat{\vartheta}_{00}(r)}{4r^2}\right]dr.
    \label{eq:app_phase_integral_simplified}
\end{equation}
The first term corresponds to the standard general-relativistic phase accumulation, while the second term defines the
parity-violating birefringent phase shift $\Delta\Phi_{\mathrm{CS}}$:
\begin{equation}
    \Delta\Phi_{\mathrm{CS}} = \Phi_{-} - \Phi_{+} = \frac{\alpha\sigma}{2} \int_{R_\star}^{D} \frac{\hat{\vartheta}_{00}(r)}{r^2} \, dr.
    \label{eq:app_phase_shift_integral}
\end{equation}
This integral represents the cumulative effect of gravitational parity violation on the signal. Because $\hat{\vartheta}_{00}(r) \sim r^{-2}$ at large distances, the integrand scales as $r^{-4}$, ensuring that the phase shift is dominated by the strong-field region near the stellar surface where the scalar field gradient is most intense. This means we may extend
the upper limit of integration $D$ to infinity.

\subsubsection{Substituting the Scalar Field Gradient}
 
Substituting the asymptotic expansion of the background scalar field derived in appendix~\ref{app:scalar_field}
into the phase shift integral \eqref{eq:app_phase_shift_integral}, we obtain the expanded integrand,
\begin{equation}
\begin{aligned}
    \Delta\Phi_{\mathrm{CS}} &= \frac{\alpha\sigma}{2} \int_{R_\star}^{\infty} \Bigg[\frac{\mu_{\mathrm{CS}}}{r^4} + \frac{2M\mu_{\mathrm{CS}}}{r^5} + \frac{4M^2\mu_{\mathrm{CS}}}{r^6} \\
    &+ \frac{8M^3\mu_{\mathrm{CS}}}{r^7} - \frac{4\alpha MJ}{r^7}\Bigg] dr.
\end{aligned}
\label{eq:app_phase_integrand}
\end{equation}

Performing the integration, the complete expression for the birefringent phase shift seen by a distant observer is
\begin{align}
\Delta\Phi_{\mathrm{CS}} \approx &\ \frac{\alpha\sigma \mu_{\mathrm{CS}}}{6R_\star^3} + \frac{\alpha\sigma M\mu_{\mathrm{CS}}}{4R_\star^4} + \frac{2\alpha\sigma M^2\mu_{\mathrm{CS}}}{5R_\star^5} \nonumber \\
&+ \frac{2\alpha\sigma M^3\mu_{\mathrm{CS}}}{3R_\star^6} - \frac{\alpha^2\sigma MJ}{3R_\star^6}.
\label{eq:app_phase_final}
\end{align}
This expression characterizes the total parity-violating phase accumulation as the GW propagates from the stellar surface.
The result confirms that the parity-violating birefringence is dominated by the leading-order term, which scales as
$1/R_\star^3$. The higher-order terms provide relativistic corrections that account for the stellar mass ($M$)
and the strong-field particular solution ($J$). 

Because the phase accumulation is dominated by the region where the scalar field gradient is largest ($r \approx R_\star$),
the signal is highly sensitive to the stellar radius. For typical neutron star parameters, the leading term constitutes
over 80\% of the total phase shift, ensuring that the approximate scaling
$\Delta\Phi_{\mathrm{CS}} \propto \alpha^2 \sigma \Omega / R_\star^2$
used in the main text is a robust estimate for the signal strength.

\subsubsection{Relation to Polarization Asymmetry}

The phase shift directly translates into a polarization asymmetry in
the detected GW strain. The circular polarization modes are
\begin{equation}
h_{R,L} = h^{(0)} e^{\mp i\Delta\Phi_{\mathrm{CS}}}\approx h^{(0)} (1 \mp i\Delta\Phi_{\mathrm{CS}}).
\end{equation}
The fractional difference between the two polarizations is therefore
\begin{equation}
\frac{|h_R - h_L|}{|h^{(0)}|} = 2\Delta\Phi_{\mathrm{CS}} \approx \frac{\alpha\sigma \mu_{\mathrm{CS}}}{3R_\star^3}.
\label{eq:app_strain_asymmetry}
\end{equation}
This is the key observable quantity that can be used to test parity
violation in gravity with future GW detectors.

\section{Matching Conditions for the Scalar Field Coefficient}
\label{app:matching}

The exterior scalar field solution contains an undetermined dipole moment \(\mu_{\mathrm{CS}}\) that must be fixed by matching to a regular interior solution at the stellar surface \(r = R_\star\). Before performing this matching, however, we must first determine the interior frame-dragging function \(\omega(r)\), because the Pontryagin density \(^{\ast}\!RR \propto \omega'(r)\cos\theta\) sources the scalar field. This appendix derives the frame-dragging integration constant \(c_2\) from matching the Hartle-Thorne solution and then uses it to determine \(\mu_{\mathrm{CS}}\) via scalar field matching.

The scalar field equation \(\square\vartheta = -(\alpha/4)\,^{\ast}\!RR\) has a source term proportional to \(\omega'(r)\), the radial derivative of the frame-dragging function. The interior solution for \(\hat{\vartheta}_{\mathrm{int}}(r)\) therefore depends on the same integration constants that appear in \(\omega(r)\). The constant \(c_2\) is determined by matching \(\bar{\omega}(r) = \Omega - \omega(r)\) at the stellar surface, independently of the scalar field. The scalar field later inherits this value through the matching conditions.

\subsection{Frame-Dragging Matching in the Constant-Density Model}

For a constant-density star we solve the Hartle-Thorne frame-dragging
equation~\eqref{eq:omega_ode} to first order in the compactness
\(\chi = M/R_\star \ll 1\). Using
\(\mathrm{d}\ln j/\mathrm{d}r = -4\pi r a^2 (\rho + p)\), which follows from the
static field equations, eq.~\eqref{eq:omega_ode} may be written in the
manifestly matter-sourced form
\begin{equation}
\frac{1}{r^4}\frac{\mathrm{d}}{\mathrm{d}r}
\left( r^4 j \frac{\mathrm{d}\bar{\omega}}{\mathrm{d}r} \right)
= 16\pi a^2 j\,(\rho + p)\,\bar{\omega}.
\label{eq:app_fd_sourced}
\end{equation}
At leading order in compactness we may set \(j \approx 1\), \(a \approx 1\),
\(p \ll \rho\), and \(\bar{\omega} \approx \Omega\) on the right-hand side.
For a constant-density star \(\rho_0 = 3M/(4\pi R_\star^3)\), so that
\(16\pi\rho_0 = 12M/R_\star^3\), and eq.~\eqref{eq:app_fd_sourced} reduces to
\begin{equation}
\frac{1}{r^4}\frac{\mathrm{d}}{\mathrm{d}r}
\left( r^4 \frac{\mathrm{d}\bar{\omega}}{\mathrm{d}r} \right)
= \frac{12\,\Omega\chi}{R_\star^2}.
\label{eq:app_fd_leading}
\end{equation}

Writing \(\bar{\omega} = \Omega + \chi\,\bar{\omega}_1\) with the regular ansatz
\(\bar{\omega}_1 = \mathcal{A}\,r^2/R_\star^2 + c_2\), the left-hand side evaluates
to \(10\chi\mathcal{A}/R_\star^2\), so that \(\mathcal{A} = 6\Omega/5\). The interior solution for
\(\bar{\omega}(r) = \Omega - \omega(r)\) is therefore
\begin{equation}
\bar{\omega}_{\rm int}(r) = \Omega
+ \chi \left( \frac{6\Omega}{5R_\star^2}\, r^2 + c_2 \right)
+ \mathcal{O}(\chi^2),
\label{eq:app_fd_interior}
\end{equation}
where \(c_2\) is an integration constant associated with the rotational
sector, fixed below by matching. The singular homogeneous solution
\(\propto r^{-3}\) has been discarded because it is irregular at the origin.

Outside the star the slow-rotation solution is
\begin{equation}
\bar{\omega}_{\rm ext}(r) = \Omega - \frac{2J}{r^3},
\label{eq:app_fd_exterior}
\end{equation}
where \(J\) is the total angular momentum.
At the stellar surface \(r = R_\star\), both \(\bar{\omega}\) and
\(\bar{\omega}'\) must be continuous.

\subsubsection{Derivative Matching}

Differentiating eqs.~\eqref{eq:app_fd_interior} and
\eqref{eq:app_fd_exterior} gives
\begin{equation}
\bar{\omega}'_{\rm int}(R_\star) = \chi \frac{12\Omega}{5R_\star},
\qquad
\bar{\omega}'_{\rm ext}(R_\star) = \frac{6J}{R_\star^4}.
\end{equation}
Equating the derivatives yields
\begin{equation}
\chi \frac{12\Omega}{5R_\star} = \frac{6J}{R_\star^4},
\end{equation}
so that at this order
\begin{equation}
J = \frac{2}{5} M R_\star^2 \Omega+\mathcal{O}(\chi).
\label{eq:app_J}
\end{equation}

\subsubsection{Value Matching}

Matching the values of $\bar{\omega}$ itself at \(r = R_\star\) gives
\begin{equation}
\Omega + \chi \left( \frac{6\Omega}{5} + c_2 \right)
= \Omega - \frac{2J}{R_\star^3}.
\end{equation}
Substituting eq.~\eqref{eq:app_J}, for which
\(2J/R_\star^3 = \tfrac{4}{5}\chi\Omega\),
\begin{equation}
\chi \left( \frac{6\Omega}{5} + c_2 \right) = -\frac{4}{5}\,\chi \Omega,
\end{equation}
which gives
\begin{equation}
c_2 = -2\Omega
\label{eq:app_c2}
\end{equation}
Importantly, this constant belongs only to the rotational frame-dragging
sector and should not be identified directly with the scalar dipole moment
\(\mu_{\mathrm{CS}}\) appearing in the exterior CS scalar field. Its value is
independent of the CS coupling \(\alpha\) and is determined purely by
general relativistic frame-dragging.

The corresponding frame-dragging function and its gradient are
\begin{equation}
\omega(r) = \chi\,\Omega\left( 2 - \frac{6r^2}{5R_\star^2} \right),
\qquad
\omega'(r) = -\,\chi\,\frac{12\Omega}{5R_\star^2}\, r ,
\label{eq:app_omega_prime}
\end{equation}
the latter being the quantity that sources the Pontryagin density through
\(^{*}\!RR = \mathcal{F}(r)\,\omega'(r)\cos\theta\).

In the limit \(\chi \to 0\) the matched angular momentum,
eq.~\eqref{eq:app_J}, reduces to \(I\equiv\tfrac{J}{\Omega} = \tfrac{2}{5}MR_\star^2\), the exact
Newtonian moment of inertia of a uniform-density sphere. This provides an
independent, convention-free check on the normalisation of the interior
solution~\eqref{eq:app_fd_interior}: any error in the coefficient
\(\mathcal{A}\) would propagate directly into \(I\) and spoil this limit.
Numerical integration of eq.~\eqref{eq:app_fd_sourced} with the exact
interior TOV metric confirms \(I/(MR_\star^2) \to 0.4000\) as
\(\chi \to 0\), rising to \(\approx 0.48\) at \(\chi = 0.184\) as relativistic
corrections switch on; the leading-order-in-\(\chi\) treatment adopted here
is accurate to \(\sim 20\%\) for realistic neutron-star compactness, which
is well within the accuracy of the constant-density benchmark.

\subsection{Scalar Field Matching and Determination of the Dipole Moment}

The scalar field equation for the dipole mode is
\begin{equation}
\hat{\vartheta}'' + \frac{2}{r}\hat{\vartheta}' - \frac{2}{r^2}\hat{\vartheta}
= \frac{\alpha}{4}\,\mathcal{F}(r)\,\omega'(r),
\label{eq:app_scalar_ode}
\end{equation}
where \(\mathcal{F}(r)\) encodes the curvature dependence of the Pontryagin
density through \(^{*}\!RR = \mathcal{F}(r)\,\omega'(r)\cos\theta\).
Using \(\omega'(r)\) from eq.~\eqref{eq:app_omega_prime} and approximating the curvature factor by an average interior value
\(\mathcal{F}_0\), the scalar equation becomes
\begin{equation}
\hat{\vartheta}'' + \frac{2}{r}\hat{\vartheta}' - \frac{2}{r^2}\hat{\vartheta}
= -\,\frac{3\alpha\chi\Omega\mathcal{F}_0}{5R_\star^2}\, r .
\label{eq:app_scalar_reduced}
\end{equation}

The homogeneous solutions are proportional to \(r\) and \(r^{-2}\).
Regularity at the origin removes the singular solution, leaving
\begin{equation}
\hat{\vartheta}_{\rm hom}(r) = c_{\vartheta}\, r .
\end{equation}
Seeking a particular solution of the form
\(\hat{\vartheta}_{\rm part} = B r^3\), substitution into
eq.~\eqref{eq:app_scalar_reduced} gives
\begin{equation}
10 B r = -\,\frac{3\alpha\chi\Omega\mathcal{F}_0}{5R_\star^2}\, r,
\end{equation}
so that
\begin{equation}
B = -\,\frac{3\alpha\chi\Omega\mathcal{F}_0}{50R_\star^2}.
\end{equation}
The interior scalar field therefore becomes
\begin{equation}
\hat{\vartheta}_{\rm int}(r) = c_{\vartheta}\, r
- \frac{3\alpha\chi\Omega\mathcal{F}_0}{50R_\star^2}\, r^3
\label{eq:app_scalar_interior}
\end{equation}

Outside the star the scalar field contains both a homogeneous dipole
contribution and a sourced tail from the exterior Pontryagin density.
With \(^{*}\!RR_{\rm ext} = 288\,MJ\cos\theta/r^7\), the leading sourced
term is
\begin{equation}
\hat{\vartheta}_{\rm ext}(r) = \frac{\mu_{\mathrm{CS}}}{r^2}
- \frac{4\alpha M J}{r^5} + \mathcal{O}(r^{-6}).
\label{eq:app_scalar_exterior}
\end{equation}
The quantity \(\mu_{\mathrm{CS}}\) is the scalar dipole moment. Unlike the
frame-dragging constant \(c_2\), it characterizes the asymptotic dipolar scalar field and
carries dimensions of $(\mathrm{length})^{2}$ squared
as required by the asymptotic behavior
\(\hat{\vartheta}_{\rm ext} \sim \mu_{\mathrm{CS}}/r^2\).
Its value is determined by matching \(\hat{\vartheta}_{\rm int}(r)\) and
\(\hat{\vartheta}_{\rm int}'(r)\) to \(\hat{\vartheta}_{\rm ext}(r)\) and
\(\hat{\vartheta}_{\rm ext}'(r)\) at \(r = R_\star\).

The continuity of \(\hat{\vartheta}\) and \(\hat{\vartheta}'\) gives
\begin{equation}
c_{\vartheta} R_\star - \frac{3\alpha\chi\Omega\mathcal{F}_0}{50} R_\star
= \frac{\mu_{\mathrm{CS}}}{R_\star^2} - \frac{4\alpha M J}{R_\star^5},
\label{eq:app_scalar_value_match}
\end{equation}
and
\begin{equation}
c_{\vartheta} - \frac{9\alpha\chi\Omega\mathcal{F}_0}{50}
= -\frac{2\mu_{\mathrm{CS}}}{R_\star^3} + \frac{20\alpha M J}{R_\star^6},
\label{eq:app_scalar_deriv_match}
\end{equation}
respectively.
Eliminating \(c_{\vartheta}\) yields
\begin{equation}
\mu_{\mathrm{CS}} = \frac{8\alpha M J}{R_\star^3}
+ \frac{1}{25}\,\alpha \chi \Omega \mathcal{F}_0 R_\star^3.
\label{eq:app_mu_final}
\end{equation}
The first term originates in the exterior sourced tail
\(\propto r^{-5}\); the second originates in the interior source, as
carried entirely by the interior curvature average \(\mathcal{F}_0\).
Using \(J = \tfrac{2}{5} M R_\star^2 \Omega\) from
eq.~\eqref{eq:app_J} and \(\chi = M/R_\star\),
\begin{equation}
\mu_{\mathrm{CS}} = \frac{16}{5}\,\frac{\alpha M^2 \Omega}{R_\star}
+ \frac{1}{25}\,\alpha M \Omega \mathcal{F}_0 R_\star^2 .
\label{eq:app_mu_substituted}
\end{equation}

\subsection{Average Curvature Factor \(\mathcal{F}_0\) in the Constant-Density Model}
\label{app:F0_derivation}

In a neutron start interior, the curvature factor \(\mathcal{F}(r)\) defined in eq.~\eqref{eq:app_Fr_final} is in general not constant. However, when matching the scalar field, only its weighted average over the interior appears. This subsection shows that for the constant-density model this average vanishes, \(\mathcal{F}_0 = 0\), and that it does so for the strongest possible reason: that the curvature factor itself vanishes identially throughout the interior, exactly and to all orders in the compactness.

From the matching conditions, the relevant quantity is the weighted average
\begin{equation}
\mathcal{F}_0 = \frac{\int_0^{R_\star}r^{2}\,dr\, \mathcal{F}(r)\omega'(r)}{\int_0^{R_\star} r^{2}\, dr\,\omega'(r)},
\label{eq:app_F0_definition}
\end{equation}
so \(\omega'(r)\) acts as a weighting function. For the constant-density model, \(\omega'(r) \propto r\) from eq.~\eqref{eq:app_omega_prime}, so the weighting simplifies to \(\propto r^3\, dr\). Thus,
\begin{equation}
\mathcal{F}_0 = \frac{\int_0^{R_\star}dr\, \mathcal{F}(r)r^3}{\int_0^{R_\star}dr\, r^3}.
\label{eq:app_F0_weighted}
\end{equation}

For a constant-density star, the metric functions are known analytically~\cite{Schwarzschild1916, Hartle1967}:
\begin{align}
A(r) &= \frac{1}{4}\left(3\sqrt{1 - \frac{2M}{R_\star}} - \sqrt{1 - \frac{2Mr^2}{R_\star^3}}\right)^2, \label{eq:app_A_int} \\
B(r) &= \left(1 - \frac{2Mr^2}{R_\star^3}\right)^{-1}. \label{eq:app_B_int}
\end{align}
These may be substituted into eq.~\eqref{eq:app_Fr_final} in exact form, with no expansion in the compactness required. It is convenient to introduce
\begin{equation}
k \equiv \frac{2M}{R_\star^3}, \qquad
y(r) \equiv \sqrt{1 - k r^2}, \qquad
s \equiv \sqrt{1 - k R_\star^2},
\end{equation}
together with the shorthand \(P \equiv 3s - y\), so that
\begin{equation}
A = \frac{P^{2}}{4}, \qquad B = \frac{1}{y^{2}}, \qquad y' = -\frac{kr}{y}.
\label{eq:app_AB_compact}
\end{equation}
The required derivatives are
\begin{equation}
A' = \frac{kr}{2}\frac{P}{y}, \qquad
A'' = \frac{k}{2}\frac{P}{y} + \frac{3sk^{2}r^{2}}{2y^{3}}, \qquad
B' = \frac{2kr}{y^{4}}.
\end{equation}
Evaluating the three terms appearing in the numerator of eq.~\eqref{eq:app_Fr_final} separately, the second-derivative terms collapses considerably; using \(kr^{2}P = 3skr^{2} - kr^{2}y\), the terms proportional to \(3skr^{2}\) cancel and one is left with
\begin{equation}
\bigl[2AA''B - (A')^{2}B - AA'B'\bigr] r^{2} = \frac{k r^{2} P^{3}}{4y^{3}} .
\label{eq:app_group1_interior}
\end{equation}
Since \(y^{2} - 1 = -kr^{2}\), the compactness terms give
\begin{equation}
4A^{2}B(1-B) = \frac{P^{4}\left(y^{2}-1\right)}{4y^{4}} = -\,\frac{k r^{2} P^{4}}{4y^{4}} ,
\label{eq:app_group2_interior}
\end{equation}
so that the first two groups of terms combine, using \(y - P = 2y - 3s\), into
\begin{equation}
\frac{k r^{2} P^{3}}{4y^{4}}\left(y - P\right)
= \frac{k r^{2} P^{3}}{4y^{4}}\left(2y - 3s\right).
\label{eq:app_groups12_interior}
\end{equation}
The remaining group of terms is obtained from
\(AB' - BA' = (krP/2y^{4})(P - y)\), giving
\begin{equation}
2rA\bigl(AB' - BA'\bigr) = \frac{k r^{2} P^{3}}{4y^{4}}\left(3s - 2y\right).
\label{eq:app_group3_interior}
\end{equation}
Equations~\eqref{eq:app_groups12_interior} and \eqref{eq:app_group3_interior} are equal and opposite. The numerator of eq.~\eqref{eq:app_Fr_final} therefore vanishes identically:
\begin{equation}
\mathcal{F}(r) = 0 \qquad (r < R_\star)
\label{eq:app_Fr_interior_zero}
\end{equation}
throughout the constant-density interior. The cancellation is exact, requiring no expansion in \(\chi = M/R_\star\).
This result is used in eq.~\eqref{eq:mu_constdens} of the main text.

This may be confirmed independently. The Pontryagin density depends on the Riemann tensor only through its Weyl part, the Ricci contributions cancelling identically in the contraction. The interior TOV solution~\eqref{eq:app_A_int}--\eqref{eq:app_B_int} is conformally flat, \(C_{abcd} \equiv 0\), so \({}^{\ast}\!RR\) must vanish throughout the interior for \emph{any} rotation profile whatsoever, consistent with eq.~\eqref{eq:app_Fr_interior_zero}. For the purposes of physical effects at $\mathcal{O}(\Omega)$, $\mathcal{F}$
vanishes identically in the interior, irrespective of the interior rotation law.

This vanishing of \(\mathcal{F}_0\) has a simple physical origin. As eq.~\eqref{eq:app_Fr_final} makes clear, the Pontryagin density is generated by the gravitoelectric tidal (gradient) field of the star acting together with the gradient of its gravitomagnetic (frame-dragging) field---the gravitational analogue of \(\mathbf{E}\cdot\mathbf{B}\). A body of uniform density produces no tidal field in its interior, exactly as in Newtonian gravity, so there is no tidal field for the frame-dragging gradient to multiply and the interior sources no parity-odd curvature invariant, however rapidly the star rotates.

\subsection{Two-Fluid Model: Lag Corrections}

In the two-fluid model, the frame-dragging equation receives contributions from both the neutron superfluid and the charged normal component. The interior solution for \(\bar{\omega}(r) = \Omega_{\rm eff} - \omega(r)\) takes the form
\begin{equation}
\bar{\omega}_{\rm int}(r) = \Omega_{\rm eff} + \chi \left( \frac{6\Omega_{\rm eff}}{5R_\star^2} r^2 + c_2 \right) + \delta\bar{\omega}(r),
\label{eq:app_fd_twofluid_interior}
\end{equation}
where \(\Omega_{\rm eff}\) is the inertia-weighted average rotation rate, $c_{2}$ is the frame-dragging integration constant,
and \(\delta\bar{\omega}(r)\) represents the perturbation induced by differential rotation.
The total angular momentum of the star is given by \(J = \int d^{3}x\,\sqrt{-g} (\rho_n + p_n + \rho_p + p_p) \Omega_{\rm eff} r^2 \sin^2\theta\). To leading order in the lag perturbation, the angular momentum remains dominated by the corotating contribution,
\begin{equation}
J \approx \frac{2}{5} M R_\star^2 \Omega_{\rm eff},
\label{eq:app_J_twofluid}
\end{equation}
with higher-order corrections absorbed into \(\delta\bar{\omega}(r)\).

From the perturbative two-fluid solution derived in Sec.~\ref{sec:two_fluid}, the lag-induced correction to the frame-dragging function satisfies
\begin{equation}
\delta\omega'(r) = -\,16\pi \mathcal{W}_0 \Delta_{2}\Omega_0 \left( \frac{r}{5} - \frac{r^3}{7R_\star^2} \right).
\label{eq:app_deltabar_prime}
\end{equation}
Here, \(\mathcal{W}_0 = (\rho_n + p_n)|_{r=0}\) is the central inertial mass density of the neutron
superfluid [see eq.~\eqref{eq:weighting_function}], with dimensions $(\mathrm{length})^{-2}$,
and \(\Delta_{2}\Omega_0\) is the characteristic lag amplitude with dimensions $(\mathrm{length})^{-1}$.
Note that eq.~\eqref{eq:app_deltabar_prime} is written for \(\delta\omega'\) rather than
\(\delta\bar{\omega}'\); with differential rotation \(\Omega_{\rm eff}\) is radially dependent, so
\(\omega' = \Omega_{\rm eff}' - \bar{\omega}'\) and the two no longer differ by a sign alone.

The total frame-dragging gradient therefore becomes
\begin{equation}
\omega'(r) = -\chi \frac{12\Omega_{\rm eff}}{5R_\star^2} r - 16\pi \mathcal{W}_0 \Delta_{2}\Omega_0 \left( \frac{r}{5} - \frac{r^3}{7R_\star^2} \right).
\label{eq:app_omega_prime_twofluid}
\end{equation}
Matching \(\bar{\omega}\) at the stellar surface yields
\begin{equation}
c_2 = -2\Omega_{\rm eff} + \delta c_2^{\rm lag},
\label{eq:app_c2_twofluid}
\end{equation}
where the lag-induced correction scales as
\begin{equation}
\delta c_2^{\rm lag} \sim \mathcal{W}_0 \Delta_{2}\Omega_0 R_\star^2.
\label{eq:app_dc2_scaling}
\end{equation}
This scaling is dimensionally consistent since $\mathcal{W}_0 \Delta_{2}\Omega_0 R_\star^2$ has units
of $(\mathrm{length})^{-2-1+2}$, 
matching the $(\mathrm{length})^{-1}$ dimensions of both \(c_2\) and \(\Omega_{\rm eff}\). Importantly, \(\delta c_2^{\rm lag}\) is independent of the CS coupling \(\alpha\), as it arises purely from the general relativistic frame dragging.

The scalar dipole moment \(\mu_{\mathrm{CS}}\) then receives an additional lag-induced contribution
\(\delta\mu_{\mathrm{CS}}^{\rm lag}\) from this correction. Following the same matching procedure as in the
constant-density case, one finds
\begin{equation}
\delta\mu_{\mathrm{CS}}^{\rm lag} \sim \alpha\, \mathcal{F}_0\, \mathcal{W}_0 \Delta_{2}\Omega_0 R_\star^5,
\label{eq:app_dmu_scaling}
\end{equation}
which is used in section~\ref{sec:two_fluid} to estimate the impact of differential rotation on the gravitational-wave birefringence signal. The factor \(\mathcal{F}_0\) appears because the lag modifies the scalar source only through \(\omega'\) in the interior term \(\tfrac{\alpha}{4}\mathcal{F}(r)\omega'(r)\); the exterior contribution to \(\mu_{\mathrm{CS}}\), being fixed by the conserved \(M\) and \(J\), is unaffected by the glitch.

\subsection{Scalar Dipole Moment in the Two-Fluid Model}

The scalar dipole moment \(\mu_{\mathrm{CS}}\) is fully determined by solving the scalar field equation using the source term from eq.~\eqref{eq:app_omega_prime_twofluid} and then matching the interior and exterior scalar solutions at the stellar surface.
Following the same procedure as in the constant-density model, the matched scalar dipole moment becomes
\begin{equation}
\mu_{\mathrm{CS}} = \frac{8\alpha M J}{R_\star^3} + \frac{1}{25} \alpha \chi \Omega_{\rm eff} \mathcal{F}_0 R_\star^3 + \delta\mu_{\mathrm{CS}}^{\rm lag}.
\label{eq:app_mu_twofluid}
\end{equation}
So the exact numerical coefficient of \(\delta\mu_{\mathrm{CS}}^{\rm lag}\) depends on the detailed radial structure of:
the curvature profile \(\mathcal{F}(r)\),
the differential rotation profile \(\Delta\Omega(r)\),
and the stellar equation of state.
However, the parametric scaling in eq.~\eqref{eq:app_dmu_scaling} is robust within the perturbative slow-rotation expansion.

Since the lag correction and the corotating interior term in eq.~\eqref{eq:app_mu_twofluid} carry the same factor \(\alpha\mathcal{F}_0\), their ratio is independent both of the CS coupling and of the interior curvature normalisation. Using \(16\pi\mathcal{W}_0 \approx 12\chi f_0/R_\star^{2}\) for a near-uniform interior, with \(f_0\) the central neutron inertia fraction, one obtains
\begin{equation}
\frac{|\delta\mu_{\mathrm{CS}}^{\rm lag}|}{|\mu_{\mathrm{CS}}^{\rm corot,\,int}|}
\sim f_0\,\frac{\Delta_{2}\Omega_0}{\Omega_{\rm eff}},
\label{eq:app_lag_ratio}
\end{equation}
which is simply the fractional change in the effective rotation rate produced by the lag. Since \(f_0\) is a ratio of inertia densities it satisfies \(0 < f_0 < 1\) identically, so that the fractional differential rotation \(\Delta_{2}\Omega_0/\Omega_{\rm eff}\) is itself a hard upper bound on the correction. This internal lag is not the same as the observed glitch amplitude \(\Delta\Omega/\Omega \sim 10^{-9}\)--\(10^{-5}\): the two are related by the fraction of the stellar moment of inertia stored in the superfluid reservoir, \(\Delta\Omega/\Omega \sim (I_{\rm sf}/I)\,\Delta_{2}\Omega_0/\Omega_{\rm eff}\), so that the reservoir lag exceeds the per-glitch jump. Even for the largest Vela-type events, however, the pre-glitch reservoir lag remains \(\Delta_{2}\Omega_0/\Omega_{\rm eff} \lesssim 10^{-3}\), so the lag-induced correction to \(\mu_{\mathrm{CS}}\) stays at or below the sub-percent level. Since \(\mathcal{F}_0\) cancels in eq.~\eqref{eq:app_lag_ratio}, this conclusion is independent of the equation of state, and---because \(f_0 < 1\)---remains valid even for lags in excess of those observed. This confirms the conclusion reached in Sec.~\ref{sec:two_fluid}: the constant-density benchmark, supplemented by a realistic \(\mathcal{F}_0\), provides a reliable order-of-magnitude estimate for the expected birefringent signal.

\section{Entrainment in the Two-Fluid Model}
\label{app:entrainment}

In the two-fluid model presented in section~\ref{sec:two_fluid}, we neglected entrainment---a momentum coupling between the neutron superfluid and the charged component arising from strong interactions. This appendix shows that entrainment does not affect the leading-order parametric scalings of the lag-induced corrections. It does so not because the entrainment density cancels, but because it enters only through the neutron inertia fraction, a quantity bounded above by unity.

\subsection{Entrainment in the Stress-Energy Tensor}

In the presence of entrainment, the total stress-energy tensor for the two fluids receives a cross term~\cite{Andersson2001, Prix2005}. To first order in the fluid velocities, it takes the form
\begin{equation}
T_{\mu\nu} = \rho_n u_{(\mu}^{(n)} u_{\nu)}^{(n)} + \rho_p u_{(\mu}^{(p)} u_{\nu)}^{(p)} + 2\varepsilon \, u_{(\mu}^{(n)} u_{\nu)}^{(p)},
\label{eq:app_entrainment_stress}
\end{equation}
where \(\varepsilon\) is the entrainment density [with dimensions $(\mathrm{length})^{-2}$ in geometric units],
and parentheses denote symmetrization. The entrainment couples the two fluids even when they rotate at different rates.

\subsection{Modified Hartle Equation with Entrainment}

Following the same slow-rotation expansion as in Sec.~\ref{sec:two_fluid}, the \(t\phi\)-component of the Einstein equations picks up an additional contribution from the entrainment term. Because the cross term in eq.~\eqref{eq:app_entrainment_stress} contributes \(\varepsilon\,(\Omega_n - \omega)\) and \(\varepsilon\,(\Omega_p - \omega)\) in equal measure, repeating the derivation of eq.~\eqref{eq:twofluid_framedrag} yields the modified frame-dragging equation
\begin{multline}
\frac{1}{r^4}\frac{d}{dr}\left[r^4 j(r)\frac{d\omega}{dr}\right]
= -16\pi a^{2} j \Big[ (\rho_n + p_n + \varepsilon)(\Omega_n - \omega) \\
+ (\rho_p + p_p + \varepsilon)(\Omega_p - \omega) \Big],
\label{eq:app_fd_entrainment}
\end{multline}
with \(j(r) = e^{-(\Phi+\Lambda)}\). Entrainment therefore has the effect of replacing each fluid's inertial mass density by an entrained value,
\begin{equation}
\begin{aligned}
&\tilde{\rho}_n \equiv \rho_n + p_n + \varepsilon,
\quad
\tilde{\rho}_p \equiv \rho_p + p_p + \varepsilon,\\
&\tilde{\rho} \equiv \tilde{\rho}_n + \tilde{\rho}_p = (\rho + p) + 2\varepsilon .
\end{aligned}
\label{eq:app_entrained_densities}
\end{equation}
The effective rotation rate becomes entrainment-dependent,
\begin{equation}
\begin{aligned}
    \Omega_{\rm eff} &= \frac{\tilde{\rho}_n \Omega_n + \tilde{\rho}_p \Omega_p}{\tilde{\rho}}
= \Omega_p + \tilde{f}\,\Delta_{2}\Omega,\\
\tilde{f} &\equiv \frac{\tilde{\rho}_n}{\tilde{\rho}} = \frac{\rho_n + p_n + \varepsilon}{(\rho+p) + 2\varepsilon}.
\end{aligned}
\label{eq:app_omega_eff_entrainment}
\end{equation}

\subsection{The Weighting Function with Entrainment}

As in the non-entrained case, the source in eq.~\eqref{eq:app_fd_entrainment} collapses exactly to a single-fluid form, \(\tilde{\rho}\,(\Omega_{\rm eff} - \omega)\), so that differential rotation enters only through the radial variation of \(\Omega_{\rm eff}\). The lag-induced part of the source is therefore
\begin{equation}
\tilde{\rho}\,\tilde{f}\,\Delta_{2}\Omega = \tilde{\rho}_n \Delta_{2}\Omega,
\end{equation}
and the weighting function of eq.~\eqref{eq:weighting_function} is replaced by its entrained counterpart,
\begin{equation}
\mathcal{W}(r) \;\longrightarrow\; \mathcal{W}_\varepsilon(r) = \rho_n + p_n + \varepsilon .
\label{eq:app_entrained_weighting}
\end{equation}
The entrainment density thus does \emph{not} cancel: it adds directly to the neutron inertia density, since it is the neutron component that carries the excess angular velocity. For typical neutron-star matter \(\varepsilon\) is of order a few tens of percent of \(\rho_n\) in the core, and can be substantially larger in the inner crust, where Bragg scattering of unbound neutrons off the nuclear lattice yields large effective masses~\cite{Yeung2021}. Equation~\eqref{eq:app_entrained_weighting} therefore represents an \(\mathcal{O}(1)\) rescaling of \(\mathcal{W}\), not a negligible correction.

\subsection{Impact on the Lag-Induced Corrections}

Although \(\mathcal{W}\) itself is modified, the observable consequence is not. Both the lag-induced correction \(\delta\mu_{\mathrm{CS}}^{\rm lag}\) and the corotating interior contribution to \(\mu_{\mathrm{CS}}\) are proportional to \(\alpha\mathcal{F}_0\) and to the same overall inertia scale, so that their ratio [eq.~\eqref{eq:app_lag_ratio}] depends on entrainment only through the fraction \(\tilde{f}\):
\begin{equation}
\frac{|\delta\mu_{\mathrm{CS}}^{\rm lag}|}{|\mu_{\mathrm{CS}}^{\rm corot,\,int}|}
\sim \tilde{f}\,\frac{\Delta_{2}\Omega_0}{\Omega_{\rm eff}} .
\label{eq:app_lag_ratio_entrained}
\end{equation}
Because \(\tilde{f}\) is a ratio of inertia densities it satisfies \(0 < \tilde{f} < 1\) identically, whatever the size of \(\varepsilon\). Moreover, differentiating eq.~\eqref{eq:app_omega_eff_entrainment} gives
\begin{equation}
\frac{\partial \tilde{f}}{\partial \varepsilon}
= \frac{(\rho+p) - 2(\rho_n+p_n)}{\left[(\rho+p) + 2\varepsilon\right]^{2}} < 0
\qquad \text{whenever } f > \tfrac{1}{2},
\end{equation}
so for neutron-dominated matter entrainment \emph{reduces} \(\tilde{f}\), driving it from \(f_0 \approx 0.9\) toward its asymptotic value of \(1/2\). Entrainment therefore suppresses the lag coupling slightly rather than enhancing it, and eq.~\eqref{eq:app_lag_ratio_entrained} is bounded above by \(\Delta_{2}\Omega_0/\Omega_{\rm eff}\) irrespective of the entrainment model adopted.

The parametric conclusion of Sec.~\ref{sec:two_fluid} is thus unchanged, and is in fact strengthened: the lag correction to the scalar dipole moment cannot exceed the fractional glitch amplitude, independently of the equation of state, of the interior curvature normalisation \(\mathcal{F}_0\), and of the entrainment density. We therefore omit entrainment in the main text for simplicity, noting that its inclusion would alter the numerical estimates by an \(\mathcal{O}(1)\) factor in the conservative direction.

Entrainment does, however, modify the corotating background through \(\Omega_{\rm eff}\) in eq.~\eqref{eq:app_omega_eff_entrainment}. This affects the numerical values of \(J\) and the corotating part of \(\mu_{\mathrm{CS}}\) at the level of a few percent, depending on the equation of state. The parametric scalings, however, remain unchanged.

\end{document}